\documentclass[12pt]{JHEP3}

\newcommand{\ex}{\mathrm{e}}

\usepackage{amsmath,epsfig}
\usepackage{amssymb,amsfonts}
\usepackage{latexsym}
\usepackage[latin1]{inputenc}
\usepackage{slashed}
\usepackage{empheq}
\numberwithin{equation}{section}
\usepackage{dsfont}
\usepackage{cancel}
\usepackage{overpic}
\usepackage{subcaption}
\usepackage{subeqnarray}
\usepackage{xcolor}
\let\normalcolor\relax
\usepackage{graphicx}
\usepackage{amsmath}
\usepackage{longtable}
\usepackage{color}
\usepackage{bbm}

\allowdisplaybreaks

\newcommand{\exclude}[1]{}
\def\nn{\nonumber}

\def\L{\mathcal{L}}

\def\d{\mathrm{d}}

\def\a#1{\alpha_{#1}}

\def\tr{\mathrm{tr}}

\def\beq{\begin{equation}}
\def\eeq{\end{equation}}
\def\be{\begin{equation}}
\def\ee{\end{equation}}
\def\bea{\begin{eqnarray}}
\def\eea{\end{eqnarray}}
\def\bal{\begin{align}}
\def\eal{\end{align}}

\def\td{\tilde}

\def\2b2[#1,#2][#3,#4]{\left( \begin{array}{cc} #1 & #2 \\ #3 & #4 \end{array}
\right)}
\def\3b3[#1,#2,#3][#4,#5,#6][#7,#8,#9]{\left( \begin{array}{ccc} #1 & #2 #3 \\
#4 & #5 & #6\\#7&#8&#9\end{array} \right)}

\newcommand{\ka}{\kappa}

\newcommand\fverb{\setbox\pippobox=\hbox\bgroup\verb}
\newcommand\fverbdo{\egroup\medskip\noindent%
                        \fbox{\unhbox\pippobox}\ }
\newcommand\fverbit{\egroup\item[\fbox{\unhbox\pippobox}]}

\newcommand{\la}{\lambda}

\newcommand{\bear}{\begin{eqnarray}}

\newcommand{\eear}{\end{eqnarray}}
\newcommand{\de}{\partial}
\newcommand{\bsea}{\begin{subeqnarray}}
\newcommand{\esea}{\end{subeqnarray}}
\newbox\pippobox

\def\f{\varphi}
\def\d{\delta}
\def\g{\gamma}

\def\6{\partial}
\def\a{\alpha}
\def\nn{\nonumber}

\def\e{\epsilon}
\def\m{\mu}

\def\s{\sigma}

\def\sp{\;\;\;,\;\;\;}

\def\sq
\def\a{\alpha}
\def\b{\beta}
\def\l{\lambda}

\def\tr{{\rm Tr}}

\def\hri#1#2{\href{http://arxiv.org/abs/#1}{[ArXiv:#1]#2}}
\def\hre#1#2{\href{http://arxiv.org/abs/#1/#2}{[ArXiv:#1/#2]}}
\def\hrj#1#2{\href{https://doi.org/#1}{#2}}

\def\e{\epsilon}

\def\d{\delta}

\def\L{\Lambda}

\def\D{\Delta}

\title{Tachyon-dependent Chern-Simons terms and  the V-QCD Baryon}
\author{
M. J\"arvinen$^\sharp$,  E. Kiritsis$^\natural$$^\flat$, F. Nitti$^\natural$, E. Pr\'eau$^\natural$
~\\
~\\
$^\natural$ \href{http://www.apc.univ-paris7.fr}{Universit\'e Paris Cit\'e, CNRS, Astroparticule et Cosmologie,  F-75006 Paris, France}\\
~\\
$^\flat$ \href{http://hep.physics.uoc.gr}{Crete Center for Theoretical Physics}, Institute for Theoretical and Computational Physics,
Department of Physics,  P.O. Box 2208,\\
University of Crete, 70013, Heraklion, Greece
~\\
~\\
$^\sharp$ \href{http://apctp.org}{Asia Pacific Center for Theoretical Physics}, Pohang 37673, Republic of Korea and Department of Physics, Pohang University of Science and Technology, Pohang 37673, Republic of Korea
}

\preprint{CCTP-2022-4\\ITCP-2022/2\\APCTP Pre2022 - 020
}

\abstract{The structure of the five-dimensional Tachyon-Chern-Simons
  action and its relevance to  single-baryon states in the
  context of  the  V-QCD models for
  holographic QCD with backreacting  flavor are analyzed.
The most general
  form of the Tachyon-Chern-Simons 5-form,  compatible with symmetries and
  flavor anomalies is determined. It is the sum of a non-trivial gauge-invariant 5-dimensional form and a non-invariant closed 5-form that reproduces the flavor anomalies.
      Single-baryon solutions of the
  gravity theory, arising from the
DBI plus Tachyon-Chern-Simons actions are considered. The baryon is realised as a bulk
axial instanton. The baryon ansatz and  the field equations are derived and the boundary conditions are determined, which ensure that the solution has
finite boundary energy and unit baryon charge. The  boundary baryon
number, which is
computed from the universal
(closed) part of the Tachyon-Chern-Simons action, is shown to coincide with the bulk axial instanton number.
}

\begin{document}
\maketitle

\section{Introduction}

The holographic correspondence, \cite{Malda,GKP,Witten98,review}, has provided a new tool as well as a new paradigm for strong coupling effects in Quantum Field Theory.
It has also provided a dual view to (quantum) gravitational phenomena.

An obvious application playground is standard QCD, a theory with a notorious strong coupling problem and a refreshing high-energy weak coupling behavior that brought the theory to the forth, in the early '70s, as the correct theory of the strong (nuclear) force.
The holographic correspondence  has provided very interesting and intuitive ways to address confinement, \cite{Witten98a,Cobi}, the calculation of Wilson loops, \cite{RY,MW}, the spectra of glueballs and mesons, \cite{Witten98a,COOT},
the origin of flavor anomalies and their relation to supergravity Chern-Simons terms, \cite{Witten98},  as well as  the related chiral symmetry breaking, \cite{B+M,SS,Bigazzi2005,Casero},  along the large-N ideas of Coleman and Witten, \cite{CW}.

Baryon bound states  are difficult to describe quantitatively in terms of quarks and gluons.
In the large-N limit however, \cite{Witten-N}, they have a nice description as solitons of the effective chiral theory, \cite{Skyrme,WWZ}.
This theory, beyond the ad-hoc higher derivative term introduced by Skyrme to stabilize the solitons, contains a topological term that
is crucial for making the solitons fermions in the case of $N_c$ odd, \cite{WWZ}.
Once the quarks are coupled to external sources that are flavor gauge fields, this topological part of the action becomes gauged and reproduces the
well-known flavor anomalies,   \cite{WWZ,Kaymakcalan,Manes}.

The strong coupling phenomena in QCD-like theories have been investigated via holography using a variety of models.
A first class contains  top-down  holographic models that have been developed to describe both YM, \cite{Witten98} as well as QCD, \cite{SS}.
Bottom up models, more or less  faithful to string theory dicta were also developed both for YM, \cite{PS,ihqcd,ihreview,gkmn,gubser} and QCD, \cite{hard,soft,Jarvinen}.
The holographic models usually attempt to describe a specific regime of interest, for example the finite temperature thermodynamics or the finite density dynamics.
Others, like \cite{ihqcd} and \cite{Jarvinen}, attempt to provide a complete description of the dynamics in several different regimes.

The different models have various advantages, that range from simplicity (but being crude in the quantitative part) to complexity (but being closer to real data).
The most complete bottom up model for YM is Improved Holographic QCD, \cite{ihqcd,gkmn} which has been shown to describe quantitatively rather well both $T=0$ physics for pure YM as well as finite temperature physics, \cite{panero}.
Its extension to V-QCD, provides the most sophisticated  bottom-up
holographic model for flavor,~\cite{Jarvinen,JR}. This  model is in the Veneziano limit, \cite{VL},
where both the number of colors $N_c$ and the number of flavors, $N_f$ are taken to be large with the ratio fixed
 \be
N_c\to\infty  \sp N_f\to\infty\sp x\equiv {N_f\over N_c}\to fixed
\ee
In this limit, and unlike the standard 't Hooft limit,  flavor can backreact on color.
V-QCD is 5-dimensional and is modeled on non-critical string theory, including branes and antibranes. Unlike the Sakai-Sugimoto model\footnote{In this context the open string tachyon appears on a stretched string in the bulk, \cite{AK}.} , there is an explicit open string tachyon field, as the branes and antibranes are space filling and overlap in five dimensions. This realizes the idea put forward in  \cite{Casero}, that the open string tachyon is the natural bulk dual of the quark mass operator\footnote{See also \cite{BSS}-\cite{Go}.}. The physics of open string tachyons has been analyzed by Sen and others, \cite{AS}. V-QCD is based on the Sen tachyon action for the flavour, appropriately modified to accommodate the dynamics of QCD.
Its physics has been analyzed in detail, including the vacuum structure both at zero quark mass, \cite{Jarvinen} and finite quark mass, \cite{JM}, the singlet and non-singlet spectra at zero temperature, \cite{spectrum}, the phase structure at
non-zero temperature, \cite{fT}, at finite baryon  density, \cite{fD,Ishii} and the vacuum structure and spectra at finite $\theta$-angle, \cite{theta}.

The issue of flavor anomalies and the associated topological terms, is an important ingredient of the low-energy dynamics of QCD.
In the effective chiral Lagrangian, such terms emanate from Witten's five-form topological term, \cite{WWZ}, whose quantized coefficient is given by the number of colors, $N_c$.
Such a term is written as a five dimensional (topological) integral
over a five-manifold with a single boundary which is identified with
the four-dimensional spacetime on which QCD is defined.
Once the theory is coupled to vector sources ${L_{\m},R_{\m}}$ for the flavor currents, the Noether procedure has given eventually the gauged WZW term, \cite{WWZ,Kaymakcalan,Manes}, that controls the anomalous variations of the effective action under flavor transformations. This action must be both P and C invariant as is expected from QCD without a $\theta$ angle\footnote{The relevant P and C transformations are defined in appendix \ref{app: conventions}.}. Higher topological terms that are gauge invariant start at six-derivatives, \cite{6,6a} and have been analyzed up to eight derivatives  \cite{8}.
However, in this context the chiral condensate and its size fluctuations are absent from the low energy dynamics of the chiral Lagrangian.

The issue of the relevant parity-odd terms in holographic models is also diverse\footnote{In this theory the combined parity transformation $P$ is the product of two transformations $P_1$ and $P_2$ that are defined in appendix \ref{app: conventions}. $P_2$ acts in the usual way on the boundary space coordinates. $P_1$ acts on the fields.
In most cases in holography, parity-odd refers to $P_2$-odd in our notation, and it always involves the Levi-Civita tensor.
From now on we will be careful in specifying the type of parity transformation we refer to, $P_1$, $P_2$ or $P=P_1P_2$.}.
In the simplest bottom-up model,  the hard wall model \cite{hard}, the relevant terms are five-dimensional Chern-Simons terms involving the flavor gauge fields ${L_{\m},R_{\m}}$.
They do not include, in particular, contributions from the quark condensate field, $T$, which is dual to the quark-mass operator and controls chiral symmetry breaking. Moreover,
in this simple model, chiral symmetry breaking occurs by fiat rather than dynamically.

In the more realistic top-down model of Sakai and Sugimoto \cite{SS},
the topological terms arise again from a five-dimensional Chern-Simons
term that  involves the (now unique) flavor gauge field $A_{\m}$. The
reason that the gauge field is unique is that in the chiral broken
case, the branes and antibranes fuse to a single brane and $A_{\m}$ is
the left-over gauge field after the fusion. However, as it was
anticipated later \cite{SS2}, this Chern-Simons term is not enough to
describe the physics of baryons and an extra contribution, localized
at some IR boundary on the flavor brane was added by hand \cite{SS2}.
Again, here the tachyon field is absent as it is a fluctuation of a non-local string.

The V-QCD theory has in general a $P_2$-odd five dimensional Tachyon-Chern-Simons term that is not only a functional of the flavor gauge fields, $A_{\m}^{L,R}$ but also of the (matrix-valued) tachyon field, \cite{Casero}.
As already shown in \cite{Casero} an analysis of a general tachyon-dependent CS term is notoriously difficult, and this is why, in that reference, it was determined based on string theory calculations for a tachyon field that is proportional to the identity matrix.
In this work we shall go a step further and we shall assume that
\be\label{intro1}
T=\tau U
\ee
 where $U$ is a unitary matrix and $\tau$ a single real field \footnote{ A chiral
rotation can transform  $U$ to  the identity. It  cannot however change the value
of $\tau$. This  means that flavor symmetry is broken to the diagonal
vector $U(N_f)$ but not further, as in the chiral limit of
QCD. The most general form of the tachyon matrix would be instead $T = {\cal
    T} U$, where $ {\cal
    T}$ is a Hermitian matrix. If the latter is non-trivial, the
  tachyon matrix cannot (in general) be brought to be proportional to
  the identity matrix by a L-R gauge transformation.}.
Both $\tau$ and the unitary matrix $U$ are five-dimensional fields.
With this assumption, we shall perform a comprehensive search for the most general tachyon-dependent CS term that is compatible with all symmetry expectations, and which reproduces the flavor anomalies of QCD.
From now one we shall call these terms the Tachyon-Chern-Simons terms or TCS for short.

The TCS term beyond anomalies is also relevant for dynamical questions.
Because of its structure, and the fact that it is proportional to the five-dimensional $\e$-tensor, it does not contribute in several relatively-uniform bulk solutions.
Solutions, with full Poincar\'e ISO(3,1) invariance, or $\mathbb{R}\times$SO(3) invariance relevant at finite temperature and density are not affected by the TCS term.
In the presence of a (weak) electromagnetic field the TCS term contributes to the two-point functions of currents, \cite{Son,Iatrakis}.

In a similar spirit, it is known that the (T)CS term can mediate mixing and translational invariance instabilities in finite density contexts, \cite{Ooguri,Donos,Bergman,Jokela2012}, leading to spatially modulated phases.
Finally, being cubic to leading order in the flavor gauge fields,  it is important in the determination of the three point function of the flavor currents.

Probably, the most important role of the (T)CS term so far, beyond anomalies, is in the determination of the baryon soliton solution in the holographic context. The baryon ansatz is sufficiently complicated so that it contributes non-trivially, in several ways. In particular, the soliton would have the tendency to shrink to a point, but the CS term is crucial in its stabilisation. The soliton is essentially an instanton of the flavor symmetry group, and because of the CS term it acquires a $U(1)_B$ charge.
This has a dual effect. First it makes the soliton a baryon. Second the
$U(1)_B$ charge repulsion in the bulk stabilizes the soliton that
would like otherwise to collapse as it happens in simpler cases,
\cite{Sugimoto,Yi}.

The goal in this paper is two-fold:
\begin{enumerate}
\item To analyze the general structure of the TCS term and to understand the relevance of different parts in dynamics and anomalies.
\item  To write the appropriate ansatz of the baryon solution, to derive the equations of motion, and to understand the boundary conditions relevant for the solution as well as the conditions for obtaining the correct baryon number.
\end{enumerate}

In this work, we fully answer these questions provided we are in
the exact chiral limit. This implies in particular  that the tachyon
field will be taken  to have the form (\ref{intro1}),
where $\tau$ is a real scalar function and $U$  is a
(spacetime-dependent) unitary matrix.
 These  facts allow us, on the one hand, to write down  the most general
 combination of tachyon-dependent Chern-Simons-like terms compatible with
 symmetries and anomalies of the boundary theory and  with (bulk)
 gauge-invariance, and on  the other hand to write the appropriate ansatz for a
  cylindrically-symmetric bulk instanton which is regular,
  horizonless,  has finite  mass  and unit  baryon charge with respect
  to the boundary theory, and is  therefore the candidate for a
  single-baryon state in V-QCD.

\subsection{Summary of results}

In the rest of this introduction, we summarise our results and
present further directions and open problems (and in particular we
briefly sketch  why our construction is limited to QCD in the chiral limit).

\subsection*{CS terms}

We call the TCS part of the bulk action anything which is
written as a bulk integral of a five-form,
\be \label{intro3}
S_{CS} = {i N_c^2 \over 4\pi^2} \int_{bulk} \Omega_5
\ee
where the normalization is chosen for convenience.
We assume\footnote{This assumption is dictated by string theory.} that $\Omega_5$ may depend only on the flavor sector fields
(the tachyon modulus $\tau$, the unitary matrix
$U$ and  the flavor gauge  fields), but {\em not} on the glue sector fields
(the metric and dilaton). Under this assumption, we write the most
general form  $\Omega_5$ which is compatible with the discrete
symmetries and such that, under a  bulk gauge transformation, the
variation of (\ref{intro3}) reduces to a UV-boundary term which
reproduces the flavor anomalies of QCD.
 These restrictions largely
(but not completely) fix the form $\Omega_5$. It must  be the sum of
three terms:
\be  \label{intro4}
\Omega_5 = \Omega_5^0 + \Omega_5^c + dG_4,
\ee
which have the following properties:
\begin{enumerate}
\item The first term $\Omega_5^0$ is  gauge-invariant (under bulk
  gauge-transformations) but not closed;  it is the sum of four terms:
\be \label{intro5}
 \Omega_5^0 = \sum_{i=1}^4 f_i(\tau) F_5^i(U,F^L, F^R),
\ee
where $f_i(\tau)$ are arbitrary functions of the tachyon modulus and
$F_5^i(U,F^L, F^R)$ are specific gauge-invariant five-forms which
depend on $U$, its gauge-covariant derivative $DU$, and the
field-strengths $F^L$ and $F^R$ of the left and right flavor gauge
fields, and are given explicitly in section \ref{Sec:cs-cons}. The
functions $f_i(\tau)$ can be constrained to some degree if one
makes further assumptions. In particular, their value  at $\tau=0$ can
be fixed by asking that  in the chirally unbroken phase with trivial
tachyon field $\tau=0$, (\ref{intro3}) matches the standard  gauge
Chern-Simons action in 5d for the left and right gauge fields.
\item The second term in (\ref{intro4}) is closed but not exact and it
  is completely fixed by the flavor anomaly (including the overall
  coefficient):
\be \label{intro6}
\Omega_5^c = -{1\over 60} Tr[(U^\dagger d U)^5]
\ee

\item The third term in (\ref{intro4}) is written in terms of a 4-form
  $G_4$, which is fixed,  up to a few arbitrary functions
  $h_i(\tau)$. However  as it only enters the action via a boundary
  term, only the values at $\tau=0$ of  $h_i(\tau)$ matter, and these values are
  completely fixed by the QCD flavor anomaly.
\end{enumerate}
To summarize, the TCS
 action is parametrized by two sets of
functions $f_i(\tau)$ and $h_i(\tau)$. The $f_i(\tau)$'s  have no
effect on anomaly matching, but enter the field equations in the
presence of non-trivial gauge-field configurations; the
$h_i(\tau)$'s on the other hand are irrelevant for the dynamics, and
their boundary value is fixed by anomaly matching (therefore this
description  has a certain redundancy).

In order to match the flavor anomaly, we have to make a further assumption of
{\em IR regularity} on the matrix $U$  as it usually happens in
holography. This can be translated into the requirement that the
matrix $U$ should go to constant fast enough in the IR, so that there is no IR
contributions to the on-shell action or to the anomalous variation of
the action under gauge symmetry\footnote{This can often be  rephrased more
rigorously in terms of normalizable vs. non-normalizable solutions of
the field equations near the IR: one then assumes that only
normalizable configurations are physical.}. The same argument (no IR
contributions to the on-shell action and anomaly) requires the
functions $f_i(\tau)$ and $h_i(\tau)$ to vanish fast enough in the IR
(which corresponds to the limit $\tau \to +\infty$ in the chirally broken
phase).

\subsection*{The V-QCD baryon}
As we mentioned earlier, the TCS action  is crucial
for correctly describing baryons in  holographic theories. The second part of
this work is devoted to lay out the general grounds for constructing the
baryon as an asymptotically-AdS  soliton solution of the bulk theory with the
following properties:
\begin{itemize}
\item It is time-independent and spherically symmetric with respect to
  the boundary spatial directions;
\item It is a  finite-energy excited state over the QCD vacuum. This
  requires in particular that all bulk fields which are turned on must
  have vev-like  asymptotics near the AdS boundary and
  must reduce to the vacuum fast enough at spatial infinity
  $|\vec{x}| \to +\infty$;
\item It has baryon-number equal to one with respect to the boundary abelian
  vector flavor symmetry $U(1)_B$;
\item It is horizonless: the baryon number is not {\it
    fractionalized}, i.e. it is not  provided by deconfined degrees of
  freedom like in a quark-gluon plasma state.
\end{itemize}

Here, we construct a bulk $SU(2)$
instanton configuration of the $U(N_f)$ flavor axial gauge fields,  depending
on the four directions $(r,\vec{x})$  together with a nontrivial function $U(\tau,|\vec{x}|)$ describing the
non-abelian part of the tachyon, which satisfies all above requirements. We argue later that  we can neglect the backreaction ofthe baryon fields on the background fields (the metric, dilaton and tachyon modulus), that we take to be fixed on the
homogeneous vacuum solution\footnote{The last requirement in the list above is trivial in this probe regime, since a horizon can only arise if the backreaction on the metric is non-negligible.}. Indeed the baryon contributes to order $N_c$, which is negligible compared to $N_c^2$ and $N_c N_f$ in the Veneziano limit.  Specifically, the construction involves the following steps:
\begin{enumerate}
\item   We derive the full equations of motion for the instanton;
\item We identify the boundary conditions such that the
finite-energy condition is satisfied. The boundaries here are 1) the
near-AdS region UV boundary $r\to 0$, where the solution should
satisfy vev-like boundary conditions for all the fields; 2) the
 boundary at  spatial infinity $|\vec{x}| \to \infty$, where the fields
 have to vanish fast enough so that one can meaningfully describe the
 solution as a localized  lump of finite energy in the boundary field theory;
\item We identify   suitable ``regularity'' conditions in the IR region of the
geometry (using both a gauge-invariant formulation, and in a
Lorenz-like gauge);
\item We  show that, under these conditions, the boundary baryon
charge coincides with the bulk instanton number of the solution.
\end{enumerate}
The last item on this list is particularly important because it shows
that baryon configurations are topologically stable in the
bulk.

We find that, remarkably, the derivation of the baryon number does not depend in any manner on the
non-closed part of the TCS action $\Omega_5^0$ and the corresponding TCS
potentials $f_i(\tau)$ in equation  (\ref{intro5}).  Rather,  the generation of the baryon number and the
contribution of the CS terms to the equations of motion (responsible
for the stabilization of the baryon size) are ensured by two distinct
parts of the TCS action (closed and non-closed, respectively).
Although slightly counter-intuitive, it is not a contradiction. The
reason is that the result for the baryon number simply tells us what
should be the boundary behavior of the tachyon field for $N_B$ to be
non-zero (it should have a Skyrmion winding from \eqref{NBS}). It does
not guarantee that a solution with such boundary conditions exists
though. In particular, it is expected that no finite size solution
should exist when $\Omega_5^0$ vanishes ($f_i(\tau)=0$).

\subsection{Further directions and open problems}
In this work we have determined the form of the
Tachyon-Chern-Simons action in V-QCD in the chiral limit and set up the
formalism  for obtaining single-baryon configurations. Several questions remain open.

The most important drawback of the TCS
 action obtained in this work is
that it is limited to the exact chiral limit. We have assumed an
ansatz for (\ref{intro4}) which, based on the factorization of the
tachyon (\ref{intro1}), is made up of terms of the form:
\be \label{intro7}
\Omega_5 \supset f(\tau) \times \text{5-form}(U, \text{gauge fields})
\ee
However  this cannot be the full story when quark masses
are non-zero: if this is the case, instead of (\ref{intro1}), the
scalar function $\tau$ must be replaced by a Hermitian matrix (see
footnote 5). Therefore, one cannot write the simple ansatz (\ref{intro7}) for
the TCS terms, which instead must have a more complicated  dependence
on the matrix $T$, such that it effectively reduces to the form
(\ref{intro7})  for zero quark masses. 

A possible way to proceed is to return to the superconnection formalism of Quillen, (see \cite{Casero} for a description) which has been used and shown to be relevant in string theory calculations of the Tachyon-Chern-Simons terms, \cite{Casero}. In a recent paper \cite{super}, it was shown that it is an appropriate formalism for anomalies.
If this formalism is used in our analysis here,  it will reduce the four unknown tachyon dependent functions that can appear to only one.
It is possible that within the superconnection formalism, the problem of writing the general TCS term may be tractable.
The difficult part is in writing the anomaly related tachyon-dependent six form as the exterior derivative of a five-form.

In this work we have established the baryon instanton-like ansatz and its
field equations, as well as the
boundary conditions it should satisfy in order for the corresponding
state in the boundary theory  to have finite  mass and unit baryon
charge. Whether or not a full solution exists satisfying these field
equations is a question which can only be addressed numerically. The corresponding  equations are a set of non-linear partial differential
equations in the $(|\vec{x}|,r)$ space, whose numerical resolution
is highly  non-trivial and goes beyond the scope of this work. Having a numerical
solution will allow to estimate the  baryon mass and bulk
profile  and to verify the
validity of the approximations used in this work (the baryon as a
probe on the color sector). Beyond the baryon mass, several static properties of the baryon can be extracted from a numerical solution. In particular, one can estimate how the chiral
condensate is affected by the  presence of the baryon and verify the
expectation that the chiral symmetry is partially restored inside the baryon. Also, the baryon form factors can be computed, following the same method as in the Skyrme \cite{Adkins1983} and WSS \cite{Sugimoto} models. The calculation of the numerical solution and of the static baryonic observables that can be extracted from it is left for future work. 

The static instanton solution gives the classical background whose excitations describe baryon states. In particular, the spectrum of the (low-lying) baryon states is obtained by quantizing the collective modes of the instanton solution. These modes include zero-modes (3-dimensional translations and isospin rotations) as well as modes associated with a non-zero potential, such as vibrational/breathing modes in the bulk. Quantizing the isospin rotational modes will produce the spectrum of isospin eigenstates, starting with the nucleons ($I=1/2$) and $\D$ isobars ($I=3/2$). The non-zero modes will generate the towers of excited state in each isospin sector ($N(1440),\D(1600),\dots$). The calculation of the baryon spectrum from the perturbed instanton solution will be addressed elsewhere. 

An interesting  question is whether this approach can produce  Regge trajectories for the baryons.
This implies in particular, baryon masses$^2$ that are asymptotically linear in spin. In the slow-rotation approximation, in which calculations of the quantization of the
low-lying modes proceeds,  the leading contribution in the mass is quadratic in the spin.
It is conceivable that  a resummation of higher corrections  would alter this behavior. That being said, one possibility is
that   fast spinning baryons that are solitons, would have instabilities.  Another possibility, if this is not the case, is that the very fast spinning baryon would resemble more and more like a rotating string. Whether any of these two pictures are correct is not clear to us.   Another possibility is that baryons are modelled directly as strings 
and then they can produce Regge trajectories in the baryon sector, along the lines of \cite{cobi}.

Having a numerical handle on the baryon solution can be of great help
for effective or simplified  descriptions of holographic baryonic
matter, such as the one used in \cite{Ishii:2019gta}, in which a
homogeneous state
filled with baryons on the boundary was modeled with a bulk
configuration having a thin-wall  discontinuity at a fixed radial
position. Having the actual single-baryon solution and its numerical
radial profile  can help to better understand  the effective
description in which the baryons sit on a thin layer at fixed $r$. \\

This paper is organised as follows.
In section \ref{Sec:VQCD_intro} we give an overview of the class of
 holographic models for QCD in the Veneziano limit
(V-QCD) under consideration, discussing in particular the DBI-part of
the action and the asymptotics of the various phenomenological
potentials.
In section \ref{Sec:CS} we determine the  form of
Tachyon-Chern-Simons terms in V-QCD.
In section \ref{Sec:ans} we give the ansatz of the baryon state as
an $SU(2)$ instanton of the bulk flavor sector. Then, in section
\ref{Sec:consB}, we write the corresponding field equations and the
expression for the baryon energy.
Finally, in section \ref{sec:baryon-bc} we discuss the boundary conditions at
the UV boundary and spatial infinity, as well as the conditions at the
IR end of the bulk spacetime, which have to be satisfied in order for
the baryon to have finite energy and unit charge. In that section we
also compute the baryon charge of the solution and  discuss its
relation to the bulk instanton number.
Several technical details are left to the Appendix.

\section{Overview of the V-QCD model}

\label{Sec:VQCD_intro}

We start by reviewing the V-QCD model \cite{Jarvinen}. This is a
bottom-up, five-dimensional holographic model for QCD with $N_c$ colors
and $N_f$ flavors,  which captures both
glue and  flavor dynamics, i.e. the flavor sector is fully
backreacting. The holographic description in terms of a 5-dimensional
theory is assumed to be valid in the Veneziano large-$N$ limit,
\be \label{v1}
N_c, N_f\to \infty, \quad x\equiv {N_f \over N_c} \quad \text{and} \quad
\lambda \equiv g^2 N_c \:\:  \text{finite} \, ,
\ee
where $g$ is the Yang-Mills coupling.

The holographic model consists of a five-dimensional
bulk theory whose dynamical fields are in one-to-one correspondence
with the lowest-dimensional gauge-invariant operators in QCD. These are:
\begin{enumerate}
\item The five-dimensional  metric $g_{ab}$, dual to the
  stress-tensor;
\item A scalar field $\lambda$ (the {\em dilaton})  dual to the operator
  $\text{Tr} F_{\mu\nu}F^{\mu\nu}$  (where $F_{\mu\nu}$ is the Yang-Mills
  field strength) and encoding the running 't Hooft
  coupling;
\item A set of $U(N_f)_L \times U(N_f)_R$ non-Abelian gauge fields,
  denoted
  $\mathbf{L}_M$, $ \mathbf{R}_M$, dual to the chiral flavor
  currents $\mathbf{J}_\mu^{(L)}$, $\mathbf{J}_\mu^{(R)}$, whose matrix
  elements are:
\be \label{v2}
 \Big(\mathbf{J}^{(L)}_\mu\Big)^i_j =
\bar{q}_{L}^i\gamma_\mu  q_{L\,j}, \qquad  \Big(\mathbf{J}^{(R)}_\mu\Big)^i_j =
\bar{q}_{R}^i\gamma_\mu  q_{R\,j} \qquad i,j = 1\ldots N_f \, ,
\ee
 where   $q_{L/R\,i}$ are the left-handed and right-handed quarks
 in the fundamental representation  of $U(N_f)_{L,R}$;
\item An $N_f\times N_f$ complex scalar field  matrix $T^i_j$ (the  {\em
    tachyon}) in the  bi-fundamental representation of  $U(N_f)_L
  \times U(N_f)_R$, dual  to the quark bilinear $\bar{q}_R^i
  q_{L\,j}$.
 \end{enumerate}
 All the above operators have dimension 4 or smaller. The
 bottom-up nature of the model resides in truncating the spectrum to
 these low-dimension operators and writing a phenomenological action
 describing the dynamics of the corresponding bulk fields. The structure of the action is obtained from string theory. The action is further parametrized by a few functions of $T$ and $\lambda$, whose
 relevant features are fixed by a mixture of  theoretical and phenomenological motivations.

The five-dimensional action for these fields takes the following form:
\be \label{v3}
S_{V-QCD}  = S_g + S_{DBI} + S_{CS} \, .
\ee
The first term $S_g$  depends on the color sector alone, while the second
and third terms describe the coupled flavor-color degrees of
freedom.  $S_{DBI}$ is a Dirac-Born-Infeld-type action which contains
the kinetic terms of the tachyon and gauge fields, whereas $S_{CS}$ is
a generalized Tachyon-Chern-Simons action whose role is crucial to correctly
recover chiral anomalies.  If we think of this model  as originating from a
(non-critical) string theory, then  the color action describes the
low-energy effective theory of closed string
fields, while the DBI and Tachyon-Chern-Simons actions describe the open
string sector living on the world-volume of $N_f$ pairs of
space-filling D4-branes and anti D4-branes.    Below we shall describe in
more detail the
first two terms in the action, leaving the TCS term for section \ref{Sec:CS}.

The action $S_g$  in  (\ref{v3}) gives a holographic description of
the pure glue sector in terms of the Einstein-dilaton theory of the
fields $\lambda$ and $g_{MN}$. It has the form:
\be \label{v4}
S_g[g,\lambda] = M^3 N_c^2 \int d^5x \sqrt{-g}\left[R - {4\over
    3\lambda^2} g^{MN}\de_M \lambda \de_N \lambda + V_g(\lambda)\right]
\sp 0 < \lambda < +\infty.
\ee
This is the action for the five-dimensional  IHQCD holographic model
for pure Yang-Mills theory \cite{ihqcd}.  The UV regime of the theory corresponds
to the small-$\lambda$ region, and the IR is reached as $\lambda \to
+\infty$.  The dilaton potential
$V_g(\lambda)$ is chosen in such a way that it allows for a
logarithmically running coupling constant in the UV, leads to color
confinement (i.e. a Wilson loop area law)  and has a qualitatively
correct glueball spectrum.  These  requirements fix the large-$\lambda$ and
small-$\lambda$ behavior of $V_g(\lambda)$ as follows \cite{ihqcd}.

For small $\lambda$, one requires:
\be \label{v5}
V_g(\lambda) = {12 \over \ell_g^2}\left[1 + V_{g,1}\lambda + O(\lambda^2)\right]  \sp \lambda \to 0 \, .
\ee
The leading term in the expansion ensures that the model admits an
asymptotically AdS$_5$ solution with AdS length $\ell_g$ and with
$\lambda \to 0$ as one approaches the boundary. The second term
 ensures that $\lambda$ (identified with the running 't Hooft coupling
 in the UV) has the correct perturbative running, and the coefficient $V_{g,1}$
is fixed by matching the first $\beta$-function coefficient of
four-dimensional Yang-Mills. Similarly, higher-order terms in the expansion
(\ref{v5})  can be fixed by matching higher-order  $\beta$-function
coefficients.

For large $\lambda$, matching the expected qualitative behavior of
Yang-Mills at   low-energy requires:
\be \label{v6}
V_g(\lambda) \sim  \lambda^{4/3} (\log \lambda)^{1/2} \sp
\lambda \to +\infty \, .
\ee
With this asymptotics, the Wilson loop follows an area law and the model has  a discrete  spectrum of glueball excitations with
masses $m_n$ obeying Regge-like asymptotics, $m_n^2 \propto n$, both
in the scalar and in the tensor sectors.

We now come to the DBI part of the action, the second term in (\ref{v3}).
In most of the previous work in the context of V-QCD
\cite{Jarvinen,JR,IKP,JM,spectrum,fT,fD,theta},  only the Abelian vector part of the
gauge fields was relevant,  and moreover the tachyon matrix was   restricted
 to be the
identity matrix times a real function $\tau$. These assumptions apply
when considering homogeneous solutions (the vacuum, or a thermal
equilibrium state) and neglecting quark mass differences. In  this
case the DBI action reduces to an action for a single Abelian gauge field
and a real scalar $\tau$.
For illustrative purposes,  we first present this simplified version of the DBI action, because it already contains the phenomenological
potentials which characterize the model. The full DBI action
including the non-Abelian structure will be presented in section \ref{sec:FullDBI}.

The simplified DBI action was taken to be of the form, \cite{Casero,Jarvinen}:
\begin{align}
\label{v7}
S_{DBI,simpl}[g_{ab}, \lambda, \tau, A_\mu] &= - M^3 N_c N_f\times \\
\nn &\times \int d^5 x V_f(\lambda,\tau)
\sqrt{-\mathrm{det}\left(g_{MN}+\kappa(\lambda) \de_M\tau \de_N \tau + w(\lambda) F_{MN}\right)} \, ,
\end{align}
where the field $\tau$ is defined by the tachyon ansatz
\be \label{v8}
T^i_j = \tau \delta^i_j \sp \tau \in {\mathbb R} \, ,
\ee
and $F_{MN}$ is the field strength of the vector $U(1)$ gauge
field $v_M$, defined  such that:
\be \label{v9}
\left(\mathbf{L}_M\right)^{i}_{j}
=\left(\mathbf{R}_M\right)^{i}_j=  v_M \delta^i_j \sp i,j
= 1\ldots N_f \, .
\ee
 The form  of the action  (\ref{v7})  is modeled after the Abelian form of
 Sen's DBI action for unstable D-branes \cite{AS}, which governs top-down
 holographic models of flavor which incorporate chiral symmetry
 breaking,  such as the Sakai-Sugimoto model \cite{SS}. Lacking a top-down
 description, the phenomenological character of the DBI action
 (\ref{v7}) is encoded in three functions of the scalars $
 V_f(\lambda, \tau), \kappa(\lambda), w(\lambda)$,
which affect both the flavor dynamics and its interaction with
color. These functions are
constrained by consistency and phenomenological requirements, which
are discussed in detail in \cite{spectrum} and that we summarize
below\footnote{The functions $\kappa$ and $w$ can in principle also depend on the tachyon field. We shall not consider such a dependence in this paper.}.
\begin{itemize}
\item $V_f(\lambda,\tau)$ controls the overall effect of flavor over the
color background, and it is crucial (for example) for the correct
description of chiral symmetry breaking. More specifically, we assume that,  at large
values $\tau$, it  behaves asymptotically as:
\be \label{v10}
V_f(\lambda, \tau) \sim e^{-a(\lambda) \tau^2 } \sp \tau \to + \infty \, ,
\ee
with $a(\lambda) >0$. The behavior (\ref{v10}) is modeled after Sen's
action  for unstable D-branes, and it implies  that as $\tau \to
\infty$ the space-filling flavor branes disappear.
\item  The
effective scalar potential  including the DBI contribution is
\be \label{v11}
V_{eff}(\lambda,\tau) = V_g(\lambda) - {N_f \over N_c} V_{f}(\lambda,\tau) \, .
\ee
As $\tau \to \infty$
chiral symmetry is broken, the flavor sector decouples due to (\ref{v10})
and $V_{eff}(\lambda)$
reduces to $V_g(\lambda)$.   On the other hand, for $\tau=0$ (which
corresponds to the UV), we denote  the effective dilaton potential by:
\be\label{v11.i}
  V_{eff}(\lambda,\tau=0) =   V_g(\lambda) -
{N_f \over N_c} V_{f}(\lambda,0) \, .
\ee
We also need the expansion coefficients of $V_f$ at small tachyon:
\be\label{v11.ii}
V_{f}(\lambda,\tau) = V_{f,0}(\lambda)\left[1-\hat a(\l)\tau^2 + \mathcal{O}\left(\tau^4\right)\right] \ .
\ee
\item For the correct UV behavior,  all functions of $\lambda$  have a regular
  power-law expansion in $\lambda$ around $\lambda=0$, similar to the
  one of $V_g$ in  (\ref{v5}):
\bea
&& V_{eff}(\lambda,\tau=0) = V_0 \left[1+  V_1\lambda +
  O(\lambda^2)\right] \, , \nonumber \\
&& V_{f,0}(\lambda) = W_0 \left[1+  W_1\lambda + O(\lambda^2)\right],
\nonumber \\
&&\label{v12}\kappa(\lambda)  = \kappa_0 \left[1+ \kappa_1\lambda + O(\lambda^2)\right],
\qquad \qquad
\lambda \to 0 \, , \\
&&w (\lambda) = w_0 \left[1+  w_1 \lambda + O(\lambda^2)\right], \nonumber \\
&& \nn \hat a(\lambda)  = 1 + a_1 \l + \mathcal{O}(\l^2) \, ,
\eea
where $W_0, \kappa_0$ etc. are constants.  The UV AdS length is now given by
the effective potential (\ref{v11}) evaluated at $\lambda = 0,\tau=0$:
\be \label{v13}
{12 \over \ell^2} = {12 \over \ell_g^2} - x W_0,
\ee
where $\ell_g^2$ was defined in (\ref{v5}) and $x$ in \eqref{v1}. The other expansion coefficients  are fixed by matching
the UV behavior of the QCD operators (dimensions, two-point function
normalization, etc.).

\item In the IR regime, thermodynamics and the qualitative features of
  the meson trajectories constrain the large-$\lambda$ behavior of the
  functions $V_{f,0}(\l)$, $\kappa(\lambda)$ and $w(\lambda)$. In
  particular, qualitatively consistent results are obtained if we
  assume that, to leading order as  $\lambda \to +\infty$ :
\be \label{v15}
V_{f,0} \sim W_{IR} \lambda^{v_p} , \quad \kappa \sim \kappa_{IR}
\lambda^{-4/3}(\log \lambda)^{1/2},
\quad w \sim w_{IR} \lambda^{-4/3} (\log \lambda), \quad a\sim a_{IR},
\ee
where $W_{IR}$, $\ka_{IR}$, $w_{IR}$, $a_{IR}$ and $v_p$ are constants, and $4/3<v_p<10/3$.
\end{itemize}

\subsection{Vacuum solution} \label{sec:vac}

The Poincar\'e-invariant vacuum of the dual field theory is described
by a bulk solution to the classical field equations in which all the gauge fields are set to
zero. Assuming the quark mass matrix is proportional to the identity\footnote{This implies that the quark masses are all the same.},
the solution is characterized by the three functions $A(r),\lambda(r)$ and $\tau(r)$:
\be \label{vac1}
ds^2 = e^{2A(r)}\left(dr^2 + \eta_{\mu\nu} dx^\mu dx^\nu\right), \quad
  \lambda = \lambda(r), \quad \tau = \tau(r) \, .
\ee
Here, $r$ is the holographic radial coordinate, and $x^\mu$ ,
$\mu=0\ldots 3$ are identified with the coordinates of 4d Minkowski
space-time, on which the dual field theory is defined.

With the gauge fields identically zero, the TCS action does not
contribute to the field equations, and the Poincar\'e-invariant solution (\ref{vac1}) is
completely determined by the first two terms in (\ref{v3}).

While in general the equations of motion have to be solved numerically, the
asymptotic behavior of the solution in the UV and IR  can be obtained
analytically and it is fixed by the asymptotic behavior of the
potentials discussed in the previous section, as we review below. This
asymptotic  behavior will be used to understand the boundary
conditions satisfied by inhomogeneous solutions such as  the baryon.

\paragraph*{UV asymptotics.}

The region of the geometry corresponding to the UV of the dual field
theory is the region where $e^{A} \to +\infty$ and $\lambda \to 0$ \cite{ihqcd}. If
the radial coordinate  is chosen as in  (\ref{vac1}), this region
corresponds to the limit $r\to 0$, and one finds (see
e.g. \cite{ihqcd,spectrum}):
\be
\label{AUV} A(r)= -\log \left(\frac{r}{\ell}\right) + \frac{4}{9 \log (r \Lambda)} + \mathcal{O}\left(\frac{1}{\log(r \Lambda)^2}\right) \, ,
\ee
\be \label{lUV}
\lambda(r)  = -{1\over V_1}\frac{8}{9 \log (r \Lambda)} +
\mathcal{O}\left(\frac{1}{\log(r \Lambda)^2}\right)\,  ,
\ee
\begin{align}
\nn \frac{1}{\ell} \tau(r) &= m r (-\log (r \Lambda))^c \left(1 + \mathcal{O}\left(\frac{1}{\log(r \Lambda)}\right)\right) \\
\label{tUV}&\hphantom{=} + \Sigma\, r^3 (-\log (r \Lambda))^{-c}  \left(1 + \mathcal{O}\left(\frac{1}{\log(r \Lambda)}\right)\right) \, .
\end{align}
In the equations above, $\ell$ is the AdS length,  $V_1$ is the first subleading coefficient in
the effective  potential (see equation (\ref{v12})), and  $\Lambda$, $m$ and
$\Sigma$ are integration constants of the field equations. In terms of
the dual field theory, $\Lambda$ is a scale which measures the
breaking of conformal invariance in the UV (it is the holographic
manifestation of the QCD scale);  $m$ is the quark
mass\footnote{Recall that we are assuming a quark mass matrix proportional to the identity.} and
$\Sigma$ is the quark condensate. The independent  leading and
subleading terms  in the tachyon expansion correspond to a field theory operator
of dimension $\Delta=3$, with the extra logarithm reproducing the QCD mass
anomalous dimension. The exponent $c$ is determined by the
$O(\lambda)$ terms in the expansions of the potentials (\ref{v12}),
see \cite{spectrum}:
\be \label{vac2}
c = {4\over 3} \left(1 + {\kappa_1 -a_1 \over V_1} \right).
\ee

\paragraph*{IR asymptotics}
The IR region of the geometry corresponds to $e^A\to 0$ and
$\lambda \to +\infty$, \cite{ihqcd}. In the chiral symmetry breaking
solution (for which $\tau \neq 0$) the tachyon also diverges in this
limit \cite{Jarvinen}. With the
glue potential behaving as in (\ref{v6}), the IR  is found in the limit
$r\to +\infty$,  and we have \cite{ihqcd,Jarvinen}:
\be
\label{laIR} \l(r) = \ex^{\frac{3r^2}{2 R^2}+ \l_c} \left(1 + \mathcal{O}\left(r^{-2}\right)\right) \, ,
\ee
\be
\label{AIR} \ex^{A(r)} = \sqrt{\frac{r}{R}}\ex^{-\frac{r^2}{R^2}+ A_c} \left(1 + \mathcal{O}\left(r^{-2}\right)\right) \, ,
\ee
\be
\label{tauIR} \tau(r) = \tau_0 \left(\frac{r}{R}\right)^{C_\tau} \left(1 + \mathcal{O}\left(r^{-2}\right)\right) \, .
\ee
Here,   $C_\tau$, $A_c$, and $\lambda_c$  are constants determined by the asymptotics of
the various potentials (\ref{v12}) and (\ref{v13}), and the first of these constant must obey
$C_\tau >1$;   $R$ and $\tau_0$ are  integration constants which are
functions of the integration constants appearing in the UV.  In
particular, $R$  plays the role of the   non-perturbative IR scale of the
theory.

\subsection{Full DBI action} \label{sec:FullDBI}
The full DBI action  is the non-Abelian generalization of
(\ref{v7}), in which the full matrix nature of the tachyon and of the
left and right gauge fields is explicit. It is again based  on Sen's
action, deformed by the same three phenomenological functions of the dilaton
and tachyon $V_f(\lambda, T),  \kappa(\lambda), w(\lambda)$
\cite{Jarvinen}:
\be
\label{SDBI} S_{\text{DBI}} = -\frac{1}{2}M^3N_c\, \mathrm{STr} \int\mathrm{d}^5x\, V_f(\l,T^{\dagger}T)\left(\sqrt{-\mathrm{det}\, \mathbf{A}^{(L)}} + \sqrt{-\mathrm{det}\, \mathbf{A}^{(R)}}\right) \, ,
\ee
where the symmetrized trace over the flavor indices STr is defined as
\be
\label{defSTr} \mathrm{STr}(M_1M_2\ldots M_n) = \frac{1}{n!}\sum_{\s\in S_n}\mathrm{Tr}(M_{\s(1)}M_{\s(2)}\ldots M_{\s(n)}) \, ,
\ee
for every integer $n$ and matrices $M_1,M_2,\ldots ,M_n$, with $S_n$ the symmetric group. The convention for the normalization of the $SU(N_f)$ generators is
\be
\label{normta}\mathrm{Tr}\left(t_a^{(L)}t_b^{(L)}\right)=\frac{1}{2}\d^{ab} \sp \mathrm{Tr}\left(t_a^{(R)}t_b^{(R)}\right)=\frac{1}{2}\d^{ab} \, .
\ee
The fields appearing in  \eqref{SDBI} are
\be
\label{defAL} \mathbf{A}_{MN}^{(L)} \equiv g_{MN} + w(\l)\mathbf{F}_{MN}^{(L)} + \frac{\ka(\l)}{2}\left[(D_MT)^{\dagger}D_NT + (D_NT)^{\dagger}D_MT\right] \, ,
\ee
\be
\label{defAR} \mathbf{A}_{MN}^{(R)} \equiv g_{MN} + w(\l)\mathbf{F}_{MN}^{(R)} + \frac{\ka(\l)}{2}\left[D_MT(D_NT)^{\dagger} + D_NT(D_MT)^{\dagger}\right] \, ,
\ee
where $\mathbf{F}_{MN}$ is the field strength for the gauge fields and the covariant derivative is such that
\be
\label{D} D_MT = \partial_MT + iT \mathbf{L}_M - i\mathbf{R}_MT \, .
\ee
The bold font indicates that the $U(1)$ part is included in the gauge fields.

There is no simple way to express $\sqrt{-\mathrm{det}\,
  \mathbf{A}^{(L/R)}}$  though, even
with the help of the permutativity of the symmetrized trace and within
the SU(2) ansatz\footnote{Note however that, with the SU(2) ansatz,
  the computation can be done in principle at any finite order in the
  non-Abelian gauge fields.} .

{
In order to make the problem tractable, in section \ref{Sec:consB},} {we consider the same kind of expansion that was considered in \cite{Ishii}.} {That is, the DBI action is expanded up to quadratic
order in the non-Abelian field strengths, where the non-Abelian part
of the tachyon covariant derivatives is considered to
be of the same order as the field strength. In the quadratic approximation,  the symmetrized trace can be replaced by a simple trace in the DBI action without ambiguity.}

We shall work in the chiral limit, with all quark masses set to
zero. In this case, in the chirally broken phase, the flavor group is
broken in the large-$N$ limit to the diagonal $U(N_f)$ subgroup
\cite{CW}. This is realized in the bulk theory by considering tachyon  field configurations of the form:
\be \label{unitary}
T = \tau\,  U \, ,
\ee
where $\tau$ is a real scalar and $U$ is a unitary matrix\footnote{The
chirally broken vacuum corresponds to a bulk solution  with a
non-zero profile for $\tau$ and $U=\mathbb{I}$ (up to chiral
transformations). This solution indeed  breaks the flavor symmetry to the diagonal $U(N_f)$. }

The quantities $\mathbf{A}^{(L/R)}$ defined in \eqref{defAL}-\eqref{defAR} can be written as
\be
\label{DBItL}  \mathbf{A}^{(L)}_{MN} = \tilde{g}_{MN}^{(L)} + w(\l) F^{(L)}_{MN} + \ka(\l)\tau^2 D_{(M}U^\dagger D_{N)}U \, ,
\ee
\be
\label{DBItR}  \mathbf{A}^{(R)}_{MN} = \tilde{g}_{MN}^{(R)} + w(\l) F^{(R)}_{MN} + \ka(\l)\tau^2 D_{(M}U D_{N)}U^\dagger \, ,
\ee
where the Abelian part $\hat{F}^{(L/R)}_{MN}$ of the field
strength and the tachyon derivatives were collected in an effective metric:
\be
\label{defgt} \tilde{g}^{(L/R)}_{MN} \equiv g_{MN} + w(\l)\hat{F}^{(L/R)}_{MN} + \ka(\l) \partial_M\tau\partial_N\tau \, .
\ee
As in \cite{Ishii}, we made the simplifying assumption that  the potentials $w$ and $\ka$ depend on the dilaton only.
2-tensor indices can be raised and lowered using the effective metric
(\ref{defgt}) according to the following convention
\be
\label{raise} M^A_{\,\,\,\,\,B} = \left((\tilde{g}^{(L/R)})^{-1}\right)^{AC}M_{CB} \sp M_A^{\,\,\,\,\,B} = M_{AC} \left((\tilde{g}^{(L/R)})^{-1}\right)^{CB} \, ,
\ee
where which of L or R should be used will depend on which component (L or R) of the DBI action we are considering.

The expansion of the DBI Lagrangian up to quadratic order in the non-Abelian field strengths is then obtained from
\begin{align}
\label{expL} \sqrt{-\mathrm{det}\, \mathbf{A}^{(L)}}& = \sqrt{-\mathrm{det}\, \tilde{g}^{(L)}}\times\\
\nn &\hphantom{=} \times \Bigg(1 + \frac{1}{2}w\, \mathrm{tr}\left((\tilde{g}^{(L)})^{-1}F^{(L)}\right) + \frac{1}{2}\ka\tau^2\mathrm{tr}\left((\tilde{g}^{(L)})^{-1} D_{(M}U^\dagger D_{N)}U\right) \\
\nonumber &\hphantom{= \times \bigg(}  -\frac{1}{4}\mathrm{tr}\bigg(\Big((\tilde{g}^{(L)})^{-1}\left(w F^{(L)} + \ka \tau^2D_{(M}U^\dagger D_{N)}U\right)\Big)^2\bigg) \\
\nonumber &\hphantom{= \times \bigg(} + \frac{1}{8}\bigg(\mathrm{tr}\Big((\tilde{g}^{(L)})^{-1}\left(w F^{(L)} + \ka \tau^2 D_{(M}U^\dagger D_{N)}U\right)\Big)\bigg)^2 \\
\nonumber &\hphantom{= \times \bigg(} + \mathcal{O}\left(\left((\tilde{g}^{(L)})^{-1}\left(w F^{(L)} + \ka \tau^2 D_{(M}U^\dagger D_{N)}U\right)\right)^3\right)\Bigg) \, ,
\end{align}
\begin{align}
\label{expR} \sqrt{-\mathrm{det}\, \mathbf{A}^{(R)}}& = \sqrt{-\mathrm{det}\, \tilde{g}^{(R)}} \times \\
\nn &\hphantom{=} \times \left(1 + \frac{1}{2}w \mathrm{tr}\left((\tilde{g}^{(R)})^{-1}F^{(R)}\right) + \frac{1}{2}\ka\tau^2\mathrm{tr}\left((\tilde{g}^{(R)})^{-1} D_{(M}U D_{N)}U^\dagger\right)\right. \\
\nonumber &\hphantom{= \times \bigg(}  -\frac{1}{4}\mathrm{tr}\left(\left((\tilde{g}^{(R)})^{-1}\left(w F^{(R)} + \ka \tau^2 D_{(M}U D_{N)}U^\dagger \right)\right)^2\right) \\
\nonumber &\hphantom{= \times \bigg(} + \frac{1}{8}\left(\mathrm{tr}\left((\tilde{g}^{(R)})^{-1}\left(w F^{(R)} + \ka \tau^2 D_{(M}U D_{N)}U^\dagger \right)\right)\right)^2 \\
\nonumber &\hphantom{= \times \bigg(} + \mathcal{O}\left(\left((\tilde{g}^{(R)})^{-1}\left(w F^{(R)} + \ka \tau^2 D_{(M}U D_{N)}U^\dagger \right)\right)^3\right)\bigg) \, ,
\end{align}
where the trace over the space-time indices is denoted by tr. The term linear in $F$ in the first line will vanish upon taking the (flavor) trace of the full expression. Also, as mentioned before, the $U$ covariant derivatives should be considered as of the same order as the field strength. The final expression we obtain for the expansion of the DBI action  \eqref{SDBI} up to quadratic order in the non-Abelian field strength is therefore
\begin{align}
\label{expS} S_{\text{DBI}} &= -M^3N_c \int\mathrm{d}^5x\, V_f(\l,\tau^2)\sqrt{-\mathrm{det}\, \tilde{g}^{(L)}} \times \\
\nn &\hphantom{= -M^3N_c \int} \times\left[\frac{1}{2} + \frac{1}{4}\ka \tau^2 \left((\tilde{g}^{(L)})^{-1}\right)^{(MN)}\, S_{MN} \right. \\
\nn &\hphantom{= -M^3N_c \int\times\bigg[} - \frac{1}{8}w^2\left((\tilde{g}^{(L)})^{-1}\right)^{MN}\left((\tilde{g}^{(L)})^{-1}\right)^{PQ}\mathrm{Tr}\, F^{(L)}_{NP}F^{(L)}_{QM} \\
\nonumber &\hphantom{= -M^3N_c \int\times\bigg[} \left. +\frac{1}{16}w^2\mathrm{Tr}\,\left(\left((\tilde{g}^{(L)})^{-1}\right)^{[MN]}F^{(L)}_{NM}\right)^2 + \mathcal{O}\left((F^{(L)})^3\right)\right]\\
\nn &\hphantom{=-M^3N_c} + (L\leftrightarrow R) \, ,
\end{align}
where we defined the symmetric 2-tensor
\be
\label{defS} S_{MN} \equiv \text{Tr} D_{(M} U^\dagger D_{N)} U = \text{Tr} D_{(M} U D_{N)} U^\dagger \, .
\ee

\section{Tachyon-Chern-Simons terms}

\label{Sec:CS}

We now discuss the TCS term in the brane action, i.e., $S_{CS}$ in~\eqref{v3}. The Tachyon-Chern-Simons term arises as  a part of the Wess-Zumino sector, which has been considered in~\cite{Bigazzi2005,Casero} and adjusted in  connection to V-QCD in~\cite{Jarvinen,spectrum,Arean2016}.

The Wess-Zumino term can be derived in flat-space boundary string
field theory,~\cite{Kennedy:1999nn,Kraus:2000nj,Takayanagi:2000rz}. We sketch here the main points of the construction,
see~\cite{Casero} for details. The expression of the full WZ term is
\be \label{WZdef}
 S_\mathrm{WZ} = T_4 \int C \wedge \mathrm{str}\ e^{i\mathcal{F}},
\ee
where $T_4$ is the tension of the flavor D4 branes, $C$ is a formal sum of the RR potentials $C = \sum_n (-1)^\frac{5-n}{2}C_n$, $\mathrm{str}$ denotes the supertrace as defined in~\cite{Casero}, and $\mathcal{F} = d\mathcal{A} - i\mathcal{A}\wedge\mathcal{A}$ is the curvature of the superconnection
\be
 i\mathcal{A} = \left(\begin{array}{cc}
                      i \mathbf{L} & T^\dagger \\
                      T & i \mathbf{R}
                     \end{array}
\right)
\ee
in terms of the gauge fields  $\mathbf{L}$, $\mathbf{R}$, and the tachyon $T$ defined above. Expanding the exponential in~\eqref{WZdef} we find four different terms
\be \label{WZexp}
 S_\mathrm{WZ} = T_4 \int C_5 \wedge Z_0 + C_3 \wedge Z_2 +C_1 \wedge Z_4 +C_{-1} \wedge Z_6
\ee
where the $Z_{2n}$'s are the coefficients arising from the expansion
of the exponential. The terms in~\eqref{WZexp} play  different roles in
QCD, \cite{Casero,dissect}.
As the theory lives in five dimensions, the first term contains the five-form flux under which the flavor branes are charged.
The second term is important for the correct
holographic implementation of the U(1)$_A$ anomaly, \cite{Casero}.
The third term controls the CP-odd interactions associated with magnetic strings.
Here, we shall only
discuss further the last term, which is the one important for
constructing the baryon solution.

The last term in~\eqref{WZexp} may be written as
\be
\label{SCS1} S_\mathrm{CS} =  \frac{iN_c}{4\pi^2} \int \Omega_5
\ee
where $d \Omega_5 = Z_6$ and we inserted an explicit expression for the constant $F_0 = dC_{-1}$. $F_0$ is the flux that
 is proportional to $N_c$ and supports the bulk geometry associated with the glue. It is the analogue of the RR five-form in the ten-dimensional $AdS_5\times S^5$ solution.
In string theory $Z_6$ can be computed straightforwardly, but it is somewhat nontrivial to carry out the integration to find explicitly the five form $\Omega_5$. This task was done in~\cite{Casero} for the case where the tachyon is proportional to the unit matrix, $T = \tau \mathbb{I}$.

In the rest of this section we derive a generalized expression for the
TCS action, which is the generalization of the results above  to V-QCD and to the more general tachyon configuration with a nontrivial $U$
 matrix in equation (\ref{unitary}).

 First, as we are following a bottom-up approach, there is no reason a priori to resort to the flat space expression of the Wess-Zumino term given in~\eqref{WZdef}. Therefore, we consider a general TCS action which satisfy known constraints from the bulk gauge symmetry as well as the  anomaly structure of QCD. Second, instead of taking the tachyon proportional to the unit matrix, we consider a more general Ansatz (already written down in section~\ref{sec:FullDBI})
\be \label{Simple-T}
 T = \tau U
\ee
where $\tau$ is scalar and $U$ is a generic SU$(N_f)$ matrix. While this Ansatz is not the most general one,
it will be sufficient for our purposes. Notice that the presence of the $U$ matrix allows one to write down expressions which are covariant in the full left and right handed flavor transformations, rather than only the vectorial transformations. Moreover, the fluctuations of the $U$ field are the pions, so that $U$ maps to the exponential $\exp(i \lambda_a \pi_a/f_\pi)$ at the boundary as the notation suggests, which makes it possible to explicitly compare to chiral effective theory.

\subsection{Constructing  the Tachyon-Chern-Simons term} \label{Sec:cs-cons}

We proceed to the construction of the TCS action. We require that it satisfies the following constraints (to be discussed in more detail below):
\begin{itemize}
 \item The TCS action has the expected behavior under discrete symmetries, i.e., it is even under both parity (P) and charge conjugation (C) whose actions are defined in appendix \ref{discrete}.
 \item Its variation under infinitesimal (bulk) gauge transformations is closed, $d \delta \Omega_5 = 0$, and therefore integrates to a boundary term.
 \item The gauge transformation of the boundary term matches with the expression for the flavor anomaly in QCD.
 \item When chiral symmetry is preserved, $\tau=0$, the result agrees with the standard CS action for D-branes.
 \item All IR contributions to observables vanish\footnote{This requirement becomes non-trivial as the type of bulk geometries that are relevant are mildly singular (but are compatible with the Gubser bound).}.
\end{itemize}

In what follows, we  construct the most general action satisfying these
properties. Details are given in
Appendix~\ref{GCS} and  here we  only sketch the main points.

The CS action is even under P if for the simple parity operation acting on the forms (P$_1$ in Appendix~\ref{app: conventions}), $\Omega_5$ is odd. The extra minus sign then comes from reversing the space-time coordinates (P$_2$ in Appendix~\ref{app: conventions}). Under the action of C on the forms, $\Omega_5$ is required to be even.
We therefore start by writing down the most general five-form which is
P$_1$ odd and C even, where the transformation properties of the
various fields are given in Appendix  \ref{app: conventions}.

The ansatz is  given as a linear combination of all possible single-trace 5-forms
composed out  of $U$, the gauge fields,  their covariant derivatives,
and the  one-form $d\tau$,   with the coefficients being functions of the only available scalar $\tau$:
\be
 \Omega_5 = \sum_{i=1}^{45} \bar f_{i}(\tau) F_5^{(i)}[U,\mathbf{L},\mathbf{R}] + d\tau\wedge\sum_{i=1}^{11} g_{i}(\tau)F_4^{(i)}[U,\mathbf{L},\mathbf{R}]
\ee
The forms $F_4^{(i)}$ and $F_5^{(i)}$ are listed explicitly in Appendix~\ref{GCS}.
Notice that we do not expect dependence on the closed string sector (i.e. on the metric or dilaton) to appear in these terms, as it happens in standard string theory.

The most drastic constraint is then the requirement that { $d\delta \Omega_5 = 0$}.
After imposing this constraint, the general solution can be written as
\be\label{Omega1}
\Omega_5 = \Omega_5^0 +\Omega_5^c + dG_4
\ee
where $\Omega_5^0$ is invariant under gauge transformations and $\Omega_5^c$ is closed, so that indeed (trivially) $d \delta \Omega_5 = 0$.

The nontrivial part of the derivation is to show that~\eqref{Omega1} is the only solution. This is discussed in appendix \ref{GCS}.
In~\eqref{Omega1}, $\Omega_5^0$ is the most general gauge covariant 5-form with the expected eigenvalues under P and C:
\begin{align}
\Omega_5^0 &=
\nn f_1(\tau) \big[\text{Tr}(DU\wedge \mathbf{F}^{(L)}\wedge \mathbf{F}^{(L)}U^\dagger)+\text{Tr}(DUU^{\dagger}\wedge \mathbf{F}^{(R)}\wedge \mathbf{F}^{(R)})\big]\\
&+f_2(\tau) \big[\text{Tr}(DU\wedge \mathbf{F}^{(L)}U^\dagger\wedge DUU^\dagger\wedge DUU^\dagger)\nn\\
&\qquad\qquad  +\text{Tr}(DUU^\dagger\wedge \mathbf{F}^{(R)}\wedge DUU^\dagger\wedge DUU^\dagger)\big]\nn\\
&+f_3(\tau) \big[\text{Tr}(DU\wedge \mathbf{F}^{(L)}U^\dagger\wedge \mathbf{F}^{(R)})+\text{Tr}(DUU^\dagger\wedge \mathbf{F}^{(R)}U\wedge \mathbf{F}^{(L)}U^\dagger)\big]\nn\\
&+f_4(\tau) \text{Tr}(DUU^\dagger\wedge DUU^\dagger\wedge DUU^\dagger\wedge DUU^\dagger\wedge DUU^\dagger) \, .
\label{O50Ud1}
\end{align}
and depends on four arbitrary functions of $\tau$.

The  closed term $\Omega_5^c$ is completely fixed (up to exact forms which we include
in $G_4$ in (\ref{Omega1}) ) to be
\be
\label{O5c1} \Omega_5^c = g_0 \text{Tr}((U^\dagger\mathrm{d}U)^5)
\ee
where $g_0$ is a constant. Lastly, $G_4$ is a generic 4-form, i.e., a linear combination of all the possible P$_1$ odd and C even 4-forms:
\be
 G_4 = \sum_{i=1}^{11} h_i(\tau) F_4^{(i)}[U,\mathbf{L},\mathbf{R}] \ .
\ee
Explicit expressions for the forms $F_4^{(i)}$ can be found in Appendix~\ref{GCS}.
Recall that $\Omega_5$ depends only on the boundary value of $G_4$ and therefore only on $h_i(\tau=0)$, so that the functional form of $h_i(\tau)$ is irrelevant for the final result, except that the functions should vanish for $\tau \to \infty$ in order to avoid undesired IR boundary terms.

The next step is to require agreement of the gauge transformation of the boundary term with the flavor anomalies of QCD. That is, we write $\delta \Omega_5^c = d \delta G_4^c$ and set
\be \label{ancond}
 \delta G_4^c + \delta G_4\Big|_\mathrm{bdry} = - \frac{1}{6}  \mathrm{Tr}\left[\L_L\left((\mathrm{d}\mathbf{L})^2 - \frac{i}{2}\mathrm{d}(\mathbf{L}^3)\right) - (L\leftrightarrow R) \right] + d(\cdots) \ .
\ee
where $\L_L$ is the generator of the left-handed gauge transformation and the right hand side encodes the flavor anomalies, see e.g.~\cite{WWZ,Kaymakcalan,Casero}. Notice that the contribution to the action from the last two terms of~\eqref{Omega1} is localized on the UV boundary, because consistency requires that IR contributions vanish.
As we detail in  Appendix \ref{GCS}, the condition~\eqref{ancond} completely fixes these terms near the boundary, and therefore also the action from these terms is fully determined. The result for this part of the  action may be written as
\be
\frac{iN_c}{4\pi^2} \int \Omega_5^c + dG_4 = -\frac{1}{60}\frac{iN_c}{4\pi^2} \int \text{Tr}((U^\dagger\mathrm{d}U)^5) + \frac{iN_c}{4\pi^2}\int G_4\Big|_\mathrm{bdry}
\ee
where
\begingroup
\allowdisplaybreaks
\begin{gather} \label{G4final}
 24\, G_4\Big|_\mathrm{bdry}\! =\Big\{ 2\Big[ \text{Tr}(\mathbf{L}\wedge \mathbf{F}^{(L)}\,U^\dagger\wedge DU)+ \text{Tr}(\mathbf{L}\,U^\dagger\wedge DU\wedge \mathbf{F}^{(L)})\Big]+\\\nonumber
+\Big[\text{Tr}(\mathbf{L}\,U^\dagger\wedge DU\,U^\dagger\wedge \mathbf{F}^{(R)}\,U)+\text{Tr}(\mathbf{L}\,U^\dagger\wedge \mathbf{F}^{(R)}\wedge DU)\Big]+ \\\nonumber
+i\Big[ \text{Tr}(\mathbf{L}\wedge \mathbf{L}\,U^\dagger\wedge \mathbf{R}\wedge DU)- \text{Tr}(\mathbf{L}\wedge \mathbf{L}\,U^\dagger\wedge DU\,U^\dagger\wedge \mathbf{R}\,U)\Big]+  \\\nonumber
+i\Big[ \text{Tr}(\mathbf{L}\wedge \mathbf{F}^{(L)}\,U^\dagger\wedge \mathbf{R}\,U)+ \text{Tr}(\mathbf{L}\,U^\dagger\wedge \mathbf{R}\,U\wedge \mathbf{F}^{(L)})\Big]+ \\\nonumber
+2 i \text{Tr}(\mathbf{L}\wedge \mathbf{L}\wedge \mathbf{L}\,U^\dagger\wedge DU) -2 \text{Tr}(\mathbf{L}\wedge \mathbf{L}\wedge \mathbf{L}\,U^\dagger\wedge \mathbf{R}\,U)+\\\nonumber
 +2\text{Tr}(\mathbf{L}\,U^\dagger\wedge DU\,U^\dagger\wedge DU\,U^\dagger\wedge \mathbf{R}\,U) -\text{Tr}(\mathbf{L}\,U^\dagger\wedge DU\wedge \mathbf{L}\,U^\dagger\wedge DU)+ \\\nonumber
-2 i \text{Tr}(\mathbf{L}\,U^\dagger\wedge \mathbf{R}\,U\wedge \mathbf{L}\,U^\dagger\wedge DU)
-2 i\text{Tr}(\mathbf{L}\,U^\dagger\wedge DU\,U^\dagger\wedge DU\,U^\dagger\wedge DU)\Big\}+ \\\nonumber
+\ \Big\{L\leftrightarrow R\Big\}+\text{Tr}(\mathbf{L}\,U^\dagger\wedge \mathbf{R}\,U\wedge \mathbf{L}\,U^\dagger\wedge \mathbf{R}\,U) \ .
\end{gather}
\endgroup
In particular, $g_0 = -1/60$.
The explicitly $L \leftrightarrow R$ symmetrized expression is given
in  Appendix \ref{GCS}. Notice that the last term in~\eqref{G4final} is already symmetric. As one can check, this expressions matches (up to a four dimensional total derivative)  the Wess-Zumino terms of chiral Lagrangians given in the literature~\cite{WWZ,Kaymakcalan}, see appendix~\ref{GCS}.

There is however a subtlety in the above derivation. Namely, we
assumed that the IR contributions from integrating the closed terms
vanish. For the contribution from $dG_4$ in~\eqref{Omega1} this can be
easily obtained by adjusting the $\tau$ dependence of the coefficient
functions as $\tau \to \infty$. But in the $\Omega_5^c$ term, the
coefficient $g_0$ is required to be a constant, which cannot be set to
zero in the IR. Therefore an IR contribution seems unavoidable. We
solve this by requiring that the solution for $U$ is such that the IR
boundary term vanishes. In Appendix \ref{GCS} we argue that this is
satisfied given relatively mild assumptions on the asymptotic IR
behavior of $U$. Such an asymptotic regularity condition for $U$ may
be expected due to the following reason: the geometry in the confined
phase ends in an IR singularity at $r \to \infty$, where all
components of the metric vanish so that the space-time shrinks to a
single point~\cite{ihqcd,Jarvinen}. In our choice of coordinates, this single point is
seemingly described by a 4-dimensional manifold at the boundary $r \to
\infty$. At this IR boundary, regularity conditions may be needed in
analogy to spherical or cylindrical coordinates in flat space, where
the single point at the origin $r=0$ is mapped to a higher dimensional
space (sphere), and regularity conditions for the angular dependence
at $r=0$ is required to ensure the regularity of the full
solution.

 More generally, even if
the space-time is singular at $r\to \infty$, the holographic
consistency of  models like the present one
relies on considering as physical only field configurations which vanish (or
more precisely, are normalizable in a precise sense given by the
radial Hamiltonian) at the IR endpoint of space-time. Often this condition
can be imposed without having to specify extra input at the
singularity  (in which case we say that the singularity is {\em
  repulsive} and the holographic model is {\em
  calculable} \cite{ihqcd,Charmousis:2010zz}). In short, our IR boundary
conditions  for holographically  acceptable configurations must be
such that
the term (\ref{O5c1}) does not contribute in the IR.
This is satisfied  in particular  by the baryon solutions we discuss in
later sections.

After the boundary term has been fixed, the only free functions in our results are the $f_i$'s in~\eqref{O50Ud1}. They are however not fully arbitrary: due to the requirement of vanishing IR contributions to the action, these functions should vanish fast enough in the IR, i.e.,  as $\tau \to \infty$. Moreover, the analysis of~\cite{Ishii} suggests that the Tachyon-Chern-Simons action should vanish faster than the DBI action in the IR.

There are also conditions at the UV boundary. For the chirally-symmetric vacuum, for which $\tau =0$ and $U= \mathbb{I}$, we expect that the Tachyon-Chern-Simons action should reduce to the standard expression
\begin{align}
 S_\mathrm{CS}(\tau=0) = \frac{N_c}{24\pi^2}\int \text{Tr}\bigg(&\mathbf{L}\wedge\mathbf{F}^{(L)}\wedge \mathbf{F}^{(L)} + \frac{i}{2}\mathbf{L}\wedge\mathbf{L}\wedge\mathbf{L}\wedge\mathbf{F}^{(L)} &\\\nonumber
& - \frac{1}{10}\mathbf{L}\wedge\mathbf{L}\wedge\mathbf{L}\wedge\mathbf{L}\wedge\mathbf{L} - (L\leftrightarrow R) \bigg) &
\end{align}
up to boundary terms. This is the case if
\be
 f_1(0) = -\frac{1}{6} \ , \quad f_2(0) = \frac{i}{12} \ , \quad f_3(0) = -\frac{1}{12} \ , \quad f_4(0) = \frac{1}{60} \ .
\ee

Finally, setting $U = \mathbb{I}$, the TCS action reduces to that of~\cite{Casero} if
\begin{align}
 f_1(\tau) &= -\frac{1}{6}e^{-\tau^2}\,,& f_2(\tau) &= \frac{i}{12}(1+\tau^2)e^{-\tau^2}\,,\nn\\
 f_3(\tau) &= -\frac{1}{12}e^{-\tau^2}\,,& f_4(\tau) &= \frac{1}{120}(2+2\tau^2+\tau^4)e^{-\tau^2} \, .
 \label{CKPftext}
\end{align}
That is, our result generalizes the expressions of~\cite{Casero} to $U \ne \mathbb{I}$ for these choices of the functions. We remark that the action with this choice satisfies all the requirements discussed above. A simple generalization of these functions, which we use below, is to allow the normalizations of the tachyon field between the DBI and TCS terms to be different. We choose
\begin{align}
 f_1(\tau) &= -\frac{1}{6}e^{-b \tau^2}\,,& f_2(\tau) &= \frac{i}{12}(1+b\tau^2)e^{-b\tau^2}\,,\nn\\
 f_3(\tau) &= -\frac{1}{12}e^{-b \tau^2}\,,& f_4(\tau) &= \frac{1}{120}(2+2b \tau^2+b^2\tau^4)e^{-b\tau^2} \, ,
 \label{CKPfgen}
\end{align}
where $b$ is a positive constant. It was argued in~\cite{Ishii}, within an approximation scheme for the baryon, that regularity of the solution requires $b>1$. We shall observe the same for the baryon solutions considered in this article.

The precise functions $f_i(\tau)$ here were chosen by modifying slightly the string theory result, based on the superconnection formalism, \cite{Casero}.
The modification is the parameter $b$ inserted in the exponent.
 However, nothing guarantees that this is the correct choice for QCD. More constraints on these functions should be derived in order to fix their form.

The TCS terms we have constructed,  are
 written in terms of $\tau$ and $U$ separately. This is general
enough, if the tachyon can be split according to equation (\ref{Simple-T}). However, as discussed in more detail in section \ref{sec:masses}, in the general case with non-zero quark masses, one has to use the more general tachyon ansatz
\be
T = H U
\ee
where now $H$ is a Hermitian matrix, and the TCS terms  must be generalized.

\section{Cylindrically symmetric ansatz for a single baryon}

\label{Sec:ans}

As we discussed in the introduction, the presence of the TCS terms is
crucial when searching for baryon solutions. Indeed, the TCS action is
responsible for stabilising the baryon size and position.

If we think of this model  as originating from a
five-dimensional non-critical string theory, a baryon is described by a D$_0$ brane, \cite{ihqcd1}.
The analogue of the $C_4$ flux sourcing the D$_3$ branes, in this case  is a zero-form field strength  $F_0\sim N_c$. Its Chern-Simons coupling on the one-dimensional world-volume of the D$_0$ brane is
\be
\int d\tau A_{\tau}F_0\sim N_c\int d\tau A_{\tau}
\ee
where $A_{\tau}$ is the world-line gauge field. This is the analogue of the $F_5\wedge A$ Chern-Simons coupling on the D$_5$ baryon brane in N=4 sYM.
 As usual, fundamental string end-points are charged under $A_{\tau}$ and $N_c$ of them are needed to screen the induced charge on the D$_0$ (baryon) brane,  as in the $N=4$ sYM case.
Moreover, on the D$_4$+$\overline {\rm D}_4$ flavor branes there is a $C_1$ coupling to the flavor instanton number. $C_1$ is the RR gauge field under which D$_0$ branes are minimally charged. Therefore D$_0$ number transmutes to flavor instanton number as in the SS model, \cite{Sugimoto,Yi}.

Therefore, in the gravity approximation that we shall be using, a single static baryon D$_0$ brane, is realised in the bulk as a  Euclidean instanton of
the non-Abelian bulk gauge fields extended in the three
spatial directions plus the holographic  direction.  In this section
we describe the ansatz that is used to compute the instanton solution in the bulk. On general grounds, it is expected that an appropriate ansatz can be determined by requiring that the solution is maximally symmetric, that is symmetric under all the symmetries of the bulk action \eqref{v3} compatible with the boundary conditions. In the case of the baryon solution, this means all the symmetries of the action compatible with the baryon number being non-zero \footnote{A non-zero baryon number breaks in particular Lorentz invariance. It also breaks charge conjugation, which should send a baryon solution to a distinct anti-baryon solution.}.

\subsection{Ansatz for the glue sector}

As explained in the next subsection, we shall consider a baryon whose flavor quantum numbers are a U(2) subgroup of the  U(N$_f$) flavor group. This implies that the flavor action (composed of the DBI \eqref{expS} and TCS \eqref{SCS1} actions) for the baryon ansatz does not depend on $N_f$\footnote{When considering that the on-shell action for the baryon solution should correspond to the classical contribution to the baryon mass, this is the expected scaling at large $N_c$. This scaling was equally reproduced in the context of the WSS model \cite{Hata07}.} and is of order $N_c$. On the other hand, the glue action \eqref{v4} is of order $N_c^2$. So, at leading order in $N_c$, the glue sector composed of the metric and dilaton is not affected by the presence of the baryon and is identical to the vacuum solution. The latter depends only on the holographic coordinate $r$
\begin{equation}
\label{ds2}\mathrm{d}s^2 = \ex^{2A(r)}(-\mathrm{d}t^2 + \mathrm{d}\textbf{x}^2 + \mathrm{d}r^2)\, ,
\end{equation}
\be
\label{ansl} \lambda = \lambda(r) \, .
\ee
Let us comment some more on this result that the baryon is a probe on the color background.  
In fact, it is well-known that SU(N$_f$) instantons are constructed from a single embedded SU(2) instanton that is then conjugated to cover the full SU(N$_f$) group.
This gives rise to many charged moduli associated to the one-instanton solution. However, here such parameters are fixed and are not moduli.
Therefore, in this case the effective (active) number  of flavors is two, and a non-backreaction approximation for a single baryon is valid. Note, that if flavor were to be gauged, this approximation would be invalidated. 

\subsection{Gauge fields ansatz}

The left and right handed gauge fields are denoted
\begin{equation}
\label{defLR}\mathbf{L} = L + \hat{L} \mathbb{I}_{N_f} \sp \mathbf{R} = R + \hat{R} \mathbb{I}_{N_f} \, ,
\end{equation}
where $L$ and $R$ correspond to the $SU(N_f)$ part of the gauge fields, and $\hat{L}$ and $\hat{R}$ to the $U(1)$ part. From these we define the vector and axial vector gauge fields as
\begin{equation}
\label{defVA}\mathbf{V} = \frac{\mathbf{L}+\mathbf{R}}{\sqrt{2}} \sp \mathbf{A} = \frac{\mathbf{L}-\mathbf{R}}{\sqrt{2}} \, .
\end{equation}
We look for a static instanton configuration for the $U(N_f)$ gauge fields.
This configuration belongs to a non-trivial class of the homotopy group of $U(N_f)$ on the 3-sphere at infinity of the 4-dimensional Euclidean space spanned by $(\vec{x},r)$. Because for any $N_f > 1$, the homotopy groups of $U(N_f)$ and $SU(2)$ are equal
\be
\label{p3SUN} \pi_3\left(U(N_f)\right) = \pi_3\left(SU(2)\right) = \mathbb{Z} \, ,
\ee
a $U(N_f)$ instanton can be constructed by embedding an $SU(2)$ instanton in $U(N_f)$ (acting with global $U(N_f)$ on an instanton for an $SU(2)$ subgroup of $U(N_f)$).

Because the TCS action \eqref{SCS1} contains cubic couplings for the gauge fields, a consistent ansatz should contain all the matrices in the Lie Algebra of U(N$_f$) that can be written as a product of two SU(2) generators
\be
\label{t1t2} \s^a \s^b = \d^{ab} \mathbb{I}_2 + i\e^{abc}\s^c \, , 
\ee
where the $\s^a$'s are the Pauli matrices and $\mathbb{I}_2$ is the 2 by 2 identity matrix in the same subsector of the chiral group as the SU(2) subgroup. This implies that the SU(2) instanton couples via the TCS term to the gauge field in the direction of $\mathbb{I}_2$. The full flavor structure of the baryon ansatz is therefore that of a U(2) subgroup
\be
\label{U2gf} \mathbf{L} = L^a \frac{\s^a_L}{2} + L^{\text{T}} \mathbb{I}_{2,L} \sp \mathbf{R} = R^a \frac{\s^a_R}{2} + R^{\text{T}} \mathbb{I}_{2,R} \, ,
\ee
where the superscript T stands for trace, as the corresponding part contains the abelian part of the gauge field. Note however that, although $\mathbb{I}_2$ generates a U(1) subgroup of the chiral group, in general\footnote{With the exception of $N_f = 2$.} it is not a subgroup of the abelian part of U(N$_f$). In other words, $\mathbb{I}_2$ is a combination of $\mathbb{I}_{N_f}$ and an SU(N$_f$) generator
\be
\label{I2exp} \mathbb{I}_2 =  \frac{2}{N_f} \mathbb{I}_{N_f} + \left(\mathbb{I}_2 - \frac{2}{N_f}\mathbb{I}_{N_f}\right)  \, .
\ee
The traceless matrix that appears in \eqref{I2exp} is not any matrix, as it is equal to minus the traceless part of the strong hypercharge
\be
\renewcommand{\arraystretch}{0.2}
Y \equiv 
\frac{1}{N_c} \mathbb{I}_{N_f} + 
\begin{pmatrix}
0 & & & &  \\
& 0 & & &  \\
 & & 1 & & \\
 & & & \ddots & \\
& & & & 1
\end{pmatrix}  
= \left(\frac{1}{N_c}+\frac{N_f-2}{N_f}\right)\mathbb{I}_{N_f} - \left(\mathbb{I}_2 - \frac{2}{N_f}\mathbb{I}_{N_f}\right) \, , 
\ee
where by convention the 2 flavors of the SU(2) subgroup are assumed to correspond to the down and up (note that at zero quark mass, this choice is arbitrary). The fact that the gauge field ansatz \eqref{U2gf} contains a part in the direction of $\mathbb{I}_2$ is therefore a sign that the baryon is charged under baryon number and hypercharge. Because a single combination of the two charges appears, the baryon number and hypercharge will not be independent for the classical soliton solution. Specifically, we find that for this ansatz
\be
\label{NBY} Y = N_B \, ,
\ee
which is the expected result for a baryon composed of the first two flavors\footnote{This does not mean that the baryon spectrum does not contain higher hypercharge states, with $Y > N_B$. The baryon spectrum is calculated by quantizing the collective modes of the soliton, including in particular the rotations in isospin space. Such rotations move the baryon around in the full chiral group, so they will imply the existence of higher hypercharge baryon states.}.

To find a relevant U(2) instanton ansatz we follow \cite{Pomarol08} and look for a configuration that is invariant under a maximal set of symmetries of the action \eqref{v3}, compatible with a finite baryon number. In particular, we look for a U(2) instanton solution that is invariant under cylindrical transformations. These correspond to rotations in the 3-dimensional space spanned by $\mathbf{x}$, up to a global SU(2) rotation.
 
As we look for a static solution, we impose in addition invariance under time-reversal\footnote{The action on the time component of the gauge fields is necessary for the TCS action to be invariant under time reversal.}
\be
\label{Tinv} t \to -t \sp L^\text{T} \to - L^\text{T} \sp R^\text{T} \to - R^\text{T} \, .
\ee
Note that the definition of time-reversal reduces to that of \cite{Pomarol08} for $N_f = 2$ flavors. For $N_f > 2$, the time component of the non-abelian gauge field will be non-zero, but proportional to the abelian part. The same was observed in the context of the WSS model in \cite{Hata07}.

Then, the static, cylindrically symmetric ansatz takes the form:
\begin{equation}
\label{ansatzSU2i} L_i^a = -\frac{1+\phi_2^L(\xi,r)}{\xi^2}\e_{iak}x_k + \frac{\phi_1^L(\xi,r)}{\xi^3}(\xi^2\d_{ia}-x_ix_a) + \frac{A_1^L(\xi,r)}{\xi^2}x_ix_a \, ,
\end{equation}
\begin{equation}
\label{ansatzSU2r} L_r^a = \frac{A_2^L(\xi,r)}{\xi} x_a \, ,
\end{equation}
\begin{equation}
\label{ansatzU1} L^\text{T}_0 = \Phi^L(r,\xi) \, ,
\end{equation}
where $i,k=1,2,3$ refer to spatial indices, $\xi \equiv \sqrt{x_1^2+x_2^2+x_3^2}$ is the 3-dimensional spatial radius and $a=1,2,3$ is the index for the components in the $SU(2)_L$ basis (the Pauli matrices $\sigma_a$ divided by 2). A similar ansatz can be written for the right-handed gauge field with corresponding fields $\phi_1^R, \phi_2^R, A_1^R, A_2^R, \Phi^R$. These are a priori independent from the left-handed degrees of freedom, as is the embedding of the corresponding $SU(2)_R$ in $SU(N_f)_R$.

The choice of the ansatz partially fixes the gauge but there is still a residual $U(1)_L\times U(1)_R$ invariance that preserves the cylindrical symmetry,  corresponding to the $SU(2)$ transformation
\be
\label{resg} g_{(L/R)} =  \exp\left(i \a_{(L/R)}(\xi, r){x \cdot \s \over
    2\xi}\right) \, ,
\ee
where
\be
\label{defxds} x\cdot \s \equiv x^a \s^a \, .
\ee
Under this gauge transformation, $\Big(A_1^{(L/R)},A_2^{(L/R)}\Big)$ is the gauge field, $\phi_1^{(L/R)} + i\phi_2^{(L/R)}$ has charge +1 and $\Phi^{(L/R)}$ is neutral.

The V-QCD action possesses another discrete parity symmetry
\be
\label{Pinv} P \,\, : \,\, \mathbf{x}\to -\mathbf{x} \sp L\leftrightarrow R \, .
\ee
{A general instanton solution is a linear combination of a P-even instanton and a P-odd instanton. As discussed in Appendix \ref{Podd}, only the P-even part can generate a finite energy solution.} So we also impose the parity symmetry \eqref{Pinv} on the ansatz \eqref{ansatzSU2i}-\eqref{ansatzU1} for the instanton solution. This relates right and left-handed quantities in the following manner:
\begin{equation}
\label{parA} A_1 \equiv A_1^L = -A_1^R \sp A_2 \equiv A_2^L = - A_2^R \, ,
\end{equation}
\begin{equation}
\label{parf} \phi_1 \equiv \phi_1^L = -\phi_1^R \sp \phi_2 \equiv \phi_2^L = \phi_2^R \, ,
\end{equation}
\begin{equation}
\label{parF} \Phi \equiv \Phi^L = \Phi^R \, .
\end{equation}
Also, the right and left-handed
gauge fields should belong to the same $SU(2)$ subgroup of
$U(N_f)$\footnote{This implies in particular that the manifest
  $SU(2)_V$ invariance of the $SU(2)$ ansatz simply maps to invariance
  under the conserved diagonal subgroup $SU(N_f)_V$ of the chiral
  symmetry when lifted to $SU(N_f)$.}. Our ansatz is now fully
specified by 5 real  functions
\be
\label{def2D} A_{\bar{\mu}} \equiv
(A_1,A_2) \sp \phi \equiv \phi_1 + i \phi_2 \;\;\; \text{and} \;\;\; \Phi \, ,
\ee
depending on
the two variables
$x^{\bar{\mu}}\equiv (\xi,r)$, that we will use as coordinates on a 2D
space. The constraints  \eqref{parA}-\eqref{parF} fix the vector part of the remaining
$U(1)_L\times U(1)_R$ gauge invariance, leaving only a residual
(axial) $U(1)$ invariance,  corresponding to the $SU(2)_A$
transformation
\be
\label{resg1}
 g_L = g_R^{\dagger} = \exp\left(i \a(\xi, r){x\cdot \s
  \over 2\xi}\right) \, .
\ee
Under the transformation \eqref{resg1}, $A_{\bar{\mu}}$ is the gauge field, $\phi$ has charge +1 and $\Phi$ is neutral, with gauge transformations
\be
\label{gtgf} A_{\bar{\mu}} \to A_{\bar{\mu}} + \partial_{\bar{\mu}}\a \sp \phi \to \ex^{i\a} \phi \sp \Phi \to \Phi \, .
\ee

\subsection{Tachyon ansatz}

\label{Sec:tans}

The most general cylindrically symmetric\footnote{Up to a global $SU(2)$ transformation.} ansatz for the tachyon field is
\be
\label{ansT} T^{ij} = \rho(r,\xi) \d^{ij} + \f(r,\xi) \frac{(x\cdot\sigma)^{ij}}{\xi} \, ,
\ee
where $\rho$ and $\f$ are two complex scalar fields and $i$ and $j$ are the indices in the fundamental representation of the right and left handed $SU(2)$ subgroup, respectively. The parity transformation acts on the tachyon field as $T(x) \to T^{\dagger}(-x)$ \cite{Casero}. Imposing the parity symmetry implies that $i$ and $j$ in  \eqref{ansT} describe the same space and constrains $\rho$ and $\f$ to obey
\be
\label{consT} \rho \in \mathbb{R} \sp \f \in i\mathbb{R} \, .
\ee
The tachyon matrix restricted to the $SU(2)$ subgroup $T^{SU(2)}$  can therefore be cast in the shape of a scalar times an $SU(2)$ matrix
\begin{align}
\nn T^{SU(2)} = \rho \mathbb{I}_2 + \f \frac{x\cdot\s}{\xi} &= \tau(r,\xi) \left(\cos\theta(r,\xi)\, \mathbb{I}_2 + i\sin\theta(r,\xi)\frac{x\cdot\s}{\xi}\right) \\
\label{TU2} &= \tau(r,\xi) \exp{\left(i\theta(r,\xi)\frac{x\cdot\s}{\xi}\right)} \, ,
\end{align}
where $\tau$ and $\theta$ are respectively the modulus and phase of
the complex number $\rho + \varphi$, where $\rho$ and $\varphi$ are defined in  \eqref{TU2},
\be
\label{deftth} \tau \equiv \sqrt{|\rho|^2+|\varphi|^2} \sp \theta \equiv \arctan{\left(-\frac{i \varphi}{\rho}\right)} \, .
\ee
Equation \eqref{TU2} reproduces the Skyrmion ansatz (equation (16) in \cite{Pomarol07}) for the unitary part of the tachyon. The $SU(N_f)$ tachyon ansatz is then obtained by embedding $T^{SU(2)}$ in $SU(N_f)$
\be
\label{TU} T = \tau(r,\xi) U(r,\xi) \, ,
\ee
where $U(r,\xi)$ is the $SU(N_f)$ matrix resulting from the embedding of $\exp{\left(i\theta(r,\xi)\frac{x\cdot\s}{\xi}\right)}$.

Note that under the residual $U(1)_A$ gauge freedom  \eqref{resg1}, the tachyon in equation \eqref{TU2} transforms as
\be
\label{gtt} \theta\to \theta - \a \, .
\ee

\paragraph*{Source and vev}

Depending on whether we set the quark mass
$m$ to 0 or not, the vev term should be identified differently from
the tachyon near-boundary expansion. In this work,  we will consider the case in which all quarks are massless. The situation with non-zero quark mass will be treated elsewhere.

In the massless quark case, the chiral condensate is simply the coefficient of the leading term of the tachyon, in the near-boundary expansion as $r\to 0$
\be
\label{vevm0} T^{SU(2)}(r,\xi) = \ell \Sigma(\xi) \exp{\left(i\theta(0,\xi)\frac{x\cdot\s}{\xi}\right)} \, r^3(-\log(r\L))^{-c} \left(1+\cdots\right) \, .
\ee
We used the fact that the near-boundary behavior of the tachyon modulus is still given by \eqref{tUV}, except that the amplitude of the chiral condensate $\Sigma$ generically depends on $\xi$ due to the presence of the baryon. From \eqref{vevm0}, the pion matrix is identified to be
\be
\label{Pmat} U_P(\xi)^{\dagger} \equiv \exp{\left(i\theta(0,\xi)\frac{x\cdot\s}{\xi}\right)} \, .
\ee
The associated near-boundary expansion of the flavor gauge fields can be found in Appendix \ref{Sec:asymUV}.

To summarize the content of this section, the ansatz for the instanton solution \eqref{ansatzSU2i}-\eqref{ansatzU1}, \eqref{parA}-\eqref{parF} and \eqref{TU2} contains 7 dynamical fields
\be
\label{sumans} \Phi(r,\xi) \sp \phi(r,\xi)\equiv \phi_1(r,\xi) + i \phi_2(r,\xi) \sp A_{\bar{\mu}}(r,\xi) \equiv  \left(A_\xi(r,\xi), A_r(r,\xi)\right) \, ,
\ee
\be
\nonumber \tau(r,\xi) \sp \theta(r,\xi) \, ,
\ee
with a $U(1)$ gauge redundancy under which the fields transform as
\be
\label{gtsum} \phi \to \ex^{i\a} \phi \sp A_{\bar{\mu}} \to A_{\bar{\mu}} + \partial_{\bar{\mu}}\a \sp \theta \to \theta - \a \, .
\ee

\subsection{Comments on non-zero quark masses} \label{sec:masses}

In the presence of a non-trivial quark mass matrix $M_{ij}$, the
near-boundary asymptotics (\ref{vevm0}) must be modified, and the
leading asymptotics  is now:
\be \label{mass1}
T_{ij} \simeq M_{ij} \, r (-\log r\Lambda)^{c}  + \ldots
\ee
Since the nontrivial matrix nature of  $T$ starts at leading order, we
cannot assume the decomposition (\ref{TU}) but the most general form:
\be \label{mass2}
T = H(r,\xi) U(r,\xi)
\ee
where the scalar field $\tau(r,\xi)$ is replaced by the Hermitian
matrix field $H(r,\xi)$. Near the boundary $r\to 0$, equation
(\ref{mass1}) requires to leading order:
\be \label{mass3}
H (r,\xi) = H_0  r (-\log r\Lambda)^{c} + O(r^3) , \quad
 U(r,\xi) =  U_0 + O(r^2) \qquad r \to 0
\ee
where $H_0$ is a  constant Hermitian matrix  and $U_0$ is a
constant unitary matrix such that $(H_0 U_0 )_{ij} = M_{ij}$.

Notice that now the matrix $U$ contributes to the source term in the
near-boundary asymptotics. As a consequence, in this case  we cannot
interpret $U(r=0)$ as the pion matrix as in equation (\ref{Pmat}) in the massless case.

Furthermore, due to  equation (\ref{mass3}), $U$ is constrained to go
to a constant matrix  as a function of $\vec{x}$, at
leading order in $r$. This is very different from the massless quark
case, in which the leading asymptotics of $U(r,\xi)$ is unconstrained,
and in particular it is allowed to be a non-trivial function of the
space-time coordinates to leading order in $r$.  This changes
drastically the on-shell asymptotics of our TCS terms\footnote{even
  assuming that the quark mass matrix is proportional to the identity,
in which case the decomposition  (\ref{TU}) still holds.}.

As a consequence, the baryon ansatz has to be modified in the massive
quark case, and we will not discuss this case further in the present work.

\section{Constructing the baryon solution}

\label{Sec:consB}

We discuss in this section the construction of the baryon solution. As mentioned in the previous section, a baryon state is realized in the bulk as an instanton solution on the 4D Euclidean space parametrized by $(r,\vec{x})$. The instanton is a configuration of the ansatz of equations \eqref{ansatzSU2i}-\eqref{ansatzU1} and \eqref{TU2} that obeys the bulk equations of motion. Constructing the baryon solution therefore requires deriving the equations of motion obeyed by the 7 fields of the
ansatz in \eqref{sumans}. These field equations are obtained by
substituting the ansatz into the general field equations  \eqref{abEoM},\eqref{UEoM} and \eqref{nabEoM}. We present here the general procedure for the computation of the equations of motion and give more details in Appendix \ref{App:EoM}.

As discussed in the previous section, the baryon can be treated as a flavor probe on the glue background. Note that the modulus of the tachyon field $\tau$ is also non-zero in vacuum \eqref{vac1}. In the Veneziano limit where $N_c$ and $N_f$ are both large and of the same order, the tachyon modulus couples to the glue sector at leading order in $N_c$. Because of that, the back-reaction of the baryon on the tachyon background is also of order $\mathcal{O}(1/N_c)$. The leading order baryon solution is therefore a probe on the vacuum background, and the back-reaction on the background can be computed order by order in $1/N_c$. In the following, we first discuss the leading order probe baryon solution and then consider the back-reaction. We do not expect the color back-reaction to be qualitatively important for the flavor structure of the baryon, which is its most important dynamical property. This motivates the approximation that we consider later, where we compute the back-reaction on the tachyon modulus alone, assuming the color background to remain unchanged. 

We start the discussion of the baryon construction by a few general results. First note that, for this particular ansatz, the effective metric
$\tilde{g}_{MN}$ in equation (\ref{defgt})  is the same in the L and R
sectors. We will simply denote it $\tilde{g}_{MN}$  and it is given by
the matrix:
\be
\label{ansg}
\begin{pmatrix}
-\ex^{2A} & -\partial_r\Phi & -\frac{x_1}{\xi}\partial_{\xi}\Phi & -\frac{x_2}{\xi}\partial_{\xi}\Phi & -\frac{x_3}{\xi}\partial_{\xi}\Phi \\
\partial_r\Phi & \ex^{2A} \! + \!\ka(\partial_r\tau)^2 & \ka\frac{x_1}{\xi}\partial_r\tau\partial_{\xi}\tau & \ka\frac{x_2}{\xi}\partial_r\tau\partial_{\xi}\tau & \ka\frac{x_3}{\xi}\partial_r\tau\partial_{\xi}\tau \\
\frac{x_1}{\xi}\partial_{\xi}\Phi & \ka\frac{x_1}{\xi}\partial_r\tau\partial_{\xi}\tau & \ex^{2A} \! + \! \ka\left(\frac{x_1}{\xi}\right)^2\!(\partial_{\xi}\tau)^2 & \ka\frac{x_1}{\xi}\frac{x_2}{\xi}(\partial_{\xi}\tau)^2 & \ka\frac{x_1}{\xi}\frac{x_3}{\xi}(\partial_{\xi}\tau)^2 \\
\frac{x_2}{\xi}\partial_{\xi}\Phi & \ka\frac{x_2}{\xi}\partial_r\tau\partial_{\xi}\tau & \ka\frac{x_1}{\xi}\frac{x_2}{\xi}(\partial_{\xi}\tau)^2 & \ex^{2A} \! + \! \ka\left(\frac{x_2}{\xi}\right)^2\!(\partial_{\xi}\tau)^2 & \ka\frac{x_2}{\xi}\frac{x_3}{\xi}(\partial_{\xi}\tau)^2\\
\frac{x_3}{\xi}\partial_{\xi}\Phi & \ka\frac{x_3}{\xi}\partial_r\tau\partial_{\xi}\tau & \ka\frac{x_1}{\xi}\frac{x_3}{\xi}(\partial_{\xi}\tau)^2 & \ka\frac{x_2}{\xi}\frac{x_3}{\xi}(\partial_{\xi}\tau)^2 & \ex^{2A} \! + \! \ka\left(\frac{x_3}{\xi}\right)^2\!(\partial_{\xi}\tau)^2
\end{pmatrix} ,
\ee
where the order of the columns (and lines) is $(0|r|1|2|3)$. Its determinant is:
\begin{align}
\label{detg} -\mathrm{det}\, \tilde{g} = \ex^{10A} - \ex^{6A}w^2\left((\partial_r\Phi)^2 + (\partial_{\xi}\Phi)^2\right) &+ \ex^{8A}\ka\left((\partial_r\tau)^2+(\partial_{\xi}\tau)^2\right) -\\
\nn &-\ex^{4A}\ka w^2\left(\partial_r\Phi\partial_{\xi}\tau - \partial_{\xi}\Phi\partial_r\tau\right)^2 \, .
\end{align}
Another useful observation is that, because the equations of motion are covariant under the residual gauge transformations \eqref{gtsum}, the phase $\theta$ in the tachyon ansatz \eqref{TU2} can be absorbed into the gauge field. By doing so, the dynamical field content \eqref{sumans} can be reduced to a set of 6 fields invariant under the residual gauge freedom. In practice, if we define
\be
\label{defgth} g(\theta) \equiv \exp{\left(i\theta \frac{x\cdot\s}{2\xi}\right)}\, ,
\ee
then we consider the following redefinition of the gauge fields
\be
\label{LtoLt} \mathbf{L}_M \to \tilde{\mathbf{L}}_M \equiv g(\theta)\mathbf{L}_M g(\theta)^\dagger + ig(\theta)\partial_M g(\theta)^\dagger  \, ,
\ee
\be
\label{RtoRt} \mathbf{R}_M \to \tilde{\mathbf{R}}_M \equiv g(\theta)^\dagger\mathbf{R}_M g(\theta) + ig(\theta)^\dagger\partial_M g(\theta) \, ,
\ee
which for the ansatz \eqref{ansatzSU2i}-\eqref{ansatzU1} is equivalent to
\be
\label{LtoLtcyl} A_{\bar{\mu}} \to \tilde{A}_{\bar{\mu}} \equiv  A_{\bar{\mu}} + \partial_{\bar{\mu}}\theta \sp \phi \to \tilde{\phi} \equiv \ex^{i\theta}\phi \sp \Phi \to \Phi \, .
\ee
From  \eqref{gtsum}, it is clear that the gauge field thus redefined is invariant under the residual gauge transformation  \eqref{resg1}.

We shall use these gauge-invariant fields in some of our future calculations in this paper. When we do that, we will always write the tildes so that it is clear that we are using the gauge-invariant fields. Working with the gauge-invariant fields is identical to working in the unitary gauge.

\subsection{Probe instanton}

\label{Sec:probe}

At leading order in $N_c$, one can consider the probe regime where the
geometry of the bulk, the dilaton and the modulus of the tachyon field are fixed to their background value
(describing the V-QCD vacuum solution) and search for an instanton solution for the gauge fields, plus an associated non-trivial solution for the unitary part of the tachyon field. In our case, there is no Abelian field strength in the background (no baryon number density in the vacuum). Because of this, it is consistent to expand the DBI action at quadratic order in the Abelian field strength also
\begin{align}
\label{expSprobe} S_{\text{DBI}} = -M^3N_c &\int\mathrm{d}^5x\, V_f(\l,\tau^2) \sqrt{-\mathrm{det}\, \tilde{g}} \,\times \\
\nn &\times \left(\left[\frac{1}{2}\! +\! \frac{1}{4}\ka \tau^2 \left(\tilde{g}^{-1}\right)^{MN}S_{MN} \! -\! \frac{1}{8}w^2\left(\tilde{g}^{-1}\right)^{MN}\left(\tilde{g}^{-1}\right)^{PQ}\text{Tr}\, \mathbf{F}^{(L)}_{NP}\mathbf{F}^{(L)}_{QM} + \right.\right. \\
\nn  &\hphantom{\times \left(\left[\frac{1}{2} + \frac{1}{4}\ka \tau^2 \left(\tilde{g}^{-1}\right)^{MN}S_{MN}\right.\right.} \left.   + \mathcal{O}\left((\mathbf{F}^{(L)})^3\right)\bigg] + (L\leftrightarrow R)\right) \, ,
\end{align}
where $S_{MN}$ was defined in \eqref{defS} and $\tilde{g}$ is now the effective background metric
\be
\label{gtprobe} \tilde{g}_{MN} \equiv g_{MN} + \ka(\l) \partial_M\tau\partial_N\tau = g_{MN} + \ka(\l(r))\,  (\tau'(r))^2 \d_M^r \d_N^r \, ,
\ee
which takes a diagonal form
\be
\label{ansgprobe} \tilde{g} =
\begin{pmatrix}
-\ex^{2A} & 0 & 0 & 0 & 0 \\
0 & \ex^{2A} + \ka(\partial_r\tau)^2 & 0 & 0 & 0 \\
0 & 0 & \ex^{2A} & 0 & 0 \\
0 & 0 & 0 & \ex^{2A} & 0\\
0 & 0 & 0 & 0 & \ex^{2A}
\end{pmatrix}
\, .
\ee
Its determinant is given by
\be
\label{detgprobe} -\mathrm{det}\,\tilde{g} = \ex^{10A}(1+\ex^{-2A}\ka (\partial_r\tau)^2) \, .
\ee

\paragraph*{SU(2) ansatz}
Substituting the ansatz \eqref{ansatzSU2i}-\eqref{ansatzU1}, \eqref{parA}-\eqref{parF} and \eqref{TU2} into the bulk action \eqref{expSprobe} and \eqref{SCS1} yields the expression for the instanton energy\footnote{Strictly speaking, this is the bulk action, which as we shall evaluate it on the solutions of our equations of motion will also be the on-shell action. As all fields considered are time-independent, the boundary energy differs from this action by a trivial $-\int dt$ factor and this is why we shall call it the ``energy" from now on.} in terms of the fields of  \eqref{sumans}
\be
\label{Etot} E = E_{\text{DBI}} + E_{\text{CS}} \, ,
\ee
\begin{align}
\nonumber E_{\text{DBI}} &= 4\pi M^3N_c \int \mathrm{d}r\mathrm{d}\xi \, V_f(\l,\tau)\ex^A\sqrt{1+\ex^{-2A}\ka (\partial_r\tau)^2} \, \times \\
\nonumber &\hphantom{=} \times \Bigg(\ex^{2A}\xi^2\ka(\l)\tau^2\left(\frac{1}{1+\ex^{-2A}\ka (\partial_r\tau)^2}\td{A}_r^2 + \td{A}_{\xi}^2 + \frac{(\td{\phi} + \td{\phi}^*)^2}{2\xi^2} \right) +  \\
\nonumber &\hphantom{=}  +\! w(\l)^2\Bigg(\frac{1}{8}\frac{1}{1+\ex^{-2A}\ka (\partial_r\tau)^2}\xi^2(F_{\bar{\mu} \bar{\nu}})^2 \!+\! \frac{1}{2}\left(\left|\mathrm{D}_{\xi}\phi\right|^2 \!+\! \frac{1}{1+\ex^{-2A}\ka (\partial_r\tau)^2} \left|\mathrm{D}_{r}\phi\right|^2\right)\! + \\
\label{EDBI1} &\hphantom{=} \left. \left. + \frac{\left(1 - |\phi|^2\right)^2}{4\xi^2} - \xi^2 \left(\frac{1}{1+\ex^{-2A}\ka (\partial_r\tau)^2}(\partial_r\Phi)^2 + (\partial_{\xi}\Phi)^2\right) \right)\right) \, ,
\end{align}
\begin{align}
\nonumber E_{\text{CS}} = \frac{4N_c}{\pi} &\int \mathrm{d}r\mathrm{d}\xi \, \e^{\bar{\mu}\bar{\nu}}\partial_{\bar{\mu}}\Phi\times \\
\nn &\times \left[(f_1(\tau)+f_3(\tau))\left(\td{A}_{\bar{\nu}} + \frac{1}{2}(-i\phi^*D_{\bar{\nu}}\phi + h.c.)+\frac{1}{4i}\partial_{\bar{\nu}}(\td{\phi}^2-(\td{\phi}^*)^2)\right)+ \right. \\
\label{ECS1} &\hphantom{\times\bigg[} \left. + \frac{1}{2}(3if_2(\tau)-f_1(\tau)-f_3(\tau))(\td{\phi} + \td{\phi}^*)^2 A_{\bar{\nu}} \right] \, .
\end{align}
The differential equations obeyed by the fields of the ansatz are then obtained by extremizing the energy \eqref{Etot} with respect to variations of the fields\footnote{
It is well-known that substituting an ansatz into the action and then varying the action to derive the equations of motion leads to wrong results.
However, symmetry sometimes can protect this procedure.
In our case,   the SU(2) instanton ansatz  \eqref{ansatzSU2i}-\eqref{ansatzU1}, \eqref{parA}-\eqref{parF} and \eqref{TU2} fixes most of the gauge invariance, and therefore
one could worry that deriving the equations of motion for the fields of the ansatz in this manner may not reproduce all the constraints obeyed by these fields.
However we can show that the constraints that one misses in this procedure are trivially satisfied by our ansatz as it has cylindrical symmetry.}. The expressions for these equations \eqref{SUCSn}-\eqref{nabCSi3n} are presented in appendix \ref{App:EoM}. It is also checked that substituting the ansatz directly into the general equations of motion \eqref{UEoMprobe}-\eqref{abEoMprobe} yields the same equations as extremizing the energy \eqref{Etot}.

\subsection{Inhomogeneous tachyon}

To go beyond the probe limit, we now consider
the back-reaction on the tachyon modulus $\tau$. We will assume that the color sector remains fixed to its vacuum value, and write the equations of motion for the tachyon modulus coupled to the baryon fields. At leading order in the Veneziano limit, the baryon fields are given by the probe baryon solution, and the correction to the tachyon background starts at order $\mathcal{O}(1/N_f)$. Note that considering such a back-reaction will imply that $\tau$ will depend on the 3-dimensional radius $\xi$. Its  EoM is \eqref{tEoM}. 

In this case the expression \eqref{expSprobe} for the expanded DBI action can still be used but the effective metric, although symmetric, is not diagonal anymore:
\be
\label{ansgtx}
\begin{pmatrix}
-\ex^{2A} & 0 & 0 & 0 & 0 \\
0 & \ex^{2A} \! +\! \ka(\partial_r\tau)^2 & \ka\frac{x_1}{\xi}\partial_r\tau\partial_{\xi}\tau & \ka\frac{x_2}{\xi}\partial_r\tau\partial_{\xi}\tau & \ka\frac{x_3}{\xi}\partial_r\tau\partial_{\xi}\tau \\
0 & \ka\frac{x_1}{\xi}\partial_r\tau\partial_{\xi}\tau & \ex^{2A} \! +\! \ka\left(\frac{x_1}{\xi}\right)^2\!(\partial_{\xi}\tau)^2 & \ka\frac{x_1}{\xi}\frac{x_2}{\xi}(\partial_{\xi}\tau)^2 & \ka\frac{x_1}{\xi}\frac{x_3}{\xi}(\partial_{\xi}\tau)^2 \\
0 & \ka\frac{x_2}{\xi}\partial_r\tau\partial_{\xi}\tau & \ka\frac{x_1}{\xi}\frac{x_2}{\xi}(\partial_{\xi}\tau)^2 & \ex^{2A} \! +\! \ka\left(\frac{x_2}{\xi}\right)^2\!(\partial_{\xi}\tau)^2 & \ka\frac{x_2}{\xi}\frac{x_3}{\xi}(\partial_{\xi}\tau)^2\\
0 & \ka\frac{x_3}{\xi}\partial_r\tau\partial_{\xi}\tau & \ka\frac{x_1}{\xi}\frac{x_3}{\xi}(\partial_{\xi}\tau)^2 & \ka\frac{x_2}{\xi}\frac{x_3}{\xi}(\partial_{\xi}\tau)^2 & \ex^{2A} \! +\! \ka\left(\frac{x_3}{\xi}\right)^2\!(\partial_{\xi}\tau)^2
\end{pmatrix} .
\ee
Its  determinant is given by:
\be
\label{detgtx} -\mathrm{det}\, \tilde{g} = \ex^{10A}\left(1 + \ex^{-2A}\ka\left((\partial_r\tau)^2+(\partial_{\xi}\tau)^2\right)\, \right) \, .
\ee

\paragraph*{Soliton energy} With the new metric \eqref{ansgtx}, substituting the ansatz of equations \eqref{ansatzSU2i}-\eqref{ansatzU1}, \eqref{parA}-\eqref{parF} and \eqref{TU2} into the DBI action  \eqref{expSprobe} and the TCS action \eqref{SCS1} yields the following result for the soliton energy
\be
\label{Etotx} E = E_{\text{DBI}} + E_{\text{CS}} \, ,
\ee
\begin{align}
\nn E_{\text{DBI}} & =  4\pi M^3N_cN_f \int \mathrm{d}r\mathrm{d}\xi\,\xi^2 \,\sqrt{1 + \ex^{-2A}\ka\left((\partial_r\tau)^2+(\partial_{\xi}\tau)^2\right)} \, V_f(\l,\tau)\,\ex^{5A} - E_{\text{DBI,vac}} + \\
\nn  &\hphantom{=} +4\pi M^3N_c \int \mathrm{d}r\mathrm{d}\xi \,\sqrt{1 + \ex^{-2A}\ka\left((\partial_r\tau)^2+(\partial_{\xi}\tau)^2\right)}\, V_f(\l,\tau)\,\ex^A \times \\
\nonumber &\hphantom{= +4\pi M^3N_c } \times \left(\ex^{2A}\xi^2\ka(\l)\tau^2\left(\ex^{2A}\D_{rr}\td{A}_r^2 + \left(1 - \ex^{2A}\D_{\xi\xi}  \right)\td{A}_{\xi}^2 + \frac{(\td{\phi} + \td{\phi}^*)^2}{2\xi^2} - \right. \right. \\
\nn &\hphantom{= +4\pi M^3N_c \times \left(\ex^{2A}\xi^2\ka(\l)\tau^2\left(\ex^{2A}\D_{rr}\td{A}_r^2 + \left(1 - \ex^{2A}\D_{\xi\xi}  \right)\td{A}_{\xi}^2 \right.\right.}  - 2\ex^{2A}\D_{\xi r}\td{A}_r\td{A}_{\xi} \Bigg) + \\
\nonumber &\hphantom{= +4\pi M^3N_c \times \Bigg( }  + w(\l)^2\left(\frac{1}{8}\ex^{2A}\left[\D_{rr}\left(1 - \ex^{2A}\D_{\xi\xi}  \right) - \ex^{2A}\D_{\xi r}^2\right]\xi^2(F_{\bar{\mu} \bar{\nu}})^2 + \right. \\
\nn &\hphantom{= +4\pi M^3N_c \times \Bigg( + w(\l)^2 \bigg( }
 + \frac{1}{2}\left(\left(1 - \ex^{2A}\D_{\xi\xi}  \right)\left|\mathrm{D}_{\xi}\phi\right|^2 + \ex^{2A}\D_{rr} \left|\mathrm{D}_{r}\phi\right|^2\right) + \\
\nn & \hphantom{= +4\pi M^3N_c \times \Bigg( + w(\l)^2 \bigg(}
 + \frac{\left(1 - |\phi|^2\right)^2}{4\xi^2} - \frac{1}{2}\ex^{2A}\D_{\xi r} (D_r\phi^*D_{\xi}\phi + h.c.) - \\
\nn & \hphantom{= +4\pi M^3N_c \times \Bigg( + w(\l)^2 \bigg(}
 - \xi^2 \Big(\ex^{2A}\D_{rr}(\partial_r\Phi)^2 + \left(1 - \ex^{2A}\D_{\xi\xi}  \right)(\partial_{\xi}\Phi)^2  - \\
\label{EDBIx} &\hphantom{= +4\pi M^3N_c \times \Bigg( + w(\l)^2 \bigg(  - \frac{N_f}{2} \xi^2 \Big(\ex^{2A}\D_{rr}(\partial_r\Phi)^2 }
 - 2\ex^{2A}\D_{\xi r}\partial_{\xi}\Phi\partial_r\Phi \Big)\bigg)\Bigg) \, ,
\end{align}
\begin{align}
\nonumber E_{\text{CS}} = \frac{4N_c}{\pi} &\int \mathrm{d}r\mathrm{d}\xi \, \e^{\bar{\mu}\bar{\nu}}\partial_{\bar{\mu}}\Phi\times \\
\nn &\times \left[(f_1(\tau)+f_3(\tau))\left(\td{A}_{\bar{\nu}} + \frac{1}{2}(-i\phi^*D_{\bar{\nu}}\phi + h.c.)+\frac{1}{4i}\partial_{\bar{\nu}}(\td{\phi}^2-(\td{\phi}^*)^2)\right)+ \right. \\
\label{ECSx} &\hphantom{\times\bigg[} \left. + \frac{1}{2}(3if_2(\tau)-f_1(\tau)-f_3(\tau))(\td{\phi} + \td{\phi}^*)^2 \td{A}_{\bar{\nu}} \right] \, ,
\end{align}
where in the DBI part, $E_{\text{DBI,vac}}$ refers to the DBI contribution to the vacuum energy and the symbol $\D$ is defined in \eqref{defDxx}-\eqref{defDrr}. The equations of motion are obtained by extremizing the energy \eqref{Etotx} with respect to small deformations of the ansatz fields \eqref{sumans}. They are presented in Appendix \ref{Sec:EoMx}.

\section{Boundary conditions} \label{sec:baryon-bc}

The field equations presented in the previous section must be subject
to appropriate boundary conditions both at spatial infinity $\xi \to
+\infty$ and at the UV boundary $r\to 0$. Moreover, certain
(generalized) regularity
conditions must be imposed at the center of the instanton $\xi=0$ and
in the bulk interior. In this section we present the conditions that
are imposed on the fields of the ansatz \eqref{sumans} for the bulk instanton solution to be the dual of
a single baryon  at the boundary.

The discussion assumes that the dynamics for the tachyon modulus $\tau$ is also solved and the equations of motion given by \eqref{SUCSx}-\eqref{EoMt} (this corresponds to what was referred to as the \emph{inhomogeneous tachyon} case in the previous section). The conditions that are presented will also apply in the probe approximation, where $\tau$ is assumed to be a background field and the equations of motion are given by \eqref{SUCSn}-\eqref{nabCSi3n}.

\subsection{Baryon charge and mass}

We start by deriving the expression for the baryon charge and mass in the boundary theory, in terms of the fields of the ansatz \eqref{sumans}. We will then use these results to determine the boundary conditions required for the charge to be equal to unity and the mass to be finite.

We first discuss the calculation of the baryon charge, whose details can be found in appendix \ref{App:charge}. The baryon current at the boundary is given by \eqref{JB4D}
\begin{align} \label{JBbc}
 N_c J^{\mu}_B\, \omega_4 &= -\frac{iN_c}{48\pi^2}\, \mathrm{d}x^\mu\!\wedge \Big[-4 i \text{Tr}(L\wedge F^{(L)})\! +\!\text{Tr}(L\wedge L\wedge L)\! +\! 4 i \text{Tr}(R\wedge F^{(R)}) - \nonumber\\
 &\hphantom{= -\frac{iN_c}{48\pi^2}\, \mathrm{d}x^\mu\!\wedge} -\text{Tr}(R\wedge R\wedge R)\! +\! 6 \text{Tr}(DU\wedge F^{(L)} U^\dagger)\! +\! 6 \text{Tr}(DU U^\dagger\wedge F^{(R)}) - \nonumber\\
 &\hphantom{= -\frac{iN_c}{48\pi^2}\, \mathrm{d}x^\mu\!\wedge} \left. -2 i \text{Tr}(DU U^\dagger\wedge DU U^\dagger\wedge DU U^\dagger)\Big]\,\right|_{UV} \,\, ,
\end{align}
where we denoted by $\omega_4$ the Minkowski volume 4-form
\be
\label{defo4} \omega_4 \equiv \mathrm{d}t\wedge\mathrm{d}x^1\wedge\mathrm{d}x^2\wedge\mathrm{d}x^3 \, .
\ee
Remarkably, as shown in the appendix, the baryon number current only arises from the closed, $G_4$ term~\eqref{G4final} in the TCS action and does not depend on the non-closed part of the CS action $\Omega_5^0$ in~\eqref{Omega1},~\eqref{O50Ud1}. That is, it is also independent of the functions $f_i(\tau)$ which are the only degrees of freedom in the CS action that were not fixed by general arguments in section~\ref{Sec:CS}.

Because no external gauge fields are present in the UV boundary theory, the non-Abelian gauge field at $r=0$ should vanish. The baryon number current is therefore simply
\be \label{JBbc2}
 J^{\mu}_B\, \omega_4 = \left. -\frac{1}{24\pi^2}\, \mathrm{d}x^\mu\!\wedge \text{Tr}(\mathrm{d}UU^\dagger\wedge \mathrm{d}UU^\dagger\wedge \mathrm{d}UU^\dagger) \,\right|_{UV} \,\, .
\ee
Then the baryon number is\footnote{Remember that the baryon density is $\rho_B = J^0$ and the raising of indices for the Levi-Civita tensor is such that $\e_{0123} = -\e^{0123} = 1$.}
\be
\label{NB} N_B = \left. \frac{1}{24\pi^2}\int \Big[\text{Tr}(\mathrm{d}UU^\dagger\wedge \mathrm{d}UU^\dagger\wedge \mathrm{d}UU^\dagger)\Big]\,\right|_{UV} \,\, ,
\ee
which is nothing but the Skyrmion number for a matrix that is
identified as the pion field at the boundary\footnote{{ As
    discussed in section \ref{sec:masses}, this identification only
    makes sense for vanishing quark masses. Moreover,  if we take
    the expression (\ref{NB}) at face value, for non-zero quark masses
    the matrix $U(r,\xi)$ has to asymptote to a constant at $r=0$, and the
    resulting baryon number vanishes identically. This suggests that for
    non-zero quark mass both the baryon ansatz (as well as the form of
  the TCS action) must be generalized.}},
\be
\label{UP} U_P(\xi) = U(r=0,\xi)^{\dagger} \, .
\ee
Substituting the tachyon ansatz \eqref{TU2} into \eqref{NB}, we finally obtain the expression of the baryon number in terms of the phase $\theta$ at the boundary
 \be
\label{NBansf} N_B = \frac{1}{\pi}\big(\theta(\xi=\infty)-\theta(0)\big) \, .
\ee

\label{Sec:NB_Ni_1}

The expression for the baryon number in \eqref{NB} is a boundary topological number but is not manifestly topologically stable in the bulk. On the other hand, a state with baryon number equal to 1 is expected to be dual to a solution of the bulk equations of motion with instanton number equal to 1 \cite{WB}. We would therefore like to relate the baryon number \eqref{NB} to the bulk instanton number
\begin{align}
\label{Ninst1} N_{\text{instanton}} &= \frac{1}{8\pi^2}\int_{\text{bulk}} \text{Tr}\, \left(\mathbf{F}^{(L)} \wedge \mathbf{F}^{(L)} - \mathbf{F}^{(R)} \wedge \mathbf{F}^{(R)} \right)\\
\nn & = \frac{1}{2\pi} \int\mathrm{dr} \mathrm{d}\xi\, \e^{\bar{\mu}\bar{\nu}}\left(F_{\bar{\mu}\bar{\nu}} + \partial_{\bar{\mu}}\left( -i\phi^*D_{\bar{\nu}}\phi + h.c. \right)\right) \, ,
\end{align}
where in the second line we substituted the instanton ansatz. The detailed calculation is presented in Appendix \ref{Sec:NB_Ni}. There it is found that, in terms of the baryon ansatz  \eqref{ansatzSU2i}-\eqref{ansatzU1} and \eqref{TU2}, the condition for the baryon number to equal the instanton number can be written as the vanishing of an IR integral
\begin{equation}
\label{cNBeNians1} \left. \int \mathrm{d}\xi\, \left(\, \tilde{A}_{\xi}(|\tilde{\phi}|^2-1) + \partial_{\xi} \tilde{\phi}_1  + \partial_{\xi} \tilde{\phi}_1  \tilde{\phi}_2 - \partial_{\xi} \tilde{\phi}_2  \tilde{\phi}_1 \,  \right)  \,\right|_{IR}  = 0 \, .
\end{equation}
This condition on the instanton solution follows from  the regular IR asymptotics presented in Table \ref{tab:bcs} that will be discussed in detail in the next section.

Another global property of the baryon solution that can be used to identify the right boundary conditions is the mass of the nucleon. The mass of the nucleon is the sum of a classical contribution, equal
to the classical instanton energy \eqref{Etot} evaluated on the solution, and quantum corrections:
\be
\label{MB} M_{\text{nucleon}} = E + \d M_Q \, .
\ee
Requiring the classical contribution to the mass to be finite sets the boundary conditions for the ansatz fields at $\xi \to \infty$, as in Table \ref{tab:bcs}.
The derivation of Table \ref{tab:bcs} is the topic of the next subsection.

In terms of the expansion in $N_c$, the classical mass is of order
$\mathcal{O}(N_c)$, whereas the quantum corrections start at order
$\mathcal{O}(1)$. Computing the quantum corrections requires to take
the sum of the ground state energies for the infinite set of bulk
excitations on the instanton background, and subtract the vacuum
energy. It is not known how to do this calculation, so we can only
assume that the classical mass gives the dominant contribution. Note
that it is correct at least in the large $N_c$ limit.

The experimental spectrum of baryons contains the nucleons but also
many excited states, such as the isobar $\D$. Here we focus for
concreteness on the nucleon mass, but the mass of the excited states
can be derived by an appropriate quantization of the perturbation
modes of the instanton solution. This will be discussed in a separate
 work.

\subsection{Boundary conditions for the gauge invariant fields}

We are now ready to derive the boundary conditions relevant to the baryon solution. We present here the general conditions for the gauge-invariant fields \eqref{LtoLtcyl} and discuss separately the 4 boundaries of the $(\xi,r)$ space.

\begin{table}[]
\centering
\begin{tabular}{|l|l|l|l|}
\hline
$\xi \to 0$ & $\xi \to \infty$ & UV & IR  \\ \hline
$\frac{\td{\phi}_1}{\xi} \to \mathfrak{f}_1(r)$ & $\xi^{1/2} \td{\phi}_1 \to 0$ & $\td{\phi}_1^2+\td{\phi}_2^2 \to 1$ & $\td{\phi}_1 \to 0 $         \\
 $\frac{1+\td{\phi}_2}{\xi}\to 0$ & $\xi^{1/2}(-1+\td{\phi}_2) \to 0$ & $\partial_{\xi}\td{\phi}_1 + \td{A}_{\xi}\td{\phi}_2 \to 0$ & $\partial_r\td{\phi}_2 \to 0 $     \\
$\td{A}_{\xi} - \frac{\td{\phi}_1}{\xi} \to 0$ & $\xi^{3/2} \td{A}_{\xi} \to 0$ & $\partial_r \td{A}_{\xi} \to 0$ & $\td{A}_{\xi} \to 0 $      \\
 $\td{A}_r \to 0$  & $\xi^{3/2} \td{A}_r \to 0$ & $ \td{A}_r \to 0$ & $ \td{A}_r \to 0$       \\
$\partial_{\xi}\Phi \to 0$  &$\xi^{3/2} \partial_{\bar{\mu}}\Phi \to 0$ & $\Phi \to 0$ &  $\partial_r\Phi \to 0$      \\
$\partial_\xi\tau \to 0$ & $\tau \to \tau_b(r)$ & $r^{-2}\, \tau \to 0$ & $\tau\to\tau_b(r_\text{IR})$ \\ \hline
\end{tabular}
\caption{Gauge invariant boundary conditions for the fields of the baryon solution \protect\eqref{sumans}.}
\label{tab:bcs}
\end{table}

\paragraph*{\underline{$\xi = 0\,:$}} The boundary conditions in the limit where the 3D radius $\xi$ goes to 0 are chosen as in the first column of Table \ref{tab:bcs}, where $\mathfrak{f}_1(r)$ is some function of the holographic coordinate. The first 4 conditions come from requiring that $\td{\mathbf{L}}$ \eqref{LtoLt} and $\td{\mathbf{R}}$ \eqref{RtoRt} are well defined 5D vectors at $\xi = 0$. The last two conditions respectively come from requiring that the Abelian field strength $\hat{F}$ \eqref{Fhat} and  the tachyon covariant derivative $D_M T$ are well defined at $\xi = 0$. The detailed asymptotics of the fields in this limit are presented in Appendix \ref{Sec:asymx0}.

\paragraph*{\underline{UV :}}  The boundary conditions  in the UV limit
$r \to 0$, are chosen as in the third column of Table \ref{tab:bcs}. The first three conditions in the column originate from requiring that there are no sources for the gauge fields at the boundary. The fourth condition for $\td{A}_r$ is a consequence of \eqref{ArUV1}, which originates from the near-boundary analysis of the constraint \eqref{SUCSn}. Finally, the condition for $\Phi$ corresponds to setting the baryon chemical potential $\mu$ to 0, and that for $\tau$ is due to our choice to work with massless quarks, as mentioned before.

The UV asymptotics of the fields are discussed in more details in Appendix \ref{Sec:asymUV}.

\paragraph*{\underline{$\xi \to  \infty \,:$}}  The boundary conditions in the limit $\xi \to \infty$ are chosen as in the second column of Table \ref{tab:bcs} and come from requiring that the instanton energy in \eqref{Etot} is finite. In particular, the condition for the tachyon modulus $\tau$ is that it goes to its value in vacuum $\tau_b(r)$. For $\Phi$ we set the additional condition that the baryon chemical potential $\mu$ is equal to 0. {Finally, for $\td{\phi}_2$, the finite energy condition  is
\be
\label{cinff2} \partial_r\td{\phi}_2(r,\xi) \underset{\xi\to\infty}{\to} 0 \, ,
\ee
or equivalently
\be
\label{cinff22} \td{\phi}_2(r,\xi) \underset{\xi\to\infty}{\to} \td{\phi}_2(0,\infty) \, .
\ee
The value of $\td{\phi}_2$ on the UV boundary at $\xi\to\infty$ is determined by the requirement that the baryon number} \eqref{NBansf} {is equal to 1. This condition is for the tachyon phase at the boundary $\theta(r=0,\xi)$ to be such that
\be
\label{conda} \theta(r=0,\infty) - \theta(r=0,0) = \pi \, .
\ee
Because the sources for the gauge fields at the boundary are required to vanish, $\td{\phi}_2(r=0,\xi) = -\cos{\theta(r=0,\xi)}$. Then, from the condition that $\td{\phi}_2 = -1$ at $\xi=0$, we deduce that $\theta(r=0,0)=0$.} \eqref{conda} {therefore implies that
\be
\label{cinff23} \td{\phi}_2(r,\xi) \underset{\xi\to\infty}{\to} -\cos{\pi} = 1 \, .
\ee
}
The detailed asymptotics of the fields in the limit $\xi\to\infty$ are presented in Appendix \ref{Sec:asymxinf}.

\paragraph*{\underline{IR :}}  The regularity conditions in the IR limit $r\to \infty$ are chosen as in the last column of Table \ref{tab:bcs}.
{The ansatz fields obey second order differential equations, whose general solutions are a linear combination of two independent solutions. These independent solutions can be chosen such that one is finite in the IR while the other is singular. The regularity conditions correspond to the choice of the IR finite solutions.  In that case, the precise IR asymptotics of the solution are presented in Appendix} \ref{bcIR}.
Note that the resulting conditions match those imposed in \cite{Pomarol07} for chiral symmetry to be broken on the IR wall:
\be
\label{IRc} \left.\left(\mathbf{L}-\mathbf{R}\right)\right|_{r_{\text{IR}}} = 0 \sp \left.\left(\mathbf{F}^{(L)}_{\mu r}+\mathbf{F}^{(R)}_{\mu r}\right)\right|_{r_{\text{IR}}} = 0 \, .
\ee
{As far as the tachyon modulus $\tau$ is concerned, it should match the vacuum solution far from the baryon center, so in particular in the IR at $r\to \infty$. The IR behavior of $\tau$ will therefore be given by} \eqref{tauIR}.

\subsection{Boundary conditions in  Lorenz gauge}

The non-linear second-order differential system of equations of motion
 \eqref{SUCSn}-\eqref{nabCSi3n} is not elliptic. This is due to
the presence of constraint equations. While this is not
problematic per se, it can lead to trouble when one tries to solve the
problem numerically (which we will eventually do in an upcoming work). This is because the gauge invariance indicates that given boundary conditions, the solution is not unique. This can be
avoided, if one works with an elliptic system instead.

Equations \eqref{SUCSn}-\eqref{nabCSi3n}  can  be recast in  elliptic form if we write them in terms of the gauge variant fields \eqref{ansatzSU2i}-\eqref{ansatzSU2r}, and then  fix the gauge with the Lorenz condition
\be
\label{Lorenzc1} \partial_r A_r + \partial_{\xi} A_{\xi} = 0 \, .
\ee
Note that this condition leaves a residual gauge freedom of the form
\be
\label{resg_Lg} A_{\bar{\mu}} \to A_{\bar{\mu}} + \partial_{\bar{\mu}}\mathfrak{f} \sp \partial_r^2 \mathfrak{f} + \partial_\xi^2 \mathfrak{f} = 0 \sp \mathfrak{f}(0,\xi)=0 \, .
\ee
The convenient choice that we present below for the fixing of the residual gauge freedom \eqref{resg_Lg} requires to introduce an IR cut-off $r_{\text{IR}}$.

The equations of motion in Lorenz gauge are listed in appendix \ref{Sec:EoMLorenz} and the relevant boundary conditions written in Table \ref{tab:bcsLg}. Because of the gauge fixing these conditions contain additional information compared with Table \ref{tab:bcs} that we discuss again separately for each boundary of the $(r,\xi)$ space.

\begin{table}[]
\centering
\begin{tabular}{|l|l|l|l|}
\hline
$\xi \to 0$ & $\xi \to \infty$ & $r \to 0$ & $r \to r_{\text{IR}} \to \infty $  \\ \hline
$\partial_{\xi}\tilde{\phi}_1 - (A_{\xi} + \partial_{\xi}\theta) \to 0$ & $\xi^{1/2} \tilde{\phi}_1 \to 0$ & $\tilde{\phi}_1 \to \sin{\theta}$ & $\tilde{\phi}_1 \to 0 $         \\
 $\frac{1+\tilde{\phi}_2}{\xi}\to 0$ & $\xi^{1/2}\left(\tilde{\phi}_2 - 1 \right) \to 0$ & $\tilde{\phi}_2 \to -\cos{\theta}$ & $\partial_r\tilde{\phi}_2 \to 0 $     \\
$\partial_{\xi}A_{\xi} \to 0$ & $\partial_{\xi}A_{\xi} \to 0$ & $A_{\xi} \to 0 $ & $A_{\xi} \to 0 $      \\
 $A_r \to 0$  & $\xi^{3/2} \left(A_r-\frac{\pi}{r_{\text{IR}}} \right) \to 0$ & $ \partial_r A_r \to 0$ & $ \partial_r A_r \to 0$       \\
$\partial_{\xi}\Phi \to 0$  & $\Phi \to 0$ & $\Phi \to 0$ &  $\partial_r\Phi \to 0$      \\
$\theta \to 0$ & $\theta \to \pi \left(1-\frac{r}{r_{\text{IR}}} \right) $ & $\partial_r\theta + A_r  \to 0$ & $ \theta \to 0$ \\
$\partial_\xi\tau \to 0$ & $\tau \to \tau_b(r)$ & $\tau \to 0$ & $\tau\to\tau_b(r_\text{IR})$ \\ \hline
\end{tabular}
\caption{Boundary conditions in Lorenz gauge.}
\label{tab:bcsLg}
\end{table}

\paragraph*{\underline{UV :}} In the UV limit $r \to 0$, the only difference with the gauge invariant conditions Table \ref{tab:bcs} is the additional condition for $A_r$, which comes from imposing the Lorenz gauge \eqref{Lorenzc1} near the boundary.

\paragraph*{\underline{$\xi \to \infty\,:$}}
For compatibility with the condition  \eqref{conda} $\partial_{\xi}\theta$ should go to 0 as $\xi \to \infty$, so $A_{\xi}$ should also tend to 0 according to the gauge-invariant condition in Table \ref{tab:bcs}. Then the Lorenz gauge condition in the limit where $\xi \to \infty$ reads
\be
\label{Lgcxinf} \partial_r A_r = 0 = -\partial_r^2 \theta \, ,
\ee
so that $\theta$ at $\xi=\infty$ should be of the form
\be
\label{cxinft} \theta \to \pi + \text{constant}\times  r \, .
\ee
Then, for a solution with $N_B=1$, we proved in section \ref{Sec:NB_Ni_1} that the instanton number \eqref{Ninst1} should also be equal to 1. The latter can be written as a boundary integral
\be
\label{Ni_b} N_{\text{instanton}} = \frac{1}{\pi} \int_0^{r_{\text{IR}}}\mathrm{d}r\left[A_r + \td{\phi}_1\partial_r\td{\phi}_2 - \td{\phi}_2\partial_r\td{\phi}_1\right]^{\xi=\infty}_{\xi=0} = \frac{1}{\pi} \int_0^{r_{\text{IR}}}\mathrm{d}r\left[A_r \right]^{\xi=\infty}_{\xi=0} \, ,
\ee
where, to obtain the first integral above we used the boundary conditions in Table \ref{tab:bcsLg},
 and to obtain the second integral above
 we used the boundary conditions in Table \ref{tab:bcs}.
  We find it convenient to choose a residual gauge \eqref{resg_Lg} such that the instanton winding occurs at $\xi=\infty$\footnote{Another possible choice would be to place it at $\xi = 0$ as in \cite{Cherman}. That choice has the advantage that $U$ is regular at $\xi\to\infty$ in the limit $r_{\text{IR}}\to\infty$, but instead it is not well defined at $\xi=0$.}. In this case the condition that $\td{A}_r = 0$ at $\xi = \infty$ implies that $\theta$  has to  go to
\be
\label{cxinft2} \theta \to \pi\left( 1 - \frac{r}{r_{\text{IR}}}
\right) \, , \quad \xi \to +\infty
\ee

\paragraph*{\underline{$\xi = 0$ and IR :}} Once \eqref{cxinft2} is imposed, what remains of the residual gauge freedom \eqref{resg_Lg} corresponds to the freedom of choosing the profile of $\theta$ on the 1-dimensional space composed of the lines at $\xi=0$ and $r = r_{\text{IR}}$, with the condition that $\theta$ goes to 0 at the two endpoints of this line. The simplest choice is to set
\be
\label{th_x0_IR} \theta(r,0) = \theta(r_{\text{IR}},\xi) = 0 \, ,
\ee
for which the tachyon field is well defined at $\xi=0$ and the stronger IR gauge condition proposed in \eqref{cUIRstr} is obeyed. Then, the boundary conditions at $\xi=0$ and $r=r_{\text{IR}}$ come from the gauge-invariant ones Table \ref{tab:bcs}, with the additional derivative constraints on $A_\xi$ and $A_r$ due to the gauge choice.

To summarize, these boundary conditions ensure that a topological
instanton in the bulk has finite bulk energy (therefore finite  boundary mass)
and unit baryon charge. The question of solving the field equation and
finding such smooth solutions must be tackled numerically, and will be addressed in  a future work.

\subsection{Further comments}

In the WSS model, the size of the instanton becomes small at large 't Hooft coupling $\l$. The flat space BPST instanton therefore gives a good approximation to the WSS instanton solution near its center. By contrast, there is no possibility  in V-QCD which would make the size of the instanton parametrically small. Instead, the size is set by the mass scale of the boundary theory which roughly corresponds to $\L_{\text{QCD}}$. The curvature of the background space-time should therefore be taken into account in the calculation of the V-QCD instanton, for which there is no simple BPST approximation.

Moreover, the size of the baryon of the construct is clearly N$_c$-independent. The reason is that it is a solution to the gravitational equations to leading order in N$_c$ and the solutions are N$_c$ independent.
On the other hand, the size of the baryon is of order  $\Lambda_{QCD}^{-1}$.

\section*{Acknowledgements}\label{ACKNOWL}
\addcontentsline{toc}{section}{Acknowledgements}

\noindent {We thank D.~Arean, P. Figueras,  A.~Krikun, J.~L.~Ma\~nes, A.~Pich,  C. Rosen, P. Sutcliffe and  P.  Yi for discussions and/or correspondence.} This work was supported in part by CNRS contract IEA 199430. The work of MJ was supported
by an appointment to the JRG Program at the APCTP through the Science and Technology Promotion Fund and Lottery Fund of the Korean Government. MJ was also  supported by the Korean Local Governments -- Gyeong\-sang\-buk-do Province and Pohang City -- and by the National Research Foundation of Korea (NRF) funded by the Korean government (MSIT) (grant number 2021R1A2C1010834).

\appendix
\renewcommand{\theequation}{\thesection.\arabic{equation}}
\addcontentsline{toc}{section}{Appendix\label{app}}
\section*{Appendix}

\section{Conventions and Symmetry Transformations}
\label{app: conventions}

\subsection{Conventions for the gauge fields}

For the $SU(N_f)$ generators $\lambda^a$, $A=1,\ldots N_f^2-1$, we take
\be\label{norm gen}
(\lambda^a)^\dagger = \lambda^a \sp \tr ( \lambda^a \lambda^b ) =\frac{1}{2}\delta^{ab} \, .
\ee
On the other hand the normalization for the $U(1)$ generator is
\be
\label{norml0} \lambda^0 = \mathbb{I} \, ,
\ee
where $\mathbb{I}$ is the $N_f \times N_f$ identity matrix.
For the $SU(N_f)$ generators $\lambda^a$, $a=1,\ldots, N_f^2-1$, we have
\be
[\lambda^a,\lambda^b]=if^{ab}_{\phantom{ab}c}\, \lambda^c \sp \tr(\lambda^a\{\lambda^b,\lambda^c\})= d^{abc} \, ,
\ee
where $f^{ab}_{\phantom{ab}c}$ and $d^{abc}$ are, respectively,  the structure constants and  the normalized anomaly Casimir for $SU(N_f)$. Because of (\ref{norm gen}) $f^{ab}_{\phantom{ab}c}$ and $d^{abc}$ are real numbers.

We define the gauge fields to be Hermitian
\be\label{A decomp}
A_\mu= A^{U(1)}_\mu\mathbb{I} + A_\mu^a \lambda^a \, ,
\ee
where $A^{U(1)}$ and $A_\mu^a$ are real. In differential form notation\footnote{We use the conventions of Appendix B of \cite{book}.}, the field strength and covariant derivative then read
\be
F=dA -i A\wedge A \sp  D\equiv d - i A \cdot
\ee
where $A\cdot$ indicates the representation-dependent action of the
gauge algebra. In particular the Bianchi identity reads
\be
DF=dF+iF\wedge A-iA\wedge F=0\;,
\ee
 and
the covariant derivative of the tachyon is given by
\be
DT=dT+i \, T A_L -i A_R T \qquad \qquad DT^\dagger =dT^\dagger
-iA_L T^\dagger + i \, T^\dagger A_R \, .
\ee

\subsection{Normalization of the chiral currents}

Currents are defined to be Hermitian. We decompose a $U(N_f)$ flavor current as
\be
J_\mu= \frac{1}{2N_f}J^{U(1)}_\mu \mathbb{I} +J_\mu^a\lambda^a.
\ee
This decomposition corresponds to the following normalization for the currents
\be
J_{L,R}^{a\;\mu}=\tr_{\text{flavor}} \left(i\bar{q} \gamma^\mu
\frac{1\pm\gamma_5}{2}\lambda^a q \right) \, ,
\ee
\be
J^{U(1)\; \mu}_{L,R}=\tr_{\text{flavor}} \left(i\bar{q} \gamma^\mu\frac{1\pm\gamma_5}{2} q \right) \, .
\ee
Notice that the normalization of the currents and gauge field \eqref{A decomp} has been chosen in such a
way that the boundary coupling of the current to the gauge field reads
\be
2\int d^4x\, \tr(J^\mu A_\mu) =\int d^4x \left( J^{U(1)\,\mu} A_\mu^{U(1)}  +J^{a\,\mu} A_\mu^a\right) \, .
\ee

\subsection{Gauge Transformations}

Under gauge transformations with parameters $(V_L,V_R)\in SU(N_f)_L\times SU(N_f)_R$,  the gauge fields and tachyon transform in the following way
\be
\nn A_L \to V_LA_LV^{\dagger}_L -idV_LV_L^{\dagger} \sp A_R\to V_RA_RV^{\dagger}_R -idV_RV_R^{\dagger} \, ,
\ee
\be
\label{gaugevar} F_{L}\to V_L~F_L~V_L^{\dagger}\sp F_{R}\to V_R~F_R~V_R^{\dagger} \, ,
\ee
\be
\nn T\to V_R TV_L^{\dagger}  \sp
T^\dagger\to V_L T^\dagger V_R^{\dagger} \, .
\ee
An infinitesimal gauge transformation is defined as $V_\epsilon (x)=e^{\epsilon \Lambda(x)
}\simeq 1+\epsilon\, \Lambda(x)$ and the gauge transformation of a field as
$A\to A + \epsilon \delta_\Lambda A$. From  \eqref{gaugevar} we have then:
\be\label{gaugevar inf}
\begin{split}
&\delta_\Lambda A=-i\, D\Lambda=-i\,d\Lambda +[\Lambda,A] \, , \\
&\delta_\Lambda F =[\Lambda,F] \, ,\\
&\delta_{\Lambda_L}T=-T\Lambda_L \, , \\
&\delta_{\Lambda_R}T=\Lambda_R T \, .
\end{split}
\ee
Notice that the generators of gauge transformations are antihermitian. When we decompose them in
their $U(1)$ and $SU(N_f)$ parts, we will write
\be\label{Lambda decomp}
\Lambda= i\alpha \mathbb{I} + i\Lambda^a \lambda^a \, ,
\ee
with $\alpha$ and $\Lambda^a$ real parameters. In particular we have, from (\ref{A decomp}) and (\ref{Lambda decomp})
\be\label{gauge var A}
\delta A_\mu^{U(1)} =\partial_\mu \alpha\qquad \mathrm{and} \qquad \delta A_\mu^a=(D_\mu \Lambda)^a \, .
\ee

\subsection{Discrete symmetries}\label{discrete}

We describe here the transformation properties of the flavor fields under parity and charge conjugation. These were presented in \cite{Casero}.

\subsubsection*{Parity}

The parity transformation is
\be
\label{PP1P2} P = P_1 \cdot P_2 \, .
\ee
where $P_2$ is the action of parity on space
\be
\label{defP2} P_2 : (x_1,x_2,x_3) \to (-x_1,-x_2,-x_3) \, .
\ee
and $P_1$ the action on the flavor fields.
\be
\label{defP1} P_1 : L \leftrightarrow R \sp T \leftrightarrow T^\dagger \, .
\ee

\subsubsection*{Charge conjugation}

The action of charge conjugation on the flavor fields is
\be
\label{defC}  C : L \to -R^t \sp R \to -L^t \sp T \to T^t \sp T^\dagger \to \left(T^\dagger\right)^t \, .
\ee

\subsection{The 2D theory for the ansatz fields}

\label{Sec:2Dconv}

We present in this subsection conventions and definitions for the theory of the fields of the instanton ansatz  \eqref{ansatzSU2i}-\eqref{ansatzU1} and  \eqref{ansT} that live on the 2D space $(\xi, r) \equiv x^{\bar{\mu}}$. The first thing to note is that we choose the following convention for the 2D Levi-Civita tensor
\be
\label{defe2D} \e^{\xi r} = 1 \, .
\ee

Before imposing the parity symmetry, the fields of the gauge-field ansatz exist in two copies L and R that each have a residual gauge freedom  \eqref{resg} under which the fields have well-defined transformation properties :
\begin{itemize}
\item $\Phi^{(L/R)}$ is neutral.
\item $\phi^{(L/R)} \equiv \phi_1^{(L/R)} + i\phi_2^{(L/R)}$ has charge 1.
\item $A_{\bar{\mu}}^{(L/R)} \equiv \Big(A_\xi^{(L/R)}, A_r^{(L/R)}\Big)$ is the gauge field.
\end{itemize}
From these, we construct the L/R covariant derivatives of the complex scalars $\phi^{(L/R)}$ under the residual gauge freedom
\be
\label{covDiv} D_{\bar{\mu}}\phi^{(L/R)} \equiv \big( \partial_{\bar{\mu}} - iA^{(L/R)}_{\bar{\mu}} \big)\phi^{(L/R)} \, ,
\ee
which in component reads
\be
\label{covDiv_12} D_{\bar{\mu}}\phi_1^{(L/R)} = \partial_{\bar{\mu}}\phi_1^{(L/R)} + A^{(L/R)}_{\bar{\mu}} \phi_2^{(L/R)} \sp D_{\bar{\mu}}\phi_2^{(L/R)} = \partial_{\bar{\mu}}\phi_2^{(L/R)} - A^{(L/R)}_{\bar{\mu}} \phi_1^{(L/R)} \, .
\ee
We define also the gauge-invariant field strength for $A_{\bar{\mu}}^{(L/R)}$
\be
\label{defFLR} F_{\bar{\mu}\bar{\nu}}^{(L/R)} = \partial_{\bar{\mu}}A_{\bar{\nu}}^{(L/R)} - \partial_{\bar{\nu}}A_{\bar{\mu}}^{(L/R)} \, .
\ee
Once the parity symmetry  \eqref{parA}-\eqref{parF} is imposed, the L/R fields collapse into a single set with the residual gauge freedom  \eqref{resg1}. Also, the ansatz for the tachyon takes the form of  \eqref{TU2} and the tachyon phase $\theta$ can be absorbed into the gauge fields to build gauge invariant quantities  \eqref{LtoLtcyl}
\be
\label{LtoLtcylApp} \td{A}_{\bar{\mu}} \equiv A_{\bar{\mu}} + \partial_{\bar{\mu}}\theta \sp \td{\phi} \equiv \ex^{i\theta}\phi \, .
\ee

\section{The P-odd instanton}

\label{Podd}

We justify in this appendix the statement made in Section \ref{Sec:ans} that a baryon state in the boundary theory corresponds to an axial instanton for the bulk gauge fields, that is an instanton solution even under parity
\be
\label{Pinv2} P \,\, : \,\, \mathbf{x}\to -\mathbf{x} \sp L\leftrightarrow R \, ,
\ee
and that a P-odd instanton cannot have a finite energy. This implies that not only a P-odd instanton cannot generate baryon number, but also does not correspond to any other finite energy state in the boundary theory.

Requiring the ansatz to be P-odd imposes the following relation between the left and right-handed ansatz fields, instead of \eqref{parA}- \eqref{parF}
\begin{equation}
\label{parAo} A_1 \equiv A_1^L = A_1^R \sp A_2 \equiv A_2^L = A_2^R \, ,
\end{equation}
\begin{equation}
\label{parfo} \phi_1 \equiv \phi_1^L = \phi_1^R \sp \phi_2 \equiv \phi_2^L = -\phi_2^R - 2 \, ,
\end{equation}
\begin{equation}
\label{parFo} \Phi \equiv \Phi^L = -\Phi^R \, .
\end{equation}
Note that, because of the condition for $\phi_2$ \eqref{parfo}, the P-odd ansatz completely fixes the residual gauge \eqref{resg}. The P-odd tachyon ansatz is
\be
\label{TU2o} T^{SU(2)} = i \tau(r,\xi) \exp{\left(i\theta(r,\xi)\frac{x\cdot\s}{\xi}\right)} \sp \tau,\theta \in \mathbb{R} \, .
\ee
We still consider the chiral limit where the quark masses are equal to 0. In this case, due to the factor $i$, the P-odd ansatz is not continuously connected to a vacuum state with tachyon UV asymptotics given by \eqref{tUV}. It is rather connected to a $U(1)_A$ rotation of the vacuum, which is not degenerate with the vacuum due to the $U(1)_A$ anomaly. There is therefore no such thing as a static P-odd instanton solution taking root on the vacuum. This result can be interpreted as the bulk equivalent of the non-conservation of the $U(1)_A$ charge.

\section{The DBI contribution to the equations of motion}

The DBI equations of motion for the flavor sector are obtained from varying  \eqref{expS} with respect to the gauge fields $\mathbf{L}$ and $\mathbf{R}$ and the tachyon field $\tau$.

\paragraph*{Abelian EoM} The left-handed Abelian equations of motion are obtained to be
\begin{align}
\nonumber & \partial_N\bigg[V_f(\l,\tau^2)\sqrt{-\mathrm{det}\, \tilde{g}^{(L)}}w \,\times  \\
\nn &\hphantom{ \partial_N \Big[} \times \left(\mathcal{L}^{(L)}\left((\tilde{g}^{(L)})^{-1}\right)^{[MN]} + \frac{1}{4} w^2\mathrm{Tr}\left(F_{(L)}^{MS}F_S^{(L)N} - F_{(L)}^{NS}F_S^{(L)M}\right)\right. - \\
\nonumber &\hphantom{\partial_N \Big[ \times \bigg(} \left.\left. -\frac{1}{4}w^2\mathrm{Tr}\left(F^{(L)[MN]}F_C^{(L)C}\right)- \frac{1}{2}\ka\tau^2 S^{[MN]}\right)\right] = \\
\nonumber &=  \frac{1}{4} V_f(\l,\tau^2)\left(\sqrt{-\mathrm{det}\, \tilde{g}^{(L)}}\left((\tilde{g}^{(L)})^{-1}\right)^{(MN)}+\sqrt{-\mathrm{det}\, \tilde{g}^{(R)}}\left((\tilde{g}^{(R)})^{-1}\right)^{(MN)}\right) \times \\
\label{abEoM} &\hphantom{=  \frac{1}{4} V_f(\l,\tau^2)\Big(\sqrt{-\mathrm{det}\, \tilde{g}^{(L)}}\left((\tilde{g}^{(L)})^{-1}\right)^{(MN)}} \times \ka\tau^2 \text{Tr}\left[iD_NU^\dagger U + h.c.\right] + \hat{J}_{CS}^{(L)} \, ,
\end{align}
where the indices are raised according to  \eqref{raise}. Note that, because $\tilde{g}$ is not symmetric, $S^{MN}$ is a priori not symmetric either. We denoted by $\mathcal{L^{(L/R)}}$ the left-handed/right-handed integrand of  \eqref{expS} divided by $V_f(\l,\tau^2)\sqrt{-\mathrm{det}\, \tilde{g}^{(L/R)}}$
\begin{align}
\label{defL} \mathcal{L}^{(L/R)} &\equiv \frac{1}{2} + \frac{1}{4}\ka \tau^2 \left((\tilde{g}^{(L/R)})^{-1}\right)^{(MN)}S_{MN}  \\
\nonumber &\hphantom{=} - \frac{1}{8}w^2\left((\tilde{g}^{(L/R)})^{-1}\right)^{MN}\left((\tilde{g}^{(L/R)})^{-1}\right)^{PQ}\mathrm{Tr}\, F^{(L/R)}_{NP}F^{(L/R)}_{QM} \\
\nonumber &\hphantom{=}  +\frac{1}{16}w^2\mathrm{Tr}\,\left(\left((\tilde{g}^{(L/R)})^{-1}\right)^{[MN]}F^{(L/R)}_{NM}\right)^2 \, ,
\end{align}
and the contribution from the TCS action to the Abelian EoMs is denoted by $\hat{J}_{CS}$
\be
\label{defJh} -\frac{1}{M^3 N_c} \d_{\hat{L}} S_{CS} = \d \hat{L}\wedge \hat{J}_{CS}^{(L)} \, ,
\ee
and likewise for R. The right-handed equations of motion are the same\footnote{Here it shouldn't be forgotten that it also changes the definition of the raising of indices.} upon $L\leftrightarrow R$ and $U\leftrightarrow U^\dagger$.

\paragraph*{Tachyon EoM} Because it also appears in the effective metric  \eqref{defgt}, the equations of motion for the modulus of the tachyon field have a form similar to the Abelian gauge field equations of motion
\begin{align}
\nn & \partial_N\bigg[V_f(\l,\tau^2)\sqrt{-\mathrm{det}\, \tilde{g}^{(L)}}\ka\partial_M\tau \times \\
\nn &\hphantom{\partial_N\bigg[} \times \left(\mathcal{L}^{(L)}\left((\tilde{g}^{(L)})^{-1}\right)^{(MN)} + \frac{1}{4}w^2\mathrm{Tr}\left(F_{(L)}^{MS}F_S^{(L)N} + F_{(L)}^{NS}F_S^{(L)M}\right)\right. \\
\nn &\hphantom{\partial_N\bigg[ \times \bigg(} \left.\left. -\frac{1}{4}w^2\mathrm{Tr}\left(F^{(L)(MN)}F_C^{(L)C}\right) - \frac{1}{2}\ka\tau^2 S^{(MN)}\right) + (L\leftrightarrow R)\right] = \\
\nonumber &=  V_f(\l,\tau^2)\left[\sqrt{-\mathrm{det}\, \tilde{g}^{(L)}}\left(\frac{1}{2}\left((\tilde{g}^{(L)})^{-1}\right)^{(MN)} \ka\tau S_{MN} + \frac{1}{V_f}\frac{\d V_f}{\d \tau}\mathcal{L}^{(L)}\right) + (L\leftrightarrow R)\right]\! + \\
\label{tEoM} &\hphantom{=} + J_{CS}^{\tau} \, ,
\end{align}
where the contribution of the TCS action to the tachyon EoM is denoted by $J_{CS}^{\tau}$
\be
\label{defJt} -\frac{1}{M^3 N_c} \d_{\tau} S_{CS} = \d \tau J_{CS}^{\tau} \, .
\ee
The equations of motion for $U$ are
\be
\label{UEoM} \frac{1}{4}\!\left[D_{(M}\!\left(V_f(\l,\tau^2)\ka\tau^2\sqrt{-\mathrm{det}\, \tilde{g}^{(L)}}\left((\tilde{g}^{(L)})^{-1}\right)^{(MN)}\! D_{N)}U^\dagger\right)\! U \! -\! h.c. \! +\! (L\leftrightarrow R) \right] = J_{CS}^U \, ,
\ee
where the contribution of the CS action to the U EoM is denoted by $J_{CS}^{U}$
\be
\label{defJU} -\frac{1}{M^3 N_c} \d_{U} S_{CS} = \text{Tr} \left(\d U J_{CS}^{U} U^\dagger\right) \, .
\ee

\paragraph*{Non-Abelian EoM} As the non-Abelian part of the gauge fields does not appear in the effective metric $\tilde{g}$, the DBI contribution to the non-Abelian equations of motion is much simpler than for the Abelian part and the tachyon
\begin{align}
\label{nabEoM} \hphantom{=}& \frac{1}{2}D_N\left[V_f(\l,\tau^2)\sqrt{-\mathrm{det}\, \tilde{g}^{(L)}}w^2\left(\frac{1}{2}F^{(L)[NM]} - \frac{1}{4}\left((\tilde{g}^{(L)})^{-1}\right)^{[NM]}F^{(L)C}_C \right) \right] = \\
\nonumber &= \frac{1}{8}V_f(\l,\tau^2)\left(\sqrt{-\mathrm{det}\, \tilde{g}^{(L)}}\left((\tilde{g}^{(L)})^{-1}\right)^{(MN)}+\sqrt{-\mathrm{det}\, \tilde{g}^{(R)}}\left((\tilde{g}^{(R)})^{-1}\right)^{(MN)}\right) \times \\
\nn &\hphantom{=} \times \ka\tau^2 \left(iD_NU^\dagger U - \frac{1}{N_f}\text{Tr}(iD_NU^\dagger U) + h.c.\right) \,\, + J_{CS}^{(L)} \, ,
\end{align}
and the right-handed equations are obtained by exchanging $(L\leftrightarrow R)$ and $U\leftrightarrow U^\dagger$. The contribution of the TCS action to the non-Abelian EoMs is denoted by $J_{CS}$
\be
\label{defJna} -\frac{1}{M^3 N_c} \d_{L} S_{CS} = \d L^a\wedge J_{CS}^{(L)\, a} \, ,
\ee
and likewise for R.

\section{Tachyon-dependent Chern-Simons terms}

\label{GCS}

In this appendix, we discuss the TCS action for a tachyon proportional to a unitary matrix ($T = \tau U$).

First, we construct the most general single-trace $\Omega_5$ form out of the fields $\tau$, $U$, $L$ and $R$, which is invariant under global $SU(N_f)_L\times SU(N_f)_R$ transformations. Moreover, the exterior derivative $F_6 = d\Omega_5$ must be  gauge invariant (i.e., invariant under local transformations also), and have the expected eigenvalues under parity and charge conjugation. That is, we require that our ansatz for $\Omega_5$ is odd under the action of the P$_1$ operator and charge conjugation even, with definitions of Appendix~\ref{app: conventions}.

In order to write down our ansatz, we first construct all possible single trace 5-forms and 4-forms built out of the non-Abelian 1-forms $\mathbf{L}/\mathbf{R}$ and $DU= dU - i \mathbf{R} U +iU \mathbf{L}$ as well as the 2-forms $F^{(L/R)}$ (recall that $DU^\dagger = - U^\dagger DU U^\dagger$ is not independent). They are then made covariant under global transformations by adding instances $U$ or $U^\dagger$ in the traces. We then apply the projector $\mathbb{I}+P_1-C-P_1C$ on the forms, which projects to the P$_1$-odd C-even subspace. This leaves us with 45 independent P odd and C even 5-forms, which we denote by $F_5^{(i)}[U,\mathbf{L},\mathbf{R}]$, and 11 independent 4-forms, which we denote by $F_4^{(i)}[U,\mathbf{L},\mathbf{R}]$. The complete list of 4-forms is given by
\begin{align}
 F_4^{(1)}&=\text{Tr}(\mathbf{L}\wedge \mathbf{F}^{(L)}U^\dagger\wedge DU)+\text{Tr}(\mathbf{L}U^\dagger\wedge DU\wedge \mathbf{F}^{(L)}) + &\nonumber\\ &\phantom{=}+\text{Tr}(\mathbf{R}\wedge DUU^\dagger\wedge \mathbf{F}^{(R)})+\text{Tr}(\mathbf{R}\wedge \mathbf{F}^{(R)}\wedge DUU^\dagger) & \nonumber\\
 F_4^{(2)}&=\text{Tr}(\mathbf{L}U^\dagger\wedge DUU^\dagger\wedge \mathbf{F}^{(R)}U)+\text{Tr}(\mathbf{L}U^\dagger\wedge \mathbf{F}^{(R)}\wedge DU)+ &\nonumber\\ &\phantom{=}+\text{Tr}(\mathbf{R}U\wedge \mathbf{F}^{(L)}U^\dagger\wedge DUU^\dagger)+\text{Tr}(\mathbf{R}\wedge DU\wedge \mathbf{F}^{(L)}U^\dagger) & \nonumber\\
 F_4^{(3)}&=-\text{Tr}(\mathbf{L}\wedge\mathbf{L}U^\dagger\wedge DUU^\dagger\wedge\mathbf{R}U)+\text{Tr}(\mathbf{L}U^\dagger\wedge DUU^\dagger\wedge \mathbf{R}\wedge\mathbf{R}U) + &\nonumber\\ &\phantom{=}+\text{Tr}(\mathbf{L}\wedge\mathbf{L}U^\dagger\wedge \mathbf{R}\wedge DU)+\text{Tr}(\mathbf{L}U^\dagger\wedge \mathbf{R}\wedge \mathbf{R}\wedge DU) & \nonumber\\
 F_4^{(4)}&=\text{Tr}(\mathbf{L}\wedge \mathbf{F}^{(L)}U^\dagger\wedge\mathbf{R}U)+\text{Tr}(\mathbf{L}U^\dagger\wedge\mathbf{R}U\wedge \mathbf{F}^{(L)})+ &\nonumber\\ &\phantom{=}+\text{Tr}(\mathbf{L}U^\dagger\wedge \mathbf{R}\wedge \mathbf{F}^{(R)}U)+\text{Tr}(\mathbf{L}U^\dagger\wedge \mathbf{F}^{(R)}\wedge\mathbf{R}U) & \nonumber\\
 F_4^{(5)}&=-\text{Tr}(\mathbf{L}\wedge \mathbf{L}\wedge\mathbf{L}U^\dagger\wedge DU)-\text{Tr}(\mathbf{R}\wedge \mathbf{R}\wedge \mathbf{R}\wedge DUU^\dagger) & \nonumber\\
 F_4^{(6)}&=\text{Tr}(\mathbf{L}\wedge \mathbf{L}\wedge\mathbf{L}U^\dagger\wedge\mathbf{R}U)+\text{Tr}(\mathbf{L}U^\dagger\wedge \mathbf{R}\wedge \mathbf{R}\wedge\mathbf{R}U) & \nonumber\\
 F_4^{(7)}&=\text{Tr}(\mathbf{L}U^\dagger\wedge DUU^\dagger\wedge DUU^\dagger\wedge\mathbf{R}U)+\text{Tr}(\mathbf{L}U^\dagger\wedge \mathbf{R}\wedge DUU^\dagger\wedge DU) & \nonumber\\
 F_4^{(8)}&=\text{Tr}(\mathbf{R}\wedge DUU^\dagger\wedge \mathbf{R}\wedge DUU^\dagger)-\text{Tr}(\mathbf{L}U^\dagger\wedge DU\wedge\mathbf{L}U^\dagger\wedge DU) & \nonumber\\
 F_4^{(9)}&=\text{Tr}(\mathbf{L}U^\dagger\wedge \mathbf{R}\wedge DUU^\dagger\wedge\mathbf{R}U)-\text{Tr}(\mathbf{L}U^\dagger\wedge\mathbf{R}U\wedge\mathbf{L}U^\dagger\wedge DU) & \nonumber\\
 F_4^{(10)}&=-\text{Tr}(\mathbf{L}U^\dagger\wedge DUU^\dagger\wedge DUU^\dagger\wedge DU)-\text{Tr}(\mathbf{R}\wedge DUU^\dagger\wedge DUU^\dagger\wedge DUU^\dagger) & \nonumber\\
 F_4^{(11)}&=\text{Tr}(\mathbf{L}U^\dagger\wedge\mathbf{R}U\wedge\mathbf{L}U^\dagger\wedge\mathbf{R}U)
\end{align}
The complete list of 5-forms is the following:
\begin{align}
F_5^{(1)} &= \mathrm{Tr}(DU\wedge \mathbf{F}^{(L)}\wedge \mathbf{F}^{(L)}U^\dagger)+\mathrm{Tr}(DUU^\dagger\wedge \mathbf{F}^{(R)}\wedge \mathbf{F}^{(R)})& \\\nonumber
F_5^{(2)} &=\mathrm{Tr}(DU\wedge \mathbf{F}^{(L)}U^\dagger\wedge DUU^\dagger\wedge DUU^\dagger)+\mathrm{Tr}(DUU^\dagger\wedge DUU^\dagger\wedge DUU^\dagger\wedge \mathbf{F}^{(R)})& \\\nonumber
F_5^{(3)} &=\mathrm{Tr}(DUU^\dagger\wedge \mathbf{F}^{(R)}U\wedge \mathbf{F}^{(L)}U^\dagger)+\mathrm{Tr}(DU\wedge \mathbf{F}^{(L)}U^\dagger\wedge \mathbf{F}^{(R)})& \\\nonumber
F_5^{(4)} &=\mathrm{Tr}(DUU^\dagger\wedge DUU^\dagger\wedge DUU^\dagger\wedge DUU^\dagger\wedge DUU^\dagger)& \\\nonumber
F_5^{(5)} &=\mathrm{Tr}(\mathbf{L}\wedge \mathbf{L}\wedge \mathbf{L}\wedge \mathbf{L}\wedge\mathbf{L})-\mathrm{Tr}(\mathbf{R}\wedge \mathbf{R}\wedge \mathbf{R}\wedge \mathbf{R}\wedge\mathbf{R})& \\\nonumber
F_5^{(6)} &=\mathrm{Tr}(\mathbf{L}\wedge \mathbf{L}\wedge \mathbf{L}\wedge\mathbf{L}U^\dagger\wedge\mathbf{R}U)-\mathrm{Tr}(\mathbf{L}U^\dagger\wedge \mathbf{R}\wedge \mathbf{R}\wedge \mathbf{R}\wedge\mathbf{R}U)& \\\nonumber
F_5^{(7)} &=\mathrm{Tr}(\mathbf{L}\wedge \mathbf{L}\wedge \mathbf{L}\wedge\mathbf{L}U^\dagger\wedge DU)+\mathrm{Tr}(\mathbf{R}\wedge \mathbf{R}\wedge \mathbf{R}\wedge \mathbf{R}\wedge DUU^\dagger)& \\\nonumber
F_5^{(8)} &=\mathrm{Tr}(\mathbf{L}\wedge \mathbf{L}\wedge\mathbf{L}U^\dagger\wedge \mathbf{R}\wedge\mathbf{R}U)-\mathrm{Tr}(\mathbf{L}\wedge\mathbf{L}U^\dagger\wedge \mathbf{R}\wedge \mathbf{R}\wedge\mathbf{R}U)& \\\nonumber
F_5^{(9)} &=\mathrm{Tr}(\mathbf{L}\wedge \mathbf{L}\wedge\mathbf{L}U^\dagger\wedge DUU^\dagger\wedge\mathbf{R}U)+\mathrm{Tr}(\mathbf{L}U^\dagger\wedge DUU^\dagger\wedge \mathbf{R}\wedge \mathbf{R}\wedge\mathbf{R}U)+& \\\nonumber &\phantom{=}+\mathrm{Tr}(\mathbf{L}\wedge \mathbf{L}\wedge\mathbf{L}U^\dagger\wedge \mathbf{R}\wedge DU)+\mathrm{Tr}(\mathbf{L}U^\dagger\wedge \mathbf{R}\wedge \mathbf{R}\wedge \mathbf{R}\wedge DU)& \\\nonumber
F_5^{(10)} &=\mathrm{Tr}(\mathbf{L}\wedge \mathbf{L}\wedge\mathbf{L}U^\dagger\wedge DUU^\dagger\wedge DU)-\mathrm{Tr}(\mathbf{R}\wedge \mathbf{R}\wedge \mathbf{R}\wedge DUU^\dagger\wedge DUU^\dagger)& \\\nonumber
F_5^{(11)} &=\mathrm{Tr}(\mathbf{L}\wedge\mathbf{L}U^\dagger\wedge\mathbf{R}U\wedge\mathbf{L}U^\dagger\wedge\mathbf{R}U)-\mathrm{Tr}(\mathbf{L}U^\dagger\wedge \mathbf{R}\wedge\mathbf{R}U\wedge\mathbf{L}U^\dagger\wedge\mathbf{R}U)& \\\nonumber
F_5^{(12)} &=\mathrm{Tr}(\mathbf{L}\wedge\mathbf{L}U^\dagger\wedge\mathbf{R}U\wedge\mathbf{L}U^\dagger\wedge DU)+\mathrm{Tr}(\mathbf{L}\wedge\mathbf{L}U^\dagger\wedge DU\wedge\mathbf{L}U^\dagger\wedge\mathbf{R}U)+& \\\nonumber &\phantom{=}+\mathrm{Tr}(\mathbf{L}U^\dagger\wedge \mathbf{R}\wedge \mathbf{R}\wedge DUU^\dagger\wedge\mathbf{R}U)+\mathrm{Tr}(\mathbf{L}U^\dagger\wedge \mathbf{R}\wedge DUU^\dagger\wedge \mathbf{R}\wedge\mathbf{R}U)& \\\nonumber
F_5^{(13)} &=\mathrm{Tr}(\mathbf{L}\wedge\mathbf{L}U^\dagger\wedge DUU^\dagger\wedge \mathbf{R}\wedge\mathbf{R}U)+\mathrm{Tr}(\mathbf{L}\wedge\mathbf{L}U^\dagger\wedge \mathbf{R}\wedge \mathbf{R}\wedge DU)& \\\nonumber
F_5^{(14)} &=\mathrm{Tr}(\mathbf{L}\wedge\mathbf{L}U^\dagger\wedge \mathbf{R}\wedge DUU^\dagger\wedge\mathbf{R}U)+\mathrm{Tr}(\mathbf{L}U^\dagger\wedge \mathbf{R}\wedge\mathbf{R}U\wedge\mathbf{L}U^\dagger\wedge DU)& \\\nonumber
F_5^{(15)} &=\mathrm{Tr}(\mathbf{L}\wedge\mathbf{L}U^\dagger\wedge DUU^\dagger\wedge DUU^\dagger\wedge\mathbf{R}U)+& \\\nonumber &\phantom{=}-\mathrm{Tr}(\mathbf{L}U^\dagger\wedge DUU^\dagger\wedge DUU^\dagger\wedge \mathbf{R}\wedge\mathbf{R}U)+& \\\nonumber &\phantom{=}+\mathrm{Tr}(\mathbf{L}\wedge\mathbf{L}U^\dagger\wedge \mathbf{R}\wedge DUU^\dagger\wedge DU)-\mathrm{Tr}(\mathbf{L}U^\dagger\wedge \mathbf{R}\wedge \mathbf{R}\wedge DUU^\dagger\wedge DU)& \\\nonumber
F_5^{(16)} &=\mathrm{Tr}(\mathbf{L}\wedge\mathbf{L}U^\dagger\wedge DU\wedge\mathbf{L}U^\dagger\wedge DU)-\mathrm{Tr}(\mathbf{R}\wedge \mathbf{R}\wedge DUU^\dagger\wedge \mathbf{R}\wedge DUU^\dagger)& \\\nonumber
F_5^{(17)} &=\mathrm{Tr}(\mathbf{L}\wedge\mathbf{L}U^\dagger\wedge DUU^\dagger\wedge \mathbf{R}\wedge DU)-\mathrm{Tr}(\mathbf{L}U^\dagger\wedge DUU^\dagger\wedge \mathbf{R}\wedge \mathbf{R}\wedge DU)& \\\nonumber
F_5^{(18)} &=\mathrm{Tr}(\mathbf{L}\wedge\mathbf{L}U^\dagger\wedge DUU^\dagger\wedge DUU^\dagger\wedge DU)+& \\\nonumber &\phantom{=}+\mathrm{Tr}(\mathbf{R}\wedge \mathbf{R}\wedge DUU^\dagger\wedge DUU^\dagger\wedge DUU^\dagger)& \\\nonumber
F_5^{(19)} &=\mathrm{Tr}(\mathbf{L}U^\dagger\wedge \mathbf{R}\wedge DU\wedge\mathbf{L}U^\dagger\wedge\mathbf{R}U)+\mathrm{Tr}(\mathbf{L}U^\dagger\wedge\mathbf{R}U\wedge\mathbf{L}U^\dagger\wedge DUU^\dagger\wedge\mathbf{R}U)& \\\nonumber
F_5^{(20)} &=\mathrm{Tr}(\mathbf{L}U^\dagger\wedge\mathbf{R}U\wedge\mathbf{L}U^\dagger\wedge DUU^\dagger\wedge DU)+& \\\nonumber &\phantom{=}-\mathrm{Tr}(\mathbf{L}U^\dagger\wedge \mathbf{R}\wedge DUU^\dagger\wedge DUU^\dagger\wedge\mathbf{R}U)& \\\nonumber
F_5^{(21)} &=\mathrm{Tr}(\mathbf{L}U^\dagger\wedge DU\wedge\mathbf{L}U^\dagger\wedge DUU^\dagger\wedge\mathbf{R}U)+& \\\nonumber &\phantom{=}-\mathrm{Tr}(\mathbf{L}U^\dagger\wedge DUU^\dagger\wedge \mathbf{R}\wedge DUU^\dagger\wedge\mathbf{R}U)+& \\\nonumber &\phantom{=}+\mathrm{Tr}(\mathbf{L}U^\dagger\wedge \mathbf{R}\wedge DU\wedge\mathbf{L}U^\dagger\wedge DU)-\mathrm{Tr}(\mathbf{L}U^\dagger\wedge \mathbf{R}\wedge DUU^\dagger\wedge \mathbf{R}\wedge DU)& \\\nonumber
F_5^{(22)} &=\mathrm{Tr}(\mathbf{L}U^\dagger\wedge DUU^\dagger\wedge DUU^\dagger\wedge DUU^\dagger\wedge\mathbf{R}U)+& \\\nonumber &\phantom{=}+\mathrm{Tr}(\mathbf{L}U^\dagger\wedge \mathbf{R}\wedge DUU^\dagger\wedge DUU^\dagger\wedge DU)& \\\nonumber
F_5^{(23)} &=\mathrm{Tr}(\mathbf{L}U^\dagger\wedge DU\wedge\mathbf{L}U^\dagger\wedge DUU^\dagger\wedge DU)+& \\\nonumber &\phantom{=}+\mathrm{Tr}(\mathbf{R}\wedge DUU^\dagger\wedge \mathbf{R}\wedge DUU^\dagger\wedge DUU^\dagger)& \\\nonumber
F_5^{(24)} &=\mathrm{Tr}(\mathbf{L}U^\dagger\wedge DUU^\dagger\wedge \mathbf{R}\wedge DUU^\dagger\wedge DU)+& \\\nonumber &\phantom{=}+\mathrm{Tr}(\mathbf{L}U^\dagger\wedge DUU^\dagger\wedge DUU^\dagger\wedge \mathbf{R}\wedge DU)& \\\nonumber
F_5^{(25)} &=\mathrm{Tr}(\mathbf{L}U^\dagger\wedge DUU^\dagger\wedge DUU^\dagger\wedge DUU^\dagger\wedge DU)+& \\\nonumber &\phantom{=}-\mathrm{Tr}(\mathbf{R}\wedge DUU^\dagger\wedge DUU^\dagger\wedge DUU^\dagger\wedge DUU^\dagger)& \\\nonumber
F_5^{(26)} &=\mathrm{Tr}(\mathbf{L}\wedge \mathbf{L}\wedge \mathbf{L}\wedge \mathbf{F}^{(L)})-\mathrm{Tr}(\mathbf{R}\wedge \mathbf{R}\wedge \mathbf{R}\wedge \mathbf{F}^{(R)})& \\\nonumber
F_5^{(27)} &=\mathrm{Tr}(\mathbf{L}\wedge \mathbf{L}\wedge\mathbf{L}U^\dagger\wedge \mathbf{F}^{(R)}U)-\mathrm{Tr}(\mathbf{R}\wedge \mathbf{R}\wedge\mathbf{R}U\wedge \mathbf{F}^{(L)}U^\dagger)& \\\nonumber
F_5^{(28)} &=\mathrm{Tr}(\mathbf{L}\wedge \mathbf{L}\wedge \mathbf{F}^{(L)}U^\dagger\wedge\mathbf{R}U)+\mathrm{Tr}(\mathbf{L}\wedge\mathbf{L}U^\dagger\wedge\mathbf{R}U\wedge \mathbf{F}^{(L)})+& \\\nonumber &\phantom{=}-\mathrm{Tr}(\mathbf{L}U^\dagger\wedge \mathbf{R}\wedge \mathbf{R}\wedge \mathbf{F}^{(R)}U)-\mathrm{Tr}(\mathbf{L}U^\dagger\wedge \mathbf{F}^{(R)}\wedge \mathbf{R}\wedge\mathbf{R}U)& \\\nonumber
F_5^{(29)} &=-\mathrm{Tr}(\mathbf{L}\wedge \mathbf{F}^{(L)}U^\dagger\wedge \mathbf{R}\wedge\mathbf{R}U)-\mathrm{Tr}(\mathbf{L}U^\dagger\wedge \mathbf{R}\wedge\mathbf{R}U\wedge \mathbf{F}^{(L)})+& \\\nonumber &\phantom{=}+\mathrm{Tr}(\mathbf{L}\wedge\mathbf{L}U^\dagger\wedge \mathbf{R}\wedge \mathbf{F}^{(R)}U)+\mathrm{Tr}(\mathbf{L}\wedge\mathbf{L}U^\dagger\wedge \mathbf{F}^{(R)}\wedge\mathbf{R}U)& \\\nonumber
F_5^{(30)} &=\mathrm{Tr}(\mathbf{L}\wedge \mathbf{L}\wedge \mathbf{F}^{(L)}U^\dagger\wedge DU)+\mathrm{Tr}(\mathbf{L}\wedge\mathbf{L}U^\dagger\wedge DU\wedge \mathbf{F}^{(L)})+& \\\nonumber &\phantom{=}+\mathrm{Tr}(\mathbf{R}\wedge \mathbf{R}\wedge DUU^\dagger\wedge \mathbf{F}^{(R)})+\mathrm{Tr}(\mathbf{R}\wedge \mathbf{R}\wedge \mathbf{F}^{(R)}\wedge DUU^\dagger)& \\\nonumber
F_5^{(31)} &=\mathrm{Tr}(\mathbf{L}\wedge\mathbf{L}U^\dagger\wedge DUU^\dagger\wedge \mathbf{F}^{(R)}U)+\mathrm{Tr}(\mathbf{L}\wedge\mathbf{L}U^\dagger\wedge \mathbf{F}^{(R)}\wedge DU)+& \\\nonumber &\phantom{=}+\mathrm{Tr}(\mathbf{R}\wedge\mathbf{R}U\wedge \mathbf{F}^{(L)}U^\dagger\wedge DUU^\dagger)+\mathrm{Tr}(\mathbf{R}\wedge \mathbf{R}\wedge DU\wedge \mathbf{F}^{(L)}U^\dagger)& \\\nonumber
F_5^{(32)} &=\mathrm{Tr}(\mathbf{L}\wedge \mathbf{F}^{(L)}\wedge\mathbf{L}U^\dagger\wedge\mathbf{R}U)-\mathrm{Tr}(\mathbf{L}U^\dagger\wedge \mathbf{R}\wedge \mathbf{F}^{(R)}\wedge\mathbf{R}U)& \\\nonumber
F_5^{(33)} &=\mathrm{Tr}(\mathbf{L}U^\dagger\wedge\mathbf{R}U\wedge\mathbf{L}U^\dagger\wedge \mathbf{F}^{(R)}U)-\mathrm{Tr}(\mathbf{L}U^\dagger\wedge\mathbf{R}U\wedge \mathbf{F}^{(L)}U^\dagger\wedge\mathbf{R}U)& \\\nonumber
F_5^{(34)} &=\mathrm{Tr}(\mathbf{L}\wedge \mathbf{F}^{(L)}U^\dagger\wedge DUU^\dagger\wedge\mathbf{R}U)+\mathrm{Tr}(\mathbf{L}U^\dagger\wedge \mathbf{R}\wedge DU\wedge \mathbf{F}^{(L)})+& \\\nonumber &\phantom{=}+\mathrm{Tr}(\mathbf{L}U^\dagger\wedge DUU^\dagger\wedge \mathbf{F}^{(R)}\wedge\mathbf{R}U)+\mathrm{Tr}(\mathbf{L}U^\dagger\wedge \mathbf{R}\wedge \mathbf{F}^{(R)}\wedge DU)& \\\nonumber
F_5^{(35)} &=\mathrm{Tr}(\mathbf{L}U^\dagger\wedge\mathbf{R}U\wedge \mathbf{F}^{(L)}U^\dagger\wedge DU)+\mathrm{Tr}(\mathbf{L}U^\dagger\wedge DU\wedge \mathbf{F}^{(L)}U^\dagger\wedge\mathbf{R}U)+& \\\nonumber &\phantom{=}+\mathrm{Tr}(\mathbf{L}U^\dagger\wedge \mathbf{R}\wedge DUU^\dagger\wedge \mathbf{F}^{(R)}U)+\mathrm{Tr}(\mathbf{L}U^\dagger\wedge \mathbf{F}^{(R)}\wedge DUU^\dagger\wedge\mathbf{R}U)& \\\nonumber
F_5^{(36)} &=\mathrm{Tr}(\mathbf{L}\wedge \mathbf{F}^{(L)}\wedge\mathbf{L}U^\dagger\wedge DU)+\mathrm{Tr}(\mathbf{R}\wedge DUU^\dagger\wedge \mathbf{R}\wedge \mathbf{F}^{(R)})& \\\nonumber
F_5^{(37)} &=\mathrm{Tr}(\mathbf{L}U^\dagger\wedge DU\wedge\mathbf{L}U^\dagger\wedge \mathbf{F}^{(R)}U)+\mathrm{Tr}(\mathbf{R}\wedge DUU^\dagger\wedge\mathbf{R}U\wedge \mathbf{F}^{(L)}U^\dagger)& \\\nonumber
F_5^{(38)} &=\mathrm{Tr}(\mathbf{L}U^\dagger\wedge DUU^\dagger\wedge\mathbf{R}U\wedge \mathbf{F}^{(L)})+\mathrm{Tr}(\mathbf{L}\wedge \mathbf{F}^{(L)}U^\dagger\wedge \mathbf{R}\wedge DU)+& \\\nonumber &\phantom{=}+\mathrm{Tr}(\mathbf{L}U^\dagger\wedge DUU^\dagger\wedge \mathbf{R}\wedge \mathbf{F}^{(R)}U)+\mathrm{Tr}(\mathbf{L}U^\dagger\wedge \mathbf{F}^{(R)}\wedge \mathbf{R}\wedge DU)& \\\nonumber
F_5^{(39)} &=\mathrm{Tr}(\mathbf{L}\wedge \mathbf{F}^{(L)}U^\dagger\wedge DUU^\dagger\wedge DU)+\mathrm{Tr}(\mathbf{L}U^\dagger\wedge DUU^\dagger\wedge DU\wedge \mathbf{F}^{(L)})+& \\\nonumber &\phantom{=}-\mathrm{Tr}(\mathbf{R}\wedge DUU^\dagger\wedge DUU^\dagger\wedge \mathbf{F}^{(R)})-\mathrm{Tr}(\mathbf{R}\wedge \mathbf{F}^{(R)}\wedge DUU^\dagger\wedge DUU^\dagger)& \\\nonumber
F_5^{(40)} &=\mathrm{Tr}(\mathbf{L}U^\dagger\wedge DUU^\dagger\wedge DUU^\dagger\wedge \mathbf{F}^{(R)}U)+\mathrm{Tr}(\mathbf{L}U^\dagger\wedge \mathbf{F}^{(R)}\wedge DUU^\dagger\wedge DU)+& \\\nonumber &\phantom{=}-\mathrm{Tr}(\mathbf{R}U\wedge \mathbf{F}^{(L)}U^\dagger\wedge DUU^\dagger\wedge DUU^\dagger)-\mathrm{Tr}(\mathbf{R}\wedge DUU^\dagger\wedge DU\wedge \mathbf{F}^{(L)}U^\dagger)& \\\nonumber
F_5^{(41)} &=\mathrm{Tr}(\mathbf{L}U^\dagger\wedge DU\wedge \mathbf{F}^{(L)}U^\dagger\wedge DU)-\mathrm{Tr}(\mathbf{R}\wedge DUU^\dagger\wedge \mathbf{F}^{(R)}\wedge DUU^\dagger)& \\\nonumber
F_5^{(42)} &=\mathrm{Tr}(\mathbf{L}U^\dagger\wedge DUU^\dagger\wedge \mathbf{F}^{(R)}\wedge DU)-\mathrm{Tr}(\mathbf{R}\wedge DU\wedge \mathbf{F}^{(L)}U^\dagger\wedge DUU^\dagger)& \\\nonumber
F_5^{(43)} &=\mathrm{Tr}(\mathbf{L}\wedge \mathbf{F}^{(L)}\wedge \mathbf{F}^{(L)})-\mathrm{Tr}(\mathbf{R}\wedge \mathbf{F}^{(R)}\wedge \mathbf{F}^{(R)})& \\\nonumber
F_5^{(44)} &=\mathrm{Tr}(\mathbf{L}\wedge \mathbf{F}^{(L)}U^\dagger\wedge \mathbf{F}^{(R)}U)+\mathrm{Tr}(\mathbf{L}U^\dagger\wedge \mathbf{F}^{(R)}U\wedge \mathbf{F}^{(L)})+& \\\nonumber &\phantom{=}-\mathrm{Tr}(\mathbf{R}\wedge \mathbf{F}^{(R)}U\wedge \mathbf{F}^{(L)}U^\dagger)-\mathrm{Tr}(\mathbf{R}U\wedge \mathbf{F}^{(L)}U^\dagger\wedge \mathbf{F}^{(R)})& \\\nonumber
F_5^{(45)} &=\mathrm{Tr}(\mathbf{L}U^\dagger\wedge \mathbf{F}^{(R)}\wedge \mathbf{F}^{(R)}U)-\mathrm{Tr}(\mathbf{R}U\wedge \mathbf{F}^{(L)}\wedge \mathbf{F}^{(L)}U^\dagger)
\end{align}
The most general single-trace ansatz for $\Omega_5$ which is covariant under the global transformations and has the desired P and C eigenvalues can therefore be written as
\be
 \Omega_5 = \sum_{i=1}^{45} \bar f_{i}(\tau) F_5^{(i)}[U,\mathbf{L},\mathbf{R}] + d\tau\wedge\sum_{i=1}^{11} g_{i}(\tau)F_4^{(i)}[U,\mathbf{L},\mathbf{R}] \label{O5Ans}
\ee
where we added the most general dependence on the scalar $\tau$. Notice that the Ansatz includes 56 arbitrary functions of $\tau$.

Requiring that $d \Omega_5$ is gauge invariant, i.e., requiring that $\delta d \Omega_5 =0$, sets 41 conditions for the functions $\bar f_i$, $g_i$, of which 29 are algebraic and 12 involve their first derivatives. This reduces the number of free functions down to $15$. Notice that the latter 12 constraints are differential equations, the solutions of which contain 12 integration constants. We comment on these integration constants below.

As it turns out, the result for $\Omega_5$ after imposing the above constraints is something one could have guessed from the start: it is given as a sum of two terms, where the first term (involving 4 of the 15 free functions) is explicitly gauge invariant and the second term is a closed form (involving 11 of the 15 free functions). For such a solution, it is immediate that indeed $\delta d\Omega_5 = 0$.
The solution may be written explicitly as
\be
\label{expG4} \Omega_5 = \Omega_5^0 + \Omega_5^c + \mathrm{d}G_4 \, ,
\ee
where
\begin{align}
\Omega_5^0 &=
 f_1(\tau) \big[\text{Tr}(DU\wedge \mathbf{F}^{(L)}\wedge \mathbf{F}^{(L)}U^\dagger)+\text{Tr}(DUU^\dagger\wedge \mathbf{F}^{(R)}\wedge \mathbf{F}^{(R)})\big]+\nn\\
&
+f_2(\tau) \big[\text{Tr}(DU\wedge \mathbf{F}^{(L)}U^\dagger\wedge DUU^\dagger\wedge DUU^\dagger)+\nn\\
&\qquad \ \ +\text{Tr}(DUU^\dagger\wedge \mathbf{F}^{(R)}\wedge DUU^\dagger\wedge DUU^\dagger)\big]+\nn\\
&+f_3(\tau) \big[\text{Tr}(DUU^\dagger\wedge \mathbf{F}^{(R)}U\wedge \mathbf{F}^{(L)}U^\dagger)+\text{Tr}(DU\wedge \mathbf{F}^{(L)}U^\dagger\wedge \mathbf{F}^{(R)})\big]+\nn\\
&+f_4(\tau) \text{Tr}(DUU^\dagger\wedge DUU^\dagger\wedge DUU^\dagger\wedge DUU^\dagger\wedge DUU^\dagger)
\label{O50}
\end{align}
is the gauge-invariant term and $\Omega_5^c + \mathrm{d}G_4$ is the closed term.
The latter was, for convenience, divided into two parts, where
\be
\label{O5c} \Omega_5^c = g_0 \text{Tr}((U^\dagger\mathrm{d}U)^5)
\ee
with $g_0$ a constant, and $G_4$ is a general $\mathrm{P_1}$ odd and C even 4-form, i.e.,
\be \label{G4form}
 G_4 = \sum_{i=1}^{11} h_i(\tau) F_4^{(i)}[U,\mathbf{L},\mathbf{R}] \ .
\ee
(Note that $g_i(\tau) = h_i'(\tau)$.)

Several comments are in order. Notice that the gauge invariant term is indeed the most general $\mathrm{P_1}$-odd C-even 5-form composed of the covariant forms $DU$, $\mathbf{F}^{(L)}$, and $\mathbf{F}^{(R)}$. Only this term contributes to the $F_6 = d\Omega_5$, which is given as
\begingroup
\allowdisplaybreaks
\begin{gather}
 F_6 =d\tau\wedge \Big\{f_1'(\tau )\big[\text{Tr}(DU\wedge \mathbf{F}^{(L)}\wedge \mathbf{F}^{(L)}U^\dagger)+\text{Tr}(DUU^\dagger\wedge \mathbf{F}^{(R)}\wedge \mathbf{F}^{(R)})\big]+ \nonumber\\
  + f_2'(\tau )\big[\text{Tr}(DU\wedge \mathbf{F}^{(L)}U^\dagger\wedge DUU^\dagger\wedge DUU^\dagger)\!+\! \text{Tr}(DUU^\dagger\wedge \mathbf{F}^{(R)}\wedge DUU^\dagger\wedge DUU^\dagger)\big]+ \nonumber\\
  + f_3'(\tau )\big[\text{Tr}(DUU^\dagger\wedge \mathbf{F}^{(R)}U\wedge \mathbf{F}^{(L)}U^\dagger)+\text{Tr}(DU\wedge \mathbf{F}^{(L)}U^\dagger\wedge \mathbf{F}^{(R)})\big] + \nonumber\\
 +f_4'(\tau ) \text{Tr}(DUU^\dagger\wedge DUU^\dagger\wedge DUU^\dagger\wedge DUU^\dagger\wedge DUU^\dagger)\Big\}+\nonumber\\
+f_1(\tau ) \big[\text{Tr}(DU\wedge \mathbf{F}^{(L)}\wedge \mathbf{F}^{(L)}U^\dagger\wedge DUU^\dagger)
+\text{Tr}(DUU^\dagger\wedge DUU^\dagger\wedge \mathbf{F}^{(R)}\wedge \mathbf{F}^{(R)})+\nonumber\\-i \text{Tr}(\mathbf{F}^{(L)}\wedge \mathbf{F}^{(L)}U^\dagger\wedge \mathbf{F}^{(R)}U)
+i \text{Tr}(\mathbf{F}^{(L)}U^\dagger\wedge \mathbf{F}^{(R)}\wedge \mathbf{F}^{(R)}U) + \nonumber\\+i \text{Tr}(\mathbf{F}^{(L)}\wedge \mathbf{F}^{(L)}\wedge \mathbf{F}^{(L)})
-i \text{Tr}(\mathbf{F}^{(R)}\wedge \mathbf{F}^{(R)}\wedge \mathbf{F}^{(R)})\big]+\nonumber\\
+f_2(\tau ) \big[-2 i \text{Tr}(DU\wedge \mathbf{F}^{(L)}U^\dagger\wedge DUU^\dagger\wedge \mathbf{F}^{(R)})+\nonumber\\
-2 i \text{Tr}(DU\wedge \mathbf{F}^{(L)}\wedge \mathbf{F}^{(L)}U^\dagger\wedge DUU^\dagger)+\nonumber\\
+\text{Tr}(DU\wedge \mathbf{F}^{(L)}U^\dagger\wedge DUU^\dagger\wedge DUU^\dagger\wedge DUU^\dagger)+\nonumber\\
-2 i \text{Tr}(DUU^\dagger\wedge DUU^\dagger\wedge \mathbf{F}^{(R)}\wedge \mathbf{F}^{(R)})+\nonumber\\
+\text{Tr}(DUU^\dagger\wedge DUU^\dagger\wedge DUU^\dagger\wedge DUU^\dagger\wedge \mathbf{F}^{(R)})\big]+\nonumber\\
+2 f_3(\tau ) \big[\text{Tr}(DU\wedge \mathbf{F}^{(L)}U^\dagger\wedge DUU^\dagger\wedge \mathbf{F}^{(R)})+\nonumber\\
+i \text{Tr}(\mathbf{F}^{(L)}\wedge \mathbf{F}^{(L)}U^\dagger\wedge \mathbf{F}^{(R)}U)-i \text{Tr}(\mathbf{F}^{(L)}U^\dagger\wedge \mathbf{F}^{(R)}\wedge \mathbf{F}^{(R)}U)\big]+\nonumber\\
-5 i f_4(\tau ) \big[\text{Tr}(DU\wedge \mathbf{F}^{(L)}U^\dagger\wedge DUU^\dagger\wedge DUU^\dagger\wedge DUU^\dagger)+\nonumber\\
+\text{Tr}(DUU^\dagger\wedge DUU^\dagger\wedge DUU^\dagger\wedge DUU^\dagger\wedge \mathbf{F}^{(R)})\big] 
\end{gather}
\endgroup
In particular, the 6-form derived by using the flat space expression in~\cite{Casero} is obtained for
\begin{align}
 f_1(\tau) &= -\frac{1}{6}e^{-\tau^2}\,,\qquad f_2(\tau) = \frac{i}{12}(1+\tau^2)e^{-\tau^2}\,,\nn\\
 f_3(\tau) &= -\frac{1}{12}e^{-\tau^2}\,,\qquad f_4(\tau) = \frac{1}{120}(2+2\tau^2+\tau^4)e^{-\tau^2} \, .
 \label{CKPf}
\end{align}
Moreover, we require that the bulk action agrees in the chirally symmetric case, $\tau=0$ and $U =\mathbb{I}$, with the standard expression
\be
 \Omega_5^0 = -\frac{i}{6}\text{Tr}\left( \mathbf{L}(\mathbf{F}^{(L)})^2 + \frac{i}{2}\mathbf{L}^3\mathbf{F}^{(L)} - \frac{1}{10}\mathbf{L}^5 - (L\leftrightarrow R) \right)
\ee
up to boundary terms. This is the case if
\be \label{fizerovals}
 f_1(0) = -\frac{1}{6}\, , \quad f_2(0) = \frac{i}{12} \, , \quad f_3(0) = -\frac{1}{12} \, , \quad f_4(0) = \frac{1}{60} \,.
\ee
Notice that the choice in~\eqref{CKPf} satisfies these conditions.

In \eqref{expG4} $\Omega_5^c$ is a closed 5-form which cannot be expressed globally as a total exterior derivative. This form is related to the fifth de Rham cohomology group $H_5(SU(N_f)_L\times SU(N_f)_R/SU(N_f)_V)$ which reduces to the class of $\text{Tr}((\mathrm{d}UU^{-1})^5)$.

We then comment on the 12 integration constants mentioned above. We find that these correspond to shifts of various linear combinations of the functions $\bar f_i$ in~\eqref{O5Ans}. In more detail, the 12 conditions involving derivatives are of the form
\be
\frac{d}{d\tau}\left(\textrm{linear combinations of } \bar f_i\right) =\textrm{linear combinations of } g_i
\ee
Instead of solving these conditions for $\bar f_i$ one can eliminate the $g_i$'s (as we have in practice done above by using the functions $h_i$ instead in~\eqref{G4form}, which are derivatives of the $g_i$'s), in which case the question of the integration constants does not arise explicitly. However notice that there are only 11 functions $g_i$ (and equivalently 11 $h_i$'s) but there are 12 conditions so only 11 of the constants can be removed this way and one of them remains.
It is tempting to identify
11 of the integration constants with the freedom of adding
\be
 \sum_{i=1}^{11} C_i F_4^{(i)}[U,\mathbf{L},\mathbf{R}]
\ee
in the closed contribution to $\Omega_5$. However explicit computation shows that this is not precisely correct: one also needs to include contributions from the gauge invariant terms~\eqref{O50}. Nevertheless, the conclusion is that 11 of the constants can be absorbed in the functions $h_i(\tau)$ and $f_i(\tau)$ without loss of generality. The remaining integration constant however cannot be absorbed in this way: it is identified as the constant $g_0$ in~\eqref{O5c} -- again explicit computation shows that it is actually a combination of $g_0$ and a constant term in the function $f_4$ in~\eqref{O50}. Despite these technical complications, the computation shows that the result in~\eqref{expG4}--\eqref{G4form} is consistent.

The result for $\Omega_5$ is then constrained by requiring that it produces the correct QCD flavor anomaly at the boundary. To do this we first compute the (linearized) gauge transformation of the pure gauge term: \be
\delta \Omega_5^c = 5 g_0 d\left\{\mathrm{Tr}[\Lambda_Ld((U^\dagger\mathrm{d}U )^3) + \Lambda_R d(( \mathrm{d}U U^\dagger)^3)] \right\} \ .
\ee
Moreover, the gauge transformation of the $dG_4$ can be written as
\be
 d \delta G_4 = d[\mathrm{Tr}(\Lambda_L dG_{3L}+\Lambda_R dG_{3R})]
\ee
where the 3-forms can be found by computing the gauge transformation of $G_4$,
\be
 \delta G_4 = -\mathrm{Tr}[d\Lambda_L\wedge G_{3L}+d \Lambda_R \wedge G_{3R}] \ .
\ee
Matching with the QCD anomalies therefore requires that, at the boundary,
\begin{align}
& \mathrm{Tr}[\Lambda_L dG_{3L}+\Lambda_R dG_{3R}] + 5 g_0 \mathrm{Tr}[\Lambda_Ld((U^\dagger\mathrm{d}U )^3) + \Lambda_R d(( \mathrm{d}U U^\dagger)^3)] &\nonumber\\
 =\ & - \frac{1}{6}  \mathrm{Tr}\left[\L_L\left((\mathrm{d}\mathbf{L})^2 - \frac{i}{2}\mathrm{d}(\mathbf{L}^3)\right) - (L\leftrightarrow R) \right] \ .&
\end{align}
This equation fixes the boundary values of all the functions $h_i$ in $G_4$ and the value of the constant $g_0$:
\begin{align}
 g_0 &= -\frac{1}{60}\ ,\quad
 h_1(0) = \frac{1}{12} \ , \quad h_2(0) = \frac{1}{24} \ , \quad h_3(0) = \frac{i}{24} \ , \quad h_4(0) = \frac{i}{24} \ , &\nonumber\\
 h_5(0) &= -\frac{i}{12} \ , \quad h_6(0) = -\frac{1}{12} \ , \quad h_7(0) = \frac{1}{12} \ , \quad h_8(0) = \frac{1}{24} \ , \quad h_9(0) = \frac{i}{12} \ ,&\nonumber\\
 h_{10}(0) &= \frac{i}{12} \ , \quad h_{11}(0) = \frac{1}{24} \ . &
\end{align}
Notice that since $d G_4$ integrates to a boundary term, $\int \Omega_5$ is therefore determined up to the four functions $f_i$, assuming that there is no IR boundary contribution. As for the $G_4$, this can be ensured by choosing the $h_i$ such that they vanish in the IR. The IR contribution from $\Omega_5^c$, however cannot be eliminated by the remaining freedom in the choice of the 5-form. Therefore the vanishing of this contribution needs to be required as an additional constraint on the solutions. We shall discuss this point in more detail below.

After imposing the constraints, the explicit form for $G_4$ at the boundary reads
\begingroup
\allowdisplaybreaks
\begin{gather}
\label{G4bdry} 24\, G_4\big|_\mathrm{bdry}\! =\ 2\Big[ \text{Tr}(\mathbf{L}\wedge \mathbf{F}^{(L)}\,U^\dagger\wedge DU)+ \text{Tr}(\mathbf{L}\,U^\dagger\wedge DU\wedge \mathbf{F}^{(L)})+ \\\nonumber
\ \ + \text{Tr}(\mathbf{R}\wedge DU\,U^\dagger\wedge \mathbf{F}^{(R)})+ \text{Tr}(\mathbf{R}\wedge \mathbf{F}^{(R)}\wedge DU\,U^\dagger)\Big]+\\\nonumber
+\Big[\text{Tr}(\mathbf{L}\,U^\dagger\wedge DU\,U^\dagger\wedge \mathbf{F}^{(R)}\,U)+\text{Tr}(\mathbf{L}\,U^\dagger\wedge \mathbf{F}^{(R)}\wedge DU)+ \\\nonumber
\ \ +\text{Tr}(\mathbf{R}\wedge DU\wedge \mathbf{F}^{(L)}\,U^\dagger)+\text{Tr}(\mathbf{R}\,U\wedge \mathbf{F}^{(L)}\,U^\dagger\wedge DU\,U^\dagger)\Big]+ \\\nonumber
+i\Big[- \text{Tr}(\mathbf{L}\wedge \mathbf{L}\,U^\dagger\wedge DU\,U^\dagger\wedge \mathbf{R}\,U)+ \text{Tr}(\mathbf{L}\wedge \mathbf{L}\,U^\dagger\wedge \mathbf{R}\wedge DU)+ \nonumber\\\nonumber
 \ \ + \text{Tr}(\mathbf{L}\,U^\dagger\wedge \mathbf{R}\wedge \mathbf{R}\wedge DU)+ \text{Tr}(\mathbf{L}\,U^\dagger\wedge DU\,U^\dagger\wedge \mathbf{R}\wedge \mathbf{R}\,U)\Big]+  \\\nonumber
+i\Big[ \text{Tr}(\mathbf{L}\wedge \mathbf{F}^{(L)}\,U^\dagger\wedge \mathbf{R}\,U)+ \text{Tr}(\mathbf{L}\,U^\dagger\wedge \mathbf{R}\,U\wedge \mathbf{F}^{(L)})+ \\\nonumber
\ \ + \text{Tr}(\mathbf{L}\,U^\dagger\wedge \mathbf{R}\wedge \mathbf{F}^{(R)}\,U)+ \text{Tr}(\mathbf{L}\,U^\dagger\wedge \mathbf{F}^{(R)}\wedge \mathbf{R}\,U)\Big]+ \\\nonumber
+2 i\Big[ \text{Tr}(\mathbf{L}\wedge \mathbf{L}\wedge \mathbf{L}\,U^\dagger\wedge DU)+ \text{Tr}(\mathbf{R}\wedge \mathbf{R}\wedge \mathbf{R}\wedge DU\,U^\dagger)\Big] +\\\nonumber
-2\Big[ \text{Tr}(\mathbf{L}\wedge \mathbf{L}\wedge \mathbf{L}\,U^\dagger\wedge \mathbf{R}\,U)+ \text{Tr}(\mathbf{L}\,U^\dagger\wedge \mathbf{R}\wedge \mathbf{R}\wedge \mathbf{R}\,U)\Big] +\\\nonumber
 +2\Big[\text{Tr}(\mathbf{L}\,U^\dagger\wedge DU\,U^\dagger\wedge DU\,U^\dagger\wedge \mathbf{R}\,U)+\text{Tr}(\mathbf{L}\,U^\dagger\wedge \mathbf{R}\wedge DU\,U^\dagger\wedge DU)\Big] +\\\nonumber
+\Big[-\text{Tr}(\mathbf{L}\,U^\dagger\wedge DU\wedge \mathbf{L}\,U^\dagger\wedge DU)+\text{Tr}(\mathbf{R}\wedge DU\,U^\dagger\wedge \mathbf{R}\wedge DU\,U^\dagger)\Big]+ \\\nonumber
+2 i\Big[- \text{Tr}(\mathbf{L}\,U^\dagger\wedge \mathbf{R}\,U\wedge \mathbf{L}\,U^\dagger\wedge DU)+ \text{Tr}(\mathbf{L}\,U^\dagger\wedge \mathbf{R}\wedge DU\,U^\dagger\wedge \mathbf{R}\,U)\Big] +\\\nonumber
-2 i\Big[ \text{Tr}(\mathbf{L}\,U^\dagger\wedge DU\,U^\dagger\wedge DU\,U^\dagger\wedge DU)
+ \text{Tr}(\mathbf{R}\wedge DU\,U^\dagger\wedge DU\,U^\dagger\wedge DU\,U^\dagger)\Big]+ \\\nonumber
+\text{Tr}(\mathbf{L}\,U^\dagger\wedge \mathbf{R}\,U\wedge \mathbf{L}\,U^\dagger\wedge \mathbf{R}\,U) \ .
\end{gather}
\endgroup

Our final result can then be compared to the chiral Lagrangian Wess-Zumino term written in \cite{Kaymakcalan} and given below in~\eqref{gaugedWZ}. We find that
\be
 \int\! \Omega_5^c + \int\! G_4\big|_\mathrm{bdry} = \frac{ 40i\, \pi^2}{N_c} S_{WZ} + \frac{1}{24} \int d[\text{Tr}(\mathbf{L}U^\dagger\wedge d\mathbf{R}U)-\text{Tr}(\mathbf{R}U\wedge d\mathbf{L}U^\dagger)]
\ee
where $S_{WZ}$ is the action from this reference, discussed below in Appendix~\ref{app:gaugedWZ}. That is, the expressions agree up to a derivative term, which is C-odd. Notice that adding such a derivative in $G_4$ would leave $\Omega_5$ unchanged. Because none of the $\mathrm{P_1}$-odd C-even forms $F_4^{(i)}$ is gauge invariant, there is no freedom of adding gauge invariant terms such as~\eqref{ad} below. Moreover, the normalization of the action also agrees (up to differences in sign conventions) with~\cite{Kaymakcalan}.

\subsubsection*{IR contribution from $\Omega_5^c$}

We investigate here the contribution to $\int \Omega_5^c$ -- where $\Omega_5^c$ is the non-exact closed part of the TCS 5-form  \eqref{O5c} -- that comes from the singular IR point at $r\to \infty$, and determine the condition for it to vanish. This will put constraints on the extreme IR behavior ($r\to \infty$) of the gauge transformations that are allowed in the bulk.

$\Omega_5^c$ is a closed form, so in principle it can be written locally as the exterior derivative of some 4-form $\omega_4$. However, as noted in \cite{WWZ}, there is no way to find an explicit expression in terms of $U$ for such an $\omega_4$. On the other hand, what can be done is to write $\omega_4$ order by order in the pion field $\Pi$ defined as
\be
\label{defPiU} U = \ex^{\frac{2i}{f_\pi}\Pi} \, .
\ee
Up to order $\mathcal{O}(\Pi^7)$, we find that locally $\Omega_5^c$ is equal to
\begin{align}
\label{05cexpPi} \Omega_5^c &= g_0 \text{Tr}\left\{ i\left(\frac{2}{f_\pi}\right)^5\mathrm{d}\left(\Pi \left(\mathrm{d}\Pi\right)^4\right)\right\} + \\
\nn &\hphantom{=} + g_0 \text{Tr}\left\{\frac{160i}{21f_\pi^7}\mathrm{d}\left(3\Pi^2\mathrm{d}\Pi\wedge\Pi\mathrm{d}\Pi^3 \!+\! 3\Pi\mathrm{d}\Pi\wedge \Pi\mathrm{d}\Pi\wedge\Pi\mathrm{d}\Pi^2 \!-\! 2\Pi^3\mathrm{d}\Pi^4\right) + \mathcal{O}(\Pi^8) \right\}\! ,
\end{align}
where all higher orders will also be written as the exterior derivative of a 4-form containing four exterior derivatives of $\Pi$. When integrating over the bulk, $\Omega_5^c$ expanded in powers of $\Pi$ can therefore be written as a boundary integral, with contributions from the UV boundary and IR singularity
\begin{align}
\label{intO5cPi} \int\Omega_5^c &= ig_0 \left(\frac{2}{f_\pi}\right)^5 \left[\int\text{Tr}\left\{ \Pi \left(\mathrm{d}\Pi\right)^4 \right\}\right]_{UV}^{IR} + \\
\nn &\hphantom{=} + ig_0\frac{160}{21f_\pi^7} \bigg[\int\text{Tr}\{3\Pi^2\mathrm{d}\Pi\wedge\Pi\mathrm{d}\Pi^3 + \\
\nn &\hphantom{= + ig_0\frac{160}{21f_\pi^7} \bigg[\int\text{Tr}} + 3\Pi\mathrm{d}\Pi\wedge \Pi\mathrm{d}\Pi\wedge\Pi\mathrm{d}\Pi^2 - 2\Pi^3\mathrm{d}\Pi^4 + \mathcal{O}(\Pi^8) \}\bigg]_{UV}^{IR} \, .
\end{align}
So the condition for the IR contribution to vanish at all orders in $\Pi$, is for the derivative of $\Pi$ (or equivalently $U$) in one of the boundary coordinates $x^\mu$ to be identically 0 in the IR. Note that it is always obeyed by the instanton ansatz  \eqref{TU2} because of its invariance under time reversal. Also, we expect that it should be possible for physically relevant solutions to impose the more natural stronger condition
\be
\label{cUIRstr} \partial_\mu U(x,r) \underset{r\to\infty}{\to} 0 \, ,
\ee
which is Lorentz invariant.

\subsection{Baryon charge} \label{App:charge}

We now check the coupling of the non-Abelian fields to the Abelian vectorial charge. To do this, we separate the vectorial Abelian term by replacing $\mathbf{L}/\mathbf{R} \to \Phi + L/R$ (see \eqref{defLR}). Notice that here $\Phi$ is a generic Abelian one-form, $\Phi = \Phi_M dx^M$, where the index $M$ runs over all the five coordinates. Consequently, the field strengths are replaced using the rules $\mathbf{F}^{(L/R)}\to d\Phi+F^{(L/R)}$, while $DU$ is unchanged. Therefore the bulk term becomes
\begin{align}\label{Bcharge1}
 \Omega_5^0 =\ &\Omega_5^0\big|_{\Phi=0} + 2d\Phi \wedge (f_1(\tau)+f_3(\tau))\mathrm{Tr}[DU\wedge (F^{(L)} U^\dagger + U^\dagger F^{(R)})]+& \\\nonumber
&+ 2d\Phi \wedge f_2(\tau)\mathrm{Tr}[(DU\, U^\dagger)^3]
 +2(f_1(\tau)+f_3(\tau)) d\Phi\wedge d\Phi \wedge \mathrm{Tr}(DU \,U^\dagger) \, . &
\end{align}
Notice that gauge invariance of these terms guarantees that the dependence on $\Phi$ is through the derivative. Moreover only the functions $f_1+f_3$ and $f_2$ appear in the coupling term. For this expression, the constraints~\eqref{fizerovals} impose that
\be \label{Baryonconstr}
 f_1(0)+f_3(0) = -\frac{1}{4}\ ,\qquad f_2(0) =\frac{i}{12}  \,.
\ee
For the $\Omega_5$ of~\cite{Casero}, we find that
\be
 f_1(\tau)+f_3(\tau) = -\frac{1}{4} e^{-\tau^2} \ , \qquad f_2(\tau) = \frac{i}{12}(1+\tau^2)e^{-\tau^2} \ .
\ee
As it turns out, it is convenient to rewrite the coupling as
\begin{align}
 \nn \Omega_5^0 = &\Omega_5^0\big|_{\Phi=0}+2(f_1(\tau)\!+\!f_3(\tau)) d\Phi\wedge d\Phi \wedge \mathrm{Tr}(DU \,U^\dagger) + &\\
 &-2\Phi \wedge d\{(f_1(\tau)\!+\!f_3(\tau))\mathrm{Tr}[DU\wedge (F^{(L)} U^\dagger + U^\dagger F^{(R)})]\!+\!f_2(\tau)\mathrm{Tr}[(DU\, U^\dagger)^3]\} + &\nonumber\\
 & + 2 d\{\Phi\wedge (f_1(\tau)\!+\!f_3(\tau))\mathrm{Tr}[DU\wedge (F^{(L)} U^\dagger + U^\dagger F^{(R)})]\!+\!f_2(\tau)\mathrm{Tr}[(DU\, U^\dagger)^3]\} \, .& \label{Bcharge2}
\end{align}

There are also boundary terms involving $\Phi$, which arise from $G_4$. They can be written as
\begin{align} \label{G4Phi}
 G_4\big|_\mathrm{bdry} =\ & G_4\big|_{\Phi=0} + \frac{1}{12}\Phi\wedge [-4 i \text{Tr}(L\wedge F^{(L)})+\text{Tr}(L\wedge L\wedge L)+4 i \text{Tr}(R\wedge F^{(R)}) + & \nonumber\\
 &-\text{Tr}(R\wedge R\wedge R)+6 \text{Tr}(DU\wedge F^{(L)}U^\dagger)+6 \text{Tr}(DUU^\dagger\wedge F^{(R)}) + & \nonumber\\
 &-2 i \text{Tr}(DUU^\dagger\wedge DUU^\dagger\wedge DUU^\dagger)] +& \\
 & +\frac{1}{12} d[\Phi \wedge  (2 i \text{Tr}(LU^\dagger\wedge RU)+3 \text{Tr}(LU^\dagger\wedge DU)+3 \text{Tr}(R\wedge DUU^\dagger))] \ . & \nonumber
\end{align}
Here, the total derivative term does not contribute to $\Omega_5$ and we drop it. Interestingly, the covariant terms in~\eqref{G4Phi} cancel against the boundary term arising from~\eqref{Bcharge2} after using the conditions~\eqref{Baryonconstr}. Therefore the full baryon coupling takes a simple form:
\begin{align}
 & \int \Omega_5 - \int \Omega_5\big|_{\Phi=0} &\nonumber\\
= & -2\! \int \Phi \wedge d\{(f_1(\tau)+f_3(\tau))\mathrm{Tr}[DU\wedge (F^{(L)} U^\dagger + U^\dagger F^{(R)})]\!+\!f_2(\tau)\mathrm{Tr}[(DU\, U^\dagger)^3]\}+ &\nonumber\\
\nn & + 2\! \int (f_1(\tau)+f_3(\tau)) d\Phi\wedge d\Phi \wedge \mathrm{Tr}(DU \,U^\dagger)+ &\\\label{Phicouplingfinal}
& +\frac{1}{12}\! \int \Phi\wedge [-4 i \text{Tr}(L\wedge F^{(L)})+\text{Tr}(L^3)+4 i \text{Tr}(R\wedge F^{(R)}) -\text{Tr}(R^3)]\big|_\mathrm{bdry} \ . &
\end{align}
Notice that here the integral on the last row is four dimensional whereas the other integrals are five dimensional.

We then extract the expression for the baryon number current and charge. {The full action contains the dilaton gravity, DBI, and TCS terms. Only the DBI and TCS terms depend on the gauge fields and are therefore relevant for the analysis of the baryon number current. As it turns out, the charge of the baryon will only arise from the TCS term. The full action may be written as}
\be
 S = S^{(5D)} + S^{(4D)} = \int dr d^4x\ \mathcal{L}^{(5D)} +  \int d^4x\ \mathcal{L}^{(4D)}\big|_{r=0}
\ee
where $r=0$ is the UV boundary and the bulk has $r>0$. {The 5D piece arises from the DBI and TCS terms whereas the boundary term arises from the CS sector only. The DBI action is given in section~\ref{sec:FullDBI}, but the expression will not be needed here. The division of the CS term into 5D and 4D pieces is not well defined per se, but for concreteness we may take the 5D term to be~\eqref{Bcharge1} so that the 4D term is~\eqref{G4Phi}.}

The variation of the on-shell action is then (after partial integration of the 5D term and using the EOM as usual)
\be \label{varSonshell}
 \delta S_\mathrm{on-shell} = -\int d^4x\ \frac{\partial\mathcal{L}^{(5D)}}{\partial\, \partial_r \Phi_M } \ \delta \Phi_M\big|_{r=0} +\int d^4x\ \frac{\partial \mathcal{L}^{(4D)}}{\partial \Phi_M}\ \delta \Phi_M\big|_{r=0} + \cdots
\ee
where the dots stand for boundary terms from the variations of the other fields and we assumed that the boundary Lagrangian $\mathcal{L}^{(4D)}$  is independent of the derivatives of $\Phi$.

Moreover, in order to obtain this expression, the term $\frac{\partial\mathcal{L}^{(5D)}}{\partial\, \partial_r \Phi_M } \ \delta \Phi_M$ must vanish in the IR. Otherwise the variation will receive an inconsistent IR contribution.  Similarly we must assume that contributions from spatial infinity in the partial integration vanish. This will be justified in the next subsection.

We can read off the (five dimensional) baryon current as
\be \label{JBdef}
 J^M_B = \frac{\partial\mathcal{L}^{(5D)}}{\partial\, \partial_r \Phi_M } \Big|_{r=0}  -\frac{\partial \mathcal{L}^{(4D)}}{\partial \Phi_M} \Big|_{r=0} \ .
\ee
We did not comment on the gauge dependence yet. Without loss of generality, we may assume that $\mathcal{L}^{(5D)}$ is gauge invariant and depends on $\Phi$ only through its derivatives, while $\mathcal{L}^{(4D)}$ is not gauge invariant. {Indeed this holds for the choices of the TCS terms specified above, i.e., that the 5D CS term is~\eqref{Bcharge1} while the 4D term is~\eqref{G4Phi} (omitting the irrelevant last line)}. Gauge invariance implies, in particular, that the variation of $S^{(5D)}$ vanishes for any infinitesimal vectorial $U(1)$ transformation depending on the space-time coordinates, i.e., for $\delta \Phi_\mu = \partial_\mu \epsilon$. Inserting this in the formula~\eqref{varSonshell} and integrating partially we observe that by gauge invariance we must have
\be
 \frac{\partial}{\partial x^\mu} \left(\frac{\partial\mathcal{L}^{(5D)}}{\partial\, \partial_r \Phi_\mu } \right)_{r=0} = 0
\ee
where $\mu$ is summed over the four Minkowski coordinates. (Recall that these terms would vanish by the EOMs for homogeneous configurations.) Notice also that
\be
 \frac{\partial\mathcal{L}^{(5D)}}{\partial\, \partial_r \Phi_r } = 0
\ee
i.e. the $5D$ action is independent of $\partial_r \Phi_r$ since the derivatives of $\Phi$ only appear either through the field strength or in the TCS term. Notice that we did not fix the gauge so far, but to make contact with field theory we should choose the gauge where $\Phi_r = 0$.

We now discuss the term arising from the 4D action. {Recall that we chose to use~\eqref{Bcharge1}}
for the 5D action so that the full $\Phi$ dependent 4D piece is given in~\eqref{G4Phi}. Because this is a boundary term at $r=0$, there is no dependence on $\Phi_r$ such that
\be
 \frac{\partial \mathcal{L}^{(4D)}}{\partial \Phi_r} = 0 \ .
\ee
The other terms may be written as
\begin{align} \label{JB4D}
 -\frac{\partial \mathcal{L}^{(4D)}}{\partial \Phi_\mu}\, \omega_4 \Big|_{r=0} \!\!&= -\frac{iN_c}{48\pi^2}\, dx^\mu\!\wedge [-4 i \text{Tr}(L\wedge F^{(L)})+\text{Tr}(L\wedge L\wedge L)+ & \nonumber\\
 &\ \ \ +4 i \text{Tr}(R\wedge F^{(R)}) -\text{Tr}(R\wedge R\wedge R)+6 \text{Tr}(DU\wedge F^{(L)}U^\dagger) + & \nonumber\\
 &\ \ \ +6 \text{Tr}(DUU^\dagger\wedge F^{(R)}) -2 i \text{Tr}(DUU^\dagger\wedge DUU^\dagger\wedge DUU^\dagger)]&
\end{align}
where $\omega_4 = dx^0\wedge dx^1\wedge dx^2\wedge dx^3$ is the four dimensional volume form. The first four terms in the square brackets reflect the mixed anomaly. They vanish for vectorial external gauge fields, $L=R$. For zero external gauge fields the last term gives the expected topological baryon current,
\be
\label{NBS} -\frac{N_c}{24 \pi^2} \epsilon^{\mu\nu\rho\sigma}\ \text{Tr}\, \left(\partial_\nu UU^\dagger\wedge \partial_\rho UU^\dagger\wedge \partial_\sigma UU^\dagger\right) \ .
\ee
The divergence of the current~\eqref{JB4D} is given by
\begin{align} \label{JB4Ddiv}
 -\partial_\mu\frac{\partial \mathcal{L}^{(4D)}}{\partial \Phi_\mu}\, \omega_4 \Big|_{r=0} \!&= -\frac{iN_c}{48\pi^2}\, d [-4 i \text{Tr}(L\wedge F^{(L)})\!+\!\text{Tr}(L\wedge L\wedge L)\!+\!4 i \text{Tr}(R\wedge F^{(R)}) + & \nonumber\\
 &\ \ \ -\text{Tr}(R\wedge R\wedge R)\!+\! 6 \text{Tr}(DU\wedge F^{(L)}U^\dagger)\!+\! 6 \text{Tr}(DUU^\dagger\wedge F^{(R)}) + & \nonumber\\
 &\ \ \ -2 i \text{Tr}(DUU^\dagger\wedge DUU^\dagger\wedge DUU^\dagger)] \ . &
\end{align}
Explicit computation of the exterior derivative gives
\begin{align}
 -\partial_\mu\frac{\partial \mathcal{L}^{(4D)}}{\partial \Phi_\mu}\, \omega_4 \Big|_{r=0} \!&= \frac{N_c}{48\pi^2}\big[2 \text{Tr}(F^{(L)}\wedge F^{(L)})-2 \text{Tr}(F^{(R)}\wedge F^{(R)}) + \nonumber \\
 &\quad +i\text{Tr}(L\wedge L\wedge F^{(L)})-i\text{Tr}(R\wedge R\wedge F^{(R)})\big] \ .&
\end{align}
This anomaly contribution vanishes for vectorial gauge fields, $L=R$. In conclusion, since we have shown earlier that the divergence of the first contribution in~\eqref{JBdef} vanishes, we have that
\be
 \partial_\mu J^\mu_B = 0 \ , \qquad J^r_B=0
\ee
in the absence of axial gauge fields at the boundary.

The contribution from the TCS action to the first term in~\eqref{JBdef} can also be computed explicitly (omitting the nonlinear term):\footnote{The minus sign arises due to consistency with the chosen coordinate system. Because we have chosen that $r=0$ is the UV boundary value of the holographic coordinate, taking $\omega_5 =dr\wedge \omega_4$, $\int_M g \omega_5 = - \int d^5x\, g$, and $\int_{\partial M} f \omega_4 =  \int d^4x\, f$ is consistent with $\int_M df = \int_{\partial M} f$ where $\partial M$ is the UV boundary. Notice that if we changed $r \mapsto 1/r$, there would be no need to include a minus sign in any of the definitions.}
\begin{align}
  &\frac{\partial\mathcal{L}^{(5D)}_\mathrm{CS}}{\partial\, \partial_r \Phi_\mu } \Big|_{r=0}\omega_4  =\\\nonumber &-\frac{iN_c}{2\pi^2} dx^\mu\wedge\{(f_1(0)+f_3(0))\mathrm{Tr}[DU\wedge (F^{(L)} U^\dagger + U^\dagger F^{(R)})]+f_2(0)\mathrm{Tr}[(DU\, U^\dagger)^3]\} \ .
\end{align}
After imposing the conditions~\eqref{Baryonconstr}, this contribution cancels the last three terms in~\eqref{JB4D}. This cancellation is however superficial: We can also use the equations of motion to write
\be
 \frac{\partial\mathcal{L}^{(5D)}}{\partial\, \partial_r \Phi_\mu } \Big|_{r=0} = \partial_\nu \int dr \frac{\partial\mathcal{L}^{(5D)}}{\partial\, \partial_\nu \Phi_\mu } \ .
\ee
Then the baryon number is, taking the configuration to be independent of time,
\be \label{NBdef}
 N_B = \int d^3 \mathbf{x} J^0_B = \int dr d^3\mathbf{x}\ \partial_k \frac{\partial\mathcal{L}^{(5D)}}{\partial\, \partial_k \Phi_0 } - \int d^3\mathbf{x}\ \frac{\partial \mathcal{L}^{(4D)}}{\partial \Phi_0} \Big|_{r=0}
\ee
where the first term, arising from the 5D bulk action, integrates into a boundary term at spatial infinity, so that the UV boundary contribution only arises from the second term in~\eqref{NBdef} and the purported cancellation is absent. Assuming that the contributions from spatial infinity vanish, the result
 may be written as the following 3D boundary integral:
\begin{align} \label{NBresult}
N_B \!&= -\frac{iN_c}{48\pi^2}\, \int [-4 i \text{Tr}(L\wedge F^{(L)})+\text{Tr}(L\wedge L\wedge L)+4 i \text{Tr}(R\wedge F^{(R)})+ & \nonumber\\
 &\ \ \ -\text{Tr}(R\wedge R\wedge R)+6 \text{Tr}(DU\wedge F^{(L)}U^\dagger)+6 \text{Tr}(DUU^\dagger\wedge F^{(R)}) + & \nonumber\\
 &\ \ \ -2 i \text{Tr}(DUU^\dagger\wedge DUU^\dagger\wedge DUU^\dagger)] \ . &
\end{align}

Notice that, in the end, the derivation of the baryon number did not require using the conditions~\eqref{Baryonconstr}. {Even more remarkably, $N_B$ does not depend in any manner on the non-closed part of the TCS action $\Omega_5^0$ and the corresponding TCS potentials $f_i(\tau)$. } {Here, the generation of the baryon number and the contribution of the TCS terms to the equations of motion (responsible for the stabilization of the baryon size) are ensured by two distinct parts of the CS action (closed and non-closed).
Although slightly counter-intuitive, it is not a contradiction though. The reason is that the result for the baryon number simply tells us what should be the boundary behavior of the tachyon field for $N_B$ to be non-zero (it should have a Skyrmion winding from} \eqref{NBS}).{It does not guarantee that a solution with such boundary conditions exists though. In particular, it is expected that no finite size solution should exist when $\Omega_5^0$ vanishes ($f_i(\tau)=0$). }

\begin{figure}[h]
\begin{center}
\includegraphics[width=0.5\textwidth]{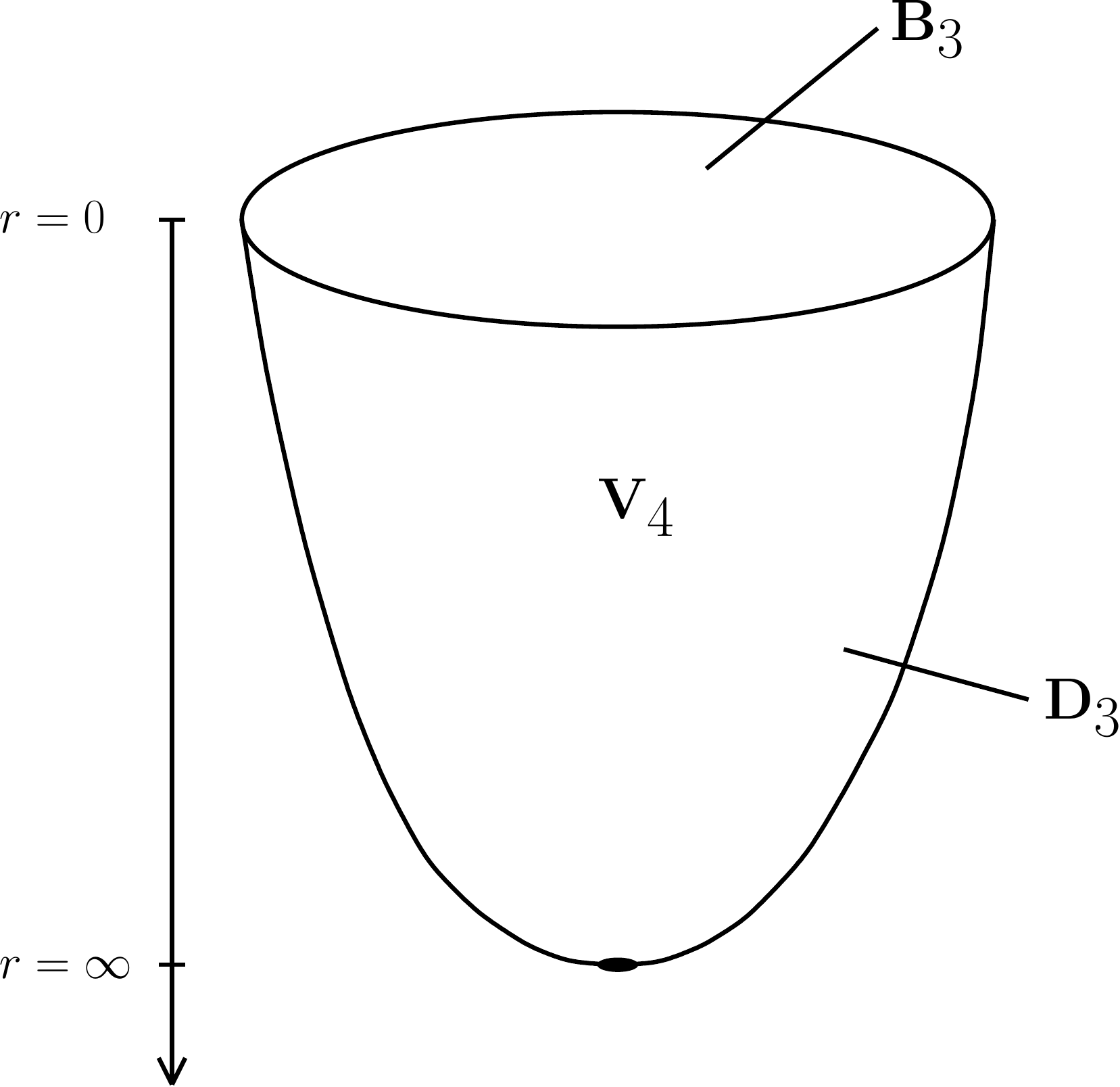}
\caption{Volume of integration $\mathbf{V}_4$ for the baryon and its three-dimensional boundaries $\mathbf{B}_3+\mathbf{D}_3$.}
\label{fig:plot_bdries}
\end{center}
\end{figure}

\subsection{Boundary terms to the baryon charge}

We finish by discussing the boundary terms at spatial infinity. That is, we consider the baryon number in a finite 3D volume. We take this volume to be the three ball $\mathbf{B}_3$, but the analysis is the same for any shape of the volume. By using the equations of motion as above, the charge may be written as
\be
 N_B = \int_{\mathbf{B}_3} d^3 \mathbf{x} J^0_B = \int dr \int_{\partial {\mathbf{B}_3}}dS \ n_k\, \frac{\partial\mathcal{L}^{(5D)}}{\partial\, \partial_k \Phi_0 } - \int_{\mathbf{B}_3} d^3\mathbf{x}\ \frac{\partial \mathcal{L}^{(4D)}}{\partial \Phi_0} 
\ee
where the unit vector $n$ is normal to the surface element $dS$ of $\partial {\mathbf{B}_3}$. We can also write down the TCS contributions explicitly:
\begin{align}
 N_B &= \int dr \int_{\partial {\mathbf{B}_3}}dS \ n_k\, \frac{\partial\mathcal{L}^{(5D)}_\mathrm{DBI}}{\partial\, \partial_k \Phi_0 }+ &\nonumber\\\nonumber
 &\ +  \frac{iN_c}{2\pi^2}\int_{\mathbf{D}_3}\!\! \big[(f_1(\tau)\!+\!f_3(\tau))\mathrm{Tr}[DU\wedge (F^{(L)} U^\dagger + U^\dagger F^{(R)})] +& \nn \\
 &\ \hphantom{= \frac{iN_c}{2\pi^2}\int_{\mathbf{D}_3}\!\! \big[} +f_2(\tau)\mathrm{Tr}[(DU\, U^\dagger)^3]\big] + & \nonumber\\
 &\ +\frac{iN_c}{\pi^2}\int_{\mathbf{D}_3}\!\! (f_1(\tau)\!+\!f_3(\tau)) d\Phi\wedge\mathrm{Tr}[DU\, U^\dagger] + & \nonumber\\
 &\ -\frac{iN_c}{48\pi^2}\, \int_{\mathbf{B}_3} \big[-4 i \text{Tr}(L\wedge F^{(L)})+\text{Tr}(L\wedge L\wedge L)+4 i \text{Tr}(R\wedge F^{(R)}) + & \nonumber\\
 &\ \ \ \ -\text{Tr}(R\wedge R\wedge R)+6 \text{Tr}(DU\wedge F^{(L)}U^\dagger)+6 \text{Tr}(DUU^\dagger\wedge F^{(R)}) + & \nonumber\\
\label{NBxinf} &\ \ \ \ -2 i \text{Tr}(DUU^\dagger\wedge DUU^\dagger\wedge DUU^\dagger)\big] 
\end{align}
where $\mathbf{D}_3$ is the boundary at spatial infinity. The boundary of the whole four-dimensional (holographic and spatial dimensions) volume therefore breaks into two terms, $\partial \mathbf{V}_4 = \mathbf{B}_3+\mathbf{D}_3$ where $\mathbf{B}_3$ lies at the holographic boundary $r=0$ and $\mathbf{D}_3$ lies at spatial infinity, see figure~\ref{fig:plot_bdries}. Notice that there is no boundary at $r = \infty$ as the geometry shrinks into a single point in the IR, where it ends in a singularity (represented by a blob in the figure). It is important to check that the boundary contributions of the first three lines vanish for soliton configurations at large volumes, i.e., when $\partial \mathbf{B}_3$ is far from the soliton.

If the conditions~\eqref{Baryonconstr} hold, the most important contributions (i.e., those that are expected to contribute to the charge of the soliton solution in the limit of large volume) can be written as a four dimensional bulk integral:
\begin{align}
 N_B =\ &   \frac{iN_c}{2\pi^2}\!\int_{\mathbf{V}_4}\!\!\! \mathrm{d}\Big\{ (f_1(\tau)\!+\!f_3(\tau))\mathrm{Tr}[DU\wedge (F^{(L)} U^\dagger + U^\dagger F^{(R)})] +& \nn \\
 & \hphantom{ \frac{iN_c}{2\pi^2}\!\int_{\mathbf{V}_4}\!\!\! \mathrm{d}\Big\{ } + f_2(\tau)\mathrm{Tr}[(DU\, U^\dagger)^3]\Big\} +& \nonumber\\
 &+ \int dr \int_{\partial \mathbf{B}_3}dS \ n_k\, \frac{\partial\mathcal{L}^{(5D)}_\mathrm{DBI}}{\partial\, \partial_k \Phi_0 } +\frac{iN_c}{\pi^2}\int_{\mathbf{D}_3} (f_1(\tau)\!+\!f_3(\tau)) d\Phi\wedge\mathrm{Tr}[DU\, U^\dagger] + & \nonumber\\
 & -\frac{iN_c}{48\pi^2}\, \int_{\mathbf{B}_3} \big[-4 i \text{Tr}(L\wedge F^{(L)})+\text{Tr}(L\wedge L\wedge L)+4 i \text{Tr}(R\wedge F^{(R)}) + & \nonumber\\
 &\ \ \ -\text{Tr}(R\wedge R\wedge R)\big]_{r=0} \ .
\end{align}
That is, while the conditions~\eqref{Baryonconstr} were not necessary to derive the main formula~\eqref{NBresult}, they are necessary to write down the charge as a compact bulk integral for finite volumes.

{We now show the vanishing of the contribution to} \eqref{NBxinf} at spatial infinity for the baryon solution. This is most conveniently done by using the gauge fields redefined to absorb the tachyon phase as in \eqref{LtoLt}. The 3-forms that are integrated on $\mathbf{D}_3$ can then be written as
\be
\label{D31} \text{Tr}\left[\td{A}\wedge \left(\td{F}^{(L)}+\td{F}^{(R)}\right)\right] \, ,
\ee
\be
\label{D32} \text{Tr}\left[\td{A}^3\right] \, ,
\ee
\be
\label{D33} \mathrm{d}\Phi\wedge\text{Tr}\left[\td{A}\right] = 0 \, ,
\ee
where $A$ is defined in \eqref{defVA}. On the other hand, the DBI action\footnote{In the probe limit. The result extends straightforwardly to the inhomogeneous tachyon case.} can be written in terms of the redefined gauge fields as
\begin{align}
\label{expSprobeT} S_{\text{DBI}} = -M^3N_c &\int\mathrm{d}^5x\, V_f(\l,\tau^2) \sqrt{-\mathrm{det}\, \tilde{g}} \,\times \\
\nn &\times \left(\left[\frac{1}{2}\! +\! \frac{1}{4}\ka \tau^2 \left(\tilde{g}^{-1}\right)^{MN}\text{Tr}\, \td{\mathbf{A}}_M\td{\mathbf{A}}_N  - \right.\right. \\
\nn  & \left. -\frac{1}{8}w^2\left(\tilde{g}^{-1}\right)^{MN}\left(\tilde{g}^{-1}\right)^{PQ}\text{Tr}\, \td{\mathbf{F}}^{(L)}_{NP}\td{\mathbf{F}}^{(L)}_{QM} + \mathcal{O}\left((\mathbf{F}^{(L)})^3\right)\bigg] + (L\leftrightarrow R)\right) ,
\end{align}
from which it is clear that both $\td{A}$ and $\td{F}^{(L/R)}$ should vanish at spatial infinity for a finite energy cylindrically symmetric baryon solution. This implies that there is no contribution from spatial infinity to the baryon number \eqref{NBxinf}.

\subsubsection{Details on the relation to the bulk instanton number}

\label{Sec:NB_Ni}

We derive in this subsection the condition for the baryon number \eqref{NB} to equal the bulk instanton number
\be
\label{Ninst} N_{\text{instanton}} = \frac{1}{8\pi^2}\int_{\text{bulk}} \text{Tr}\, \left(\mathbf{F}^{(L)} \wedge \mathbf{F}^{(L)} - \mathbf{F}^{(R)} \wedge \mathbf{F}^{(R)} \right) \, ,
\ee
or in terms of the ansatz fields  \eqref{sumans}
\be
\label{Ni_ans} N_{\text{instanton}} = \frac{1}{2\pi} \int\mathrm{dr} \mathrm{d}\xi\, \e^{\bar{\mu}\bar{\nu}}\left(F_{\bar{\mu}\bar{\nu}} + \partial_{\bar{\mu}}\left( -i\phi^*D_{\bar{\nu}}\phi + h.c. \right)\right) \, .
\ee
We first notice that  \eqref{NB} can be rewritten in terms of the redefined gauged fields of \eqref{LtoLt} at the boundary as
\be
\label{NBLt} N_B = \left. \frac{1}{24\pi^2}\int \left[-i\text{Tr}(\tilde{\mathbf{L}}^3 - \tilde{\mathbf{R}}^3 + 3\tilde{\mathbf{L}}\wedge \tilde{\mathbf{R}}^2 - 3 \tilde{\mathbf{L}}^2\wedge \tilde{\mathbf{R}})\right]\,\right|_{UV} \,\, .
\ee
Because the redefined gauge fields at the boundary take a pure gauge form, $\tilde{\mathbf{F}}^{(L/R)}=0$, which implies that $\tilde{\mathbf{L}}^2 = -i \mathrm{d}\,\tilde{\mathbf{L}}$ and likewise for $\tilde{\mathbf{R}}$.  \eqref{NBLt} can then be rewritten as
\begin{align}
\nn N_B &= \left. \frac{1}{24\pi^2}\int \left[-i\text{Tr}(\tilde{\mathbf{L}}^3 - \tilde{\mathbf{R}}^3 - 3\, \mathrm{d}(\tilde{\mathbf{L}}\wedge \tilde{\mathbf{R}}) )\right]\,\right|_{UV} \\
\nn &=  \left. \frac{1}{24\pi^2}\int \left[-i\text{Tr}(\tilde{\mathbf{L}}^3 - \tilde{\mathbf{R}}^3 )\right]\,\right|_{UV} \\
\label{NBLt2} &=  \left. -\frac{1}{8\pi^2}\int \left[\text{Tr}\left(\tilde{\mathbf{F}}^{(L)}\wedge\tilde{\mathbf{L}}+\frac{1}{3}i\, \tilde{\mathbf{L}}^3 - \left(\tilde{\mathbf{F}}^{(R)}\wedge\tilde{\mathbf{R}} + \frac{1}{3}i\, \tilde{\mathbf{R}}^3\right) \right)\right]\,\right|_{UV} \,\, .
\end{align}
On the other hand $\text{Tr}(\tilde{\mathbf{F}}^{(L)}\wedge \tilde{\mathbf{F}}^{(L)}) = \mathrm{d}\left[\text{Tr}\left(\tilde{\mathbf{F}}^{(L)}\wedge\tilde{\mathbf{L}}+\frac{1}{3}i\, \tilde{\mathbf{L}}^3\right)\right] \equiv \mathrm{d}\, \omega_3(\tilde{\mathbf{L}})$, and likewise for R, where $\omega_3$ is the Chern-Simons 3-form. So, using Stoke's theorem, we find that
\be
\label{NBLt3} N_B =  N_{\text{instanton}} - \left. \frac{1}{8\pi^2}\int \left[ \omega_3(\tilde{\mathbf{L}}) - \omega_3(\tilde{\mathbf{R}})\right]\,\right|_{IR} \,\, ,
\ee
where the contribution from spatial boundaries was dropped because of the boundary conditions Table \ref{tab:bcs}. Therefore, the condition for the baryon number to equal the instanton number is that
\begin{equation}
\label{cNBeNi} \left. \int \left[ \omega_3(\tilde{\mathbf{L}}) - \omega_3(\tilde{\mathbf{R}})\right]\,\right|_{IR} = 0 \,  .
\end{equation}
In terms of the baryon ansatz of equations \eqref{ansatzSU2i}-\eqref{ansatzU1} and \eqref{TU2} the condition reads
\begin{equation}
\label{cNBeNians} \left. \int \mathrm{d}\xi\, \left(\, \tilde{A}_{\xi}(|\tilde{\phi}|^2-1) + \partial_{\xi} \tilde{\phi}_1  + \partial_{\xi} \tilde{\phi}_1  \tilde{\phi}_2 - \partial_{\xi} \tilde{\phi}_2  \tilde{\phi}_1 \,  \right)  \,\right|_{IR}  = 0 \, .
\end{equation}

\subsection{Gauged WZ term} \label{app:gaugedWZ}

The gauged WZ term was first written down fully by Witten in \cite{WWZ} (equation (24)) where he also derived the chiral anomalies (equation (25)). It was then noted that this expression has some typos and that it is defined up to a gauge invariant term by \cite{Kaymakcalan} and \cite{Manes}. First note that the authors all have different conventions for the gauge fields:
\be
\label{convA} A_{L/R}^{\text{Witten}} = - A_{L/R}^{\text{Kaymakcalan}} = -i (L/R)^{\text{Ma\~nes}} \, .
\ee
Upon these conventions, the expressions derived in \cite{Kaymakcalan} (equation (4.18)) and \cite{Manes} (equation (69)) are exactly the same and differ from Witten's expression by the following:
\begin{itemize}
\item As mentioned above, Witten's expression should be corrected, by adding $i A_L^3$ to $A_L\mathrm{d}A_L + \mathrm{d}A_L A_L$ in the second line, interchanging $R \leftrightarrow L$ in the second term of the third line and replacing the $\frac{1}{2}$ in front of $A_L^2U A_R^2 U^{-1}$ by 1 in the last line.
\item They add to Witten's $\tilde{\Gamma}$
\be
\label{ad} \frac{i}{48\pi^2} \mathrm{Tr} \int (\mathrm{d}(A_LU\mathrm{d}A_RU^{-1}) - \mathbf{F}^{(L)}U\mathbf{F}^{(R)}U^{-1}) \, ,
\ee
where the second term is such that the gauge-invariant form $\mathbf{F}^{(L)}U\mathbf{F}^{(R)}U^{-1}$ does not appear in the final expression and the first one yields a 3-dimensional boundary term that vanishes .
\end{itemize}
We write here their expression using the convention of \cite{Kaymakcalan}
\begin{align}
\label{gaugedWZ}\nonumber S_{\text{WZ}}(U,A_L,A_R) &= C\int \mathrm{Tr}\Big[\b^5\Big] + 5Ci \int \mathrm{Tr}\Big[A_L\a^3 + A_R\b^3\Big] -\\
\nonumber &\hphantom{=} - 5C \int \mathrm{Tr}\Big[(\mathrm{d}A_L A_L + A_L \mathrm{d}A_L)\a + (\mathrm{d}A_R A_R + A_R \mathrm{d}A_R)\b\Big] +\\
\nonumber &\hphantom{=} + 5C \int \mathrm{Tr}\Big[\mathrm{d}A_L\a UA_RU^{-1} + \mathrm{d}A_R\b U^{-1}A_LU\Big] +\\
\nonumber &\hphantom{=} + 5C \int \mathrm{Tr}\Big[A_RU^{-1}A_LU\b^2 - A_LUA_RU^{-1}\a^2\Big] + \\
\nn &\hphantom{=} + \frac{5C}{2} \int \mathrm{Tr}\Big[(A_L\a)^2-(A_R\b)^2\Big] + 5Ci \int \mathrm{Tr}\Big[A_L^3\a + A_R^3\b\Big] +\\
\nonumber &\hphantom{=} + 5Ci \int \mathrm{Tr}\Big[(\mathrm{d}A_R A_R + A_R \mathrm{d}A_R)U^{-1}A_LU - \\
\nn &\hphantom{= +  5Ci \int\mathrm{Tr}\Big[\,\,} -(\mathrm{d}A_L A_L + A_L \mathrm{d}A_L)UA_RU^{-1}\Big]+ \\
\nonumber &\hphantom{=} + 5Ci \int \mathrm{Tr}\Big[A_LUA_RU^{-1}A_L\a + A_RU^{-1}A_LUA_R\b\Big] +\\
&\hphantom{=} + 5C \int \mathrm{Tr}\Big[A_R^3U^{-1}A_LU - A_L^3UA_RU^{-1} + \frac{1}{2}(UA_RU^{-1}A_L)^2\Big] \, ,
\end{align}
where $C = \frac{-iN_c}{240\pi^2}$, $\a = \mathrm{d}UU^{-1}$ and $\b = U^{-1}\mathrm{d}U$. Note that all terms are $\mathrm{P_1}$-odd. They are also C even, apart from the term in the third line which is the sum of a C-even non-closed 4-form and a C-odd exact 4-form.

\paragraph*{Gauge transformations} It can be checked that  \eqref{gaugedWZ} reproduces the correct chiral anomalies
\be
\label{Xanom} \d S_{\text{WZ}}(U,A_L,A_R) = -10 C  \int \mathrm{Tr}\left[\L_L\left((\mathrm{d}A_L)^2 - \frac{i}{2}\mathrm{d}(A_L^3)\right) - (L\leftrightarrow R) \right] \, ,
\ee
upon a general infinitesimal gauge transformation, where the various fields transform as
\be
\label{gtU} \d U = \L_L U - U\L_R \, ,
\ee
\be
\label{gtA} \d A_{L/R} = -i \mathrm{d}\L_{L/R} + [\L_{L/R},A_{L/R}]  \, .
\ee

\section{The TCS contribution to the equations of motion}

\label{Sec:CS_EoM}

The TCS action reads
\be
\label{SCS} S_{\text{CS}} = \frac{iN_c}{4\pi^2} \int \Omega_5 \, ,
\ee
where $\Omega_5$ is the TCS 5-form, which is decomposed as in  \eqref{Omega1}
\be\label{Omega}
\Omega_5 = \Omega_5^0 +\Omega_5^c + dG_4 \, .
\ee
The expressions for each term are given in Appendix \ref{GCS}. Upon a small variation of the fields, $\Omega_5^c$ and $\mathrm{d}G_4$ yield boundary terms which do not contribute to the equations of motion. The TCS contribution to the equations of motion is therefore given by varying the first term $\Omega_5^0$  \eqref{O50}.

\paragraph*{Gauge field EoM} Upon a little variation of the gauge field $\mathbf{L}\to \mathbf{L} + \d \mathbf{L}$, $\Omega_5^0$ changes as
\begin{align}
\nonumber \d_{\mathbf{L}}\Omega_5^0 &= \mathrm{Tr}\bigg\{ \delta\mathbf{L}\wedge \bigg(\! -if_1(\tau)\Big(\mathbf{F}^{(L)}\wedge \mathbf{F}^{(L)} + U^\dagger\mathbf{F}^{(R)}\wedge \mathbf{F}^{(R)} U - \mathbf{L}\wedge \mathbf{F}^{(L)}\wedge U^\dagger DU -  \\
\nonumber &\hphantom{= \mathrm{Tr}\bigg\{ \delta\mathbf{L}\wedge\Big( -if_1(\tau)}  \, - \mathbf{F}^{(L)}\wedge U^\dagger DU\wedge \mathbf{L} - \mathbf{L}\wedge U^\dagger DU\wedge \mathbf{F}^{(L)} -  \\
\nn &\hphantom{= \mathrm{Tr}\bigg\{ \delta\mathbf{L}\wedge\Big( -if_1(\tau)} \, - U^\dagger DU\wedge \mathbf{F}^{(L)}\wedge \mathbf{L} \Big) - \\
\nonumber &\hphantom{= \mathrm{Tr}\bigg\{ \delta\mathbf{L}\wedge\Big(}  - \mathrm{d}\Big[f_1(\tau)(U^\dagger DU\wedge \mathbf{F}^{(L)} + \mathbf{F}^{(L)}\wedge U^\dagger DU ) \Big] - \\
\nonumber &\hphantom{= \mathrm{Tr}\bigg\{ \delta\mathbf{L}\wedge\Big(}  -if_2(\tau) \Big(\mathbf{F}^{(L)}\wedge U^\dagger DU\wedge U^\dagger DU + U^\dagger DU\wedge U^\dagger DU\wedge \mathbf{F}^{(L)} +\\
\nn &\hphantom{= \mathrm{Tr}\bigg\{ \delta\mathbf{L}\wedge\Big( -if_2(\tau) \Big(} + U^\dagger DU\wedge \mathbf{F}^{(L)}\wedge U^\dagger DU -  \\
\nn &\hphantom{= \mathrm{Tr}\bigg\{ \delta\mathbf{L}\wedge\Big( -if_2(\tau) \Big(} - \mathbf{L}\wedge U^\dagger DU \wedge U^\dagger DU \wedge U^\dagger DU - \\
\nn &\hphantom{= \mathrm{Tr}\bigg\{ \delta\mathbf{L}\wedge\Big( -if_2(\tau) \Big(} - U^\dagger DU \wedge U^\dagger DU \wedge U^\dagger DU \wedge \mathbf{L} + \\
\nonumber &\hphantom{= \mathrm{Tr}\bigg\{ \delta\mathbf{L}\wedge\Big( -if_2(\tau) \Big(} +\mathbf{F}^{(R)}\wedge DUU^\dagger\wedge DUU^\dagger + DUU^\dagger\wedge DUU^\dagger\wedge \mathbf{F}^{(R)} + \\
\nn &\hphantom{= \mathrm{Tr}\bigg\{ \delta\mathbf{L}\wedge\Big( -if_2(\tau) \Big(} + DUU^\dagger\wedge \mathbf{F}^{(R)}\wedge DUU^\dagger \Big)- \\
\nonumber &\hphantom{= \mathrm{Tr}\bigg\{ \delta\mathbf{L}\wedge\Big(}  - \mathrm{d}\Big[f_2(\tau)\, U^\dagger DU\wedge U^\dagger DU\wedge U^\dagger DU\Big]- \\
\nonumber &\hphantom{= \mathrm{Tr}\bigg\{ \delta\mathbf{L}\wedge\Big(} -if_3(\tau)\Big(\mathbf{F}^{(L)}\wedge U^\dagger  \mathbf{F}^{(R)} U \!+\! U^\dagger\mathbf{F}^{(R)}U\wedge \mathbf{F}^{(L)} \!-\! \mathbf{L}\wedge U^\dagger\mathbf{F}^{(R)}\wedge DU - \\
\nn &\hphantom{= \mathrm{Tr}\bigg\{ \delta\mathbf{L}\wedge\Big( -if_3(\tau) \Big(} - U^\dagger\mathbf{F}^{(R)}\wedge DU\wedge \mathbf{L} - \mathbf{L}\wedge U^\dagger DU\wedge U^\dagger\mathbf{F}^{(R)}U - \\
\nonumber &\hphantom{= \mathrm{Tr}\bigg\{ \delta\mathbf{L}\wedge\Big( -if_3(\tau) \Big(} - U^\dagger DU\wedge U^\dagger\mathbf{F}^{(R)}U\wedge \mathbf{L}\Big)- \\
\nonumber &\hphantom{= \mathrm{Tr}\bigg\{ \delta\mathbf{L}\wedge\Big(}  - \mathrm{d}\Big[f_3(\tau)(U^\dagger DU\wedge U^\dagger\mathbf{F}^{(R)}U + U^\dagger\mathbf{F}^{(R)}\wedge  DU) \Big] - \\
\label{dO50L}  &\hphantom{= \mathrm{Tr}\bigg\{ \delta\mathbf{L}\wedge\Big(} -4if_4(\tau) U^\dagger DU\wedge U^\dagger DU\wedge U^\dagger DU\wedge U^\dagger DU \,\bigg) \bigg\} \, ,
\end{align}
and $\delta_{\mathbf{R}}\Omega_5^0$ is obtained by exchanging $L\leftrightarrow R$, $U\leftrightarrow U^\dagger$ and multiplying by $-1$. From the definition of the CS currents \eqref{defJh} and \eqref{defJna}, \eqref{dO50L} can be written as
\be
\label{dO5LJCS} \d_{\mathbf{L}}\Omega_5^0 = 4\pi^2 iM^3\, \mathrm{Tr}\left\{2\d L\wedge J_{CS}^{(L)} + \frac{1}{N_f}\d \hat{L}\wedge \hat{J}_{CS}^{(L)}\,\mathbb{I}_{N_f}\right\} \, ,
\ee
where
\begin{align}
\nn \hat{J}_{CS}^{(L)} &= \frac{1}{4!}\e_{MNPQR}\hat{J}_{CS}^{(L),M} \mathrm{d}x^N\wedge\mathrm{d}x^P\wedge\mathrm{d}x^Q\wedge\mathrm{d}x^R \, ,  \\
\label{expJCSL} J_{CS}^{(L)\, a} &= \frac{1}{4!}\e_{MNPQR}\hat{J}_{CS}^{(L)\,a,\, M} \mathrm{d}x^N\wedge\mathrm{d}x^P\wedge\mathrm{d}x^Q\wedge\mathrm{d}x^R \, ,
\end{align}
with the quantities $\hat{J}_{CS}^{(L),M}$ and $J_{CS}^{(L)\,a,\, M}$ appearing respectively in  \eqref{abEoM} and \eqref{nabEoM}.

\newlength{\wda}
\settowidth{\wda}{$= \mathrm{Tr}\bigg\{ \delta U \bigg($}

\newlength{\wdb}
\settowidth{\wdb}{$= \mathrm{Tr}\bigg\{ \delta U \bigg(\! -f_1(\tau)\Big($}

\newlength{\wdc}
\settowidth{\wdc}{$\hspace{\wda} + d\Big[f_2(\tau) \Big($}

\paragraph*{U EoM} Upon a little variation of $U$, $U\to U + \d U$, $\Omega_5^0$ changes as
\begin{align}
\nonumber \d_{U}\Omega_5^0 &= \mathrm{Tr}\bigg\{ \delta U \bigg(\! -f_1(\tau)\Big(i\mathbf{L}\wedge\mathbf{F}^{(L)}\wedge \mathbf{F}^{(L)}U^\dagger - i \mathbf{F}^{(L)}\wedge \mathbf{F}^{(L)}U^\dagger\wedge \mathbf{R} - \\
\nonumber & \hspace{\wdb} - U^\dagger DU\wedge\mathbf{F}^{(L)}\wedge \mathbf{F}^{(L)}U^\dagger + i\mathbf{L}\wedge U^\dagger\mathbf{F}^{(R)}\wedge \mathbf{F}^{(R)} - \\
\nn &\hspace{\wdb} - i U^\dagger\mathbf{F}^{(R)}\wedge \mathbf{F}^{(R)}\wedge \mathbf{R} - U^\dagger \mathbf{F}^{(R)}\wedge \mathbf{F}^{(R)}\wedge DUU^\dagger\Big) + \\
\nonumber &\hspace{\wda}  + \mathrm{d}\Big[f_1(\tau)(\mathbf{F}^{(L)}\wedge\mathbf{F}^{(L)}U^\dagger  + U^\dagger\mathbf{F}^{(R)}\wedge \mathbf{F}^{(R)} ) \Big] - \\
\nonumber &\hspace{\wda} -f_2(\tau)\Big(i\mathbf{L}\wedge\mathbf{F}^{(L)}\wedge U^\dagger DU\wedge U^\dagger DUU^\dagger - \\
\nn &\hspace{\wdb} -i \mathbf{F}^{(L)}\wedge U^\dagger DU\wedge U^\dagger DUU^\dagger\wedge \mathbf{R} - \\
\nonumber &\hspace{\wdb} - U^\dagger DU\wedge \mathbf{F}^{(L)}\wedge U^\dagger DU \wedge U^\dagger DUU^\dagger + \\
\nonumber &\hspace{\wdb} +i\mathbf{L}\wedge U^\dagger DU\wedge\mathbf{F}^{(L)}\wedge  U^\dagger DUU^\dagger -\\
\nn &\hspace{\wdb}-i U^\dagger DU\wedge\mathbf{F}^{(L)}\wedge U^\dagger DUU^\dagger\wedge \mathbf{R} - \\
\nonumber &\hspace{\wdb} - U^\dagger DU\wedge U^\dagger DU\wedge \mathbf{F}^{(L)}\wedge U^\dagger DUU^\dagger + \\
\nonumber &\hspace{\wdb} + i\mathbf{L}\wedge U^\dagger DU\wedge U^\dagger DU\wedge\mathbf{F}^{(L)}U^\dagger -\\
\nn &\hspace{\wdb}-i U^\dagger DU\wedge U^\dagger DU\wedge\mathbf{F}^{(L)}U^\dagger\wedge \mathbf{R} - \\
\nn &\hspace{\wdb} - U^\dagger DU\wedge U^\dagger DU \wedge U^\dagger DU\wedge \mathbf{F}^{(L)} U^\dagger + \\
\nn &\hspace{\wdb} +i\mathbf{L}\wedge U^\dagger\mathbf{F}^{(R)}\wedge  DUU^\dagger\wedge DUU^\dagger - \\
\nn &\hspace{\wdb} -i U^\dagger\mathbf{F}^{(R)}\wedge DUU^\dagger\wedge DUU^\dagger\wedge \mathbf{R} - \\
\nonumber &\hspace{\wdb} - U^\dagger DUU^\dagger\wedge \mathbf{F}^{(R)}\wedge DUU^\dagger \wedge DUU^\dagger + \\
\nonumber &\hspace{\wdb} +i\mathbf{L}\wedge U^\dagger DUU^\dagger\wedge\mathbf{F}^{(R)}\wedge  DUU^\dagger -\\
\nn &\hspace{\wdb} -i U^\dagger DUU^\dagger\wedge\mathbf{F}^{(R)}\wedge DUU^\dagger\wedge \mathbf{R} - \\
\nn &\hspace{\wdb} - U^\dagger DUU^\dagger\wedge DUU^\dagger\wedge \mathbf{F}^{(R)}\wedge DUU^\dagger + \\
\nn &\hspace{\wdb} + i\mathbf{L}\wedge U^\dagger DUU^\dagger\wedge DUU^\dagger\wedge\mathbf{F}^{(R)}-\\
\nn &\hspace{\wdb} -i U^\dagger DUU^\dagger\wedge DUU^\dagger\wedge\mathbf{F}^{(R)}\wedge \mathbf{R} - \\
\nn &\hspace{\wdb} - U^\dagger DUU^\dagger\wedge DUU^\dagger \wedge DUU^\dagger\wedge \mathbf{F}^{(R)} \Big) + \\
\nn &\hspace{\wda}  + \mathrm{d}\Big[f_2(\tau) \Big(\mathbf{F}^{(L)}\wedge U^\dagger DU\wedge U^\dagger DUU^\dagger + U^\dagger DU\wedge\mathbf{F}^{(L)}\wedge  U^\dagger DUU^\dagger +\\
\nn &\hspace{\wdc} +\! U^\dagger DU\wedge U^\dagger DU\wedge\mathbf{F}^{(L)}U^\dagger \!+\! U^\dagger\mathbf{F}^{(R)}\wedge DUU^\dagger\wedge DUU^\dagger +\\
\nn &\hspace{\wdc} +\! U^\dagger DUU^\dagger\wedge\mathbf{F}^{(R)}\wedge DUU^\dagger +\\
\nn &\hspace{\wdc} +\! U^\dagger DUU^\dagger\wedge DUU^\dagger\wedge \mathbf{F}^{(R)} \Big)\Big]- \\
\nonumber &\hspace{\wda} -f_3(\tau)\Big(i\mathbf{L}\wedge\mathbf{F}^{(L)}\wedge U^\dagger\mathbf{F}^{(R)} - i \mathbf{F}^{(L)}\wedge U^\dagger\mathbf{F}^{(R)}\wedge \mathbf{R} -\\
\nn &\hspace{\wdb} - U^\dagger \mathbf{F}^{(R)} DU\wedge\mathbf{F}^{(L)} U^\dagger + i\mathbf{L}\wedge U^\dagger\mathbf{F}^{(R)}\wedge U\mathbf{F}^{(L)}U^\dagger - \\
\nn &\hspace{\wdb} - i U^\dagger\mathbf{F}^{(R)}\wedge U\mathbf{F}^{(L)}U^\dagger\wedge \mathbf{R} - U^\dagger \mathbf{F}^{(R)}\wedge U\mathbf{F}^{(L)}U^\dagger\wedge DUU^\dagger + \\
\nn &\hspace{\wdb} + \mathbf{F}^{(L)}\wedge U^\dagger DUU^\dagger\wedge \mathbf{F}^{(R)} - U^\dagger DUU^\dagger\wedge \mathbf{F}^{(R)}\wedge U\mathbf{F}^{(L)}U^\dagger\Big)+ \\
\nn &\hspace{\wda} + \mathrm{d}\Big[f_3(\tau)\Big(\mathbf{F}^{(L)}\wedge U^\dagger\mathbf{F}^{(R)}  + U^\dagger\mathbf{F}^{(R)}U\wedge \mathbf{F}^{(L)}U^\dagger \Big) \Big] - \\
\nn &\hspace{\wda}  -4f_4(\tau)\Big(i\mathbf{L}\wedge U^\dagger DU\wedge U^\dagger DU\wedge U^\dagger DU\wedge U^\dagger DUU^\dagger - \\
\nn &\hspace{\wdb}  - i U^\dagger DU\wedge U^\dagger DU\wedge U^\dagger DU\wedge U^\dagger DUU^\dagger\wedge \mathbf{R} - \\
\nn &\hspace{\wdb}  - U^\dagger DU\wedge U^\dagger DU\wedge U^\dagger DU\wedge U^\dagger DU\wedge U^\dagger DUU^\dagger \Big) +\\
\label{dO50U} &\hspace{\wda} +4\, \mathrm{d}\Big[f_4(\tau)\, U^\dagger DU\wedge U^\dagger DU\wedge U^\dagger DU\wedge U^\dagger DUU^\dagger\Big]  \bigg)  \bigg\} \, .
\end{align}
From  \eqref{defJU},  \eqref{dO50U} can be written as
\be
\label{dO5UJCS} \d_{U}\Omega_5^0 = 4\pi^2 iM^3\, \text{Tr} \left(\d U \mathbf{J}_{CS}^{U} U^\dagger\right) \, ,
\ee
where
\be
\mathbf{J}_{CS}^{U} \equiv J_{CS}^{U}\, \mathrm{d}t\wedge \mathrm{d}r\wedge\mathrm{d}x_1\wedge\mathrm{d}x_2\wedge\mathrm{d}x_3 \, ,
\ee
and where $J_{CS}^{U}$  can be read off from  \eqref{UEoM}.

\section{The ansatz field equations}

\label{App:EoM}

We present in this appendix the derivation of the differential equations obeyed by the fields of the ansatz  \eqref{sumans}. We start with some general results which are useful to derive those equations and then specify to the two different regimes of approximation considered in this paper, the probe instanton and the inhomogeneous tachyon.

We first write down the expression for the inverse of the effective metric $\tilde{g}$  \eqref{ansg}, which is found to take a relatively simple form
\be
\label{invg}-\mathrm{det}\,\tilde{g}\,\, \tilde{g}^{-1}=\left(
\begin{array}{cc|c}
\a_0 & -X_r & -\vec{X}^t \\
X_r & \a_r & \vec{Y}^t \\
\hline
\vec{X} & \vec{Y} & g_S
\end{array}\right)
\, ,
\ee
where we defined 2 scalars
\be
\label{defa0} \a_0 \equiv \ex^{8A} + \ex^{6A}\ka\left((\partial_r\tau)^2+(\partial_{\xi}\tau)^2\right) \, ,
\ee
\be
\label{defar} \a_r \equiv \ex^{4A} \left(\ex^{4A} f - w^2(\partial_{\xi}\Phi)^2 + \ka (\partial_{\xi}\tau)^2\ex^{2A} \right) \, ,
\ee
the 4-vector $X$ with components
\be
\label{defXr} X_r \equiv \ex^{4A}w\left(\ex^{2A}\partial_r\Phi + \ka\partial_{\xi}\tau(\partial_r\Phi\partial_{\xi}\tau - \partial_{\xi}\Phi\partial_r\tau)\right) \, ,
\ee
\be
\label{defXi}X_i \equiv \ex^{4A}\frac{x_i}{\xi}w\left(\ex^{2A}\partial_{\xi}\Phi - \ka \partial_r\tau(\partial_r\Phi\partial_{\xi}\tau - \partial_{\xi}\Phi\partial_r\tau)\right) \, ,
\ee
the 3-vector $\vec{Y}$ with components
\be
\label{defYi} Y_i \equiv -\ex^{4A}  \frac{x_i}{\xi} (-w^2 \partial_r\Phi\partial_{\xi}\Phi + \ex^{2A} \ka \partial_r\tau\partial_{\xi}\tau)) \, ,
\ee
and the symmetric 3-matrix $g_S$ with components
\begin{align}
\label{defgsii} (g_S)_{ii} &= \ex^{8A}  - \ex^{4A} w^2 \left( (\partial_r\Phi)^2 + (\partial_{\xi}\Phi)^2 \left(1-\frac{x_i^2}{\xi^2}\right)\right) -\\
\nonumber &\hphantom{=} - \ex^{2A} \ka w^2 (\partial_{\xi}\tau\partial_r\Phi - \partial_r\tau \partial_{\xi}\Phi)^2 \left(1-\frac{x_i^2}{\xi^2}\right)+ \\
\nn &\hphantom{=} + \ex^{6A} \ka \left((\partial_r\tau)^2  + (\partial_{\xi}\tau)^2 \left(1-\frac{x_i^2}{\xi^2}\right)\right) \, ,
\end{align}
\be
\label{defgsij} (g_S)_{ij} = -\ex^{2A} \frac{x_i x_j}{\xi^2} (-\ex^{2A} w^2 (\partial_{\xi}\Phi)^2 + \ex^{4A} \ka (\partial_{\xi}\Phi)^2 - \ka w^2 (\partial_{\xi}\tau \partial_r\Phi - \partial_r\tau \partial_{\xi}\Phi)^2 )\, .
\ee
Provided these results, the equations of motion  \eqref{abEoM},\eqref{UEoM} and \eqref{nabEoM} are simplified in the following ways:
\begin{itemize}
\item The superscripts $^{(L)}$ and $^{(R)}$ can be removed from $\tilde{g}$.
\item $F_C^{(L/R)C} = (\tilde{g}^{-1})^{[MN]}F^{(L/R)}_{NM} = 0$ because $F_{0M}^{(L/R)}=0$.
\item \be
\label{TrDUUd0} \text{Tr}\, D_NU^\dagger U = 0\, .
\ee
\item In terms of the redefined gauge fields  \eqref{LtoLt}, the following replacements can be performed
\be
\label{repnab} iD_NU^\dagger U \to \sqrt{2}\tilde{\mathbf{A}}_N \, ,
\ee
\begin{align}
\nonumber & \frac{1}{4}\left[D_{(M}\left(V_f(\l,\tau^2)\ka\tau^2\sqrt{-\mathrm{det}\, \tilde{g}^{(L)}}\left((\tilde{g}^{(L)})^{-1}\right)^{(MN)}D_{N)}U^\dagger\right)U - h.c. +(L\leftrightarrow R) \right] \\
\nonumber & \to -i \sqrt{2}\partial_M\left(V_f(\l,\tau^2)\ka\tau^2\sqrt{-\mathrm{det}\, \tilde{g}}\left(\tilde{g}^{-1}\right)^{(MN)}\tilde{\mathbf{A}}_N\right)- \\
\label{repU} &\hphantom{\to} - V_f(\l,\tau^2)\ka\tau^2\sqrt{-\mathrm{det}\, \tilde{g}}\left(\tilde{g}^{-1}\right)^{(MN)}\left[\tilde{\mathbf{V}}_M, \tilde{\mathbf{A}}_N\right] \, ,
\end{align}
where $\mathbf{A}$ and $\mathbf{V}$ are defined in  \eqref{defVA}.

\item Within the ansatz the matrices $S_{MN}$ and $F^{(L/R)}_{MN}$ can be written
\be
\label{F4} F^{(L/R)}_{MN} = \left(
\begin{array}{c|c}
0 & 0 \\
\hline
0 & F^{(L/R)}_{(4)}
\end{array}\right)
\, ,
\ee
\be
\label{S4} S_{MN} = \left(
\begin{array}{c|c}
0 & 0 \\
\hline
0 & S_{(4)}
\end{array}\right)
\, ,
\ee
where $F^{(L/R)}_{(4)}$ is an antisymmetric 4-matrix
\begin{align}
\label{Fij} F_{(4),ij}^{(L/R) \, a} &= \e^{aij} \frac{\left(\phi_1^{(L/R)}\right)^2-2\left(1+\phi_2^{(L/R)}\right)}{\xi^2} + \\
\nonumber &\hphantom{=} +(x_i\e^{ajk}-x_j\e^{aik})x_k\frac{\xi D_\xi\phi_2^{(L/R)} + \left(\phi_1^{(L/R)}\right)^2-2\left(1+\phi_2^{(L/R)}\right)}{\xi^4} + \\
\nonumber &\hphantom{=} +(x_i\d_{ja}-x_j\d_{ia})\frac{D_\xi\phi_1^{(L/R)}}{\xi^2} + x_a\e^{ijk}x_k \frac{\left(1+\phi_2^{(L/R)}\right)^2}{\xi^4} \, ,
\end{align}
\be
\label{Fir} F_{(4),ir}^{(L/R) \, a} = -\e^{aik}x_k \frac{D_r\phi_2^{(L/R)}}{\xi^2} -(\xi^2\d_{ia} - x_ix_a) \frac{D_r\phi_1^{(L/R)}}{\xi^3}  + x_ix_a \frac{F_{\xi r}^{(L/R)}}{\xi^2}  \, ,
\ee
where the covariant derivative $D\phi$ is defined in  \eqref{covDiv} and the field strength $F_{\bar{\mu}\bar{\nu}}$ in  \eqref{defFLR}. We recall that the abelian part of the field strength is included in the effective metric $\td{g}$ and has the components
\be
\label{Fhat} \hat{F}_{(4),r0}^{(L/R)} = \partial_r\Phi(r,\xi) \sp \hat{F}_{i 0}^{(L/R)} = \frac{x_i}{\xi}\partial_{\xi}\Phi(r,\xi) \, .
\ee 
$S_{(4)}$ is a symmetric 4-matrix
\be
\label{S4rr} (S_{(4)})_{rr} = 2(\partial_r\theta + A_r)^2 = 2\tilde{A}_r^2 \, ,
\ee
\be
\label{S4ri} (S_{(4)})_{ri} = 2(A_r+\partial_r\theta)(A_{\xi}+\partial_{\xi}\theta)\frac{x_i}{\xi} = 2 \tilde{A}_r\tilde{A}_{\xi}\frac{x_i}{\xi} \, ,
\ee
\begin{align}
\nonumber (S_{(4)})_{ij} &= 2\left(\frac{(\ex^{i\theta}\phi + \ex^{-i\theta}\phi^*)^2}{4\xi^2}\left(\d_{ij}-\frac{x_ix_j}{\xi^2}\right)+(A_{\xi}+\partial_{\xi}\theta)^2\frac{x_ix_j}{\xi^2}\right)\\
\label{S4ij} &= 2\left(\frac{(\tilde{\phi} + \tilde{\phi}^*)^2}{4\xi^2}\left(\d_{ij}-\frac{x_ix_j}{\xi^2}\right)+\tilde{A}_{\xi}^2\frac{x_ix_j}{\xi^2}\right) \, ,
\end{align}
where we recall that we defined the 2-vector $(A_{\xi},A_r) = (A_1,A_2) $. Then, noting that $\tilde{g}^{-1}$ can be written
\be
\label{gSig} -\mathrm{det}\,\tilde{g}\,\, \tilde{g}^{-1} = \left(
\begin{array}{c|c}
\ex^{8A} + \ex^{6A}\ka\left((\partial_r\tau)^2+(\partial_{\xi}\tau)^2\right) & -X^t \\
\hline
X & \mathcal{S}
\end{array}\right)
\, ,
\ee
with $\mathcal{S}$ the symmetric 4-matrix
\be
\label{defSig}\mathcal{S} \equiv \left(
\begin{array}{c|c}
 \ex^{4A}  \left(\ex^{4A}  - w^2(\partial_{\xi}\Phi)^2 + \ka (\partial_{\xi}\tau)^2\ex^{2A} \right) & \vec{Y}^t \\
\hline
\vec{Y} & g_S
\end{array}\right)
\, ,
\ee
the following expressions can be obtained for the matrices with raised indices
\be
\label{Fup} \left(F^{(L/R)}\right)^{MN} = (-\mathrm{det}\,\tilde{g})^{-2}\, \left(
\begin{array}{c|c}
0 & X^tF_{(4)}^{(L/R)}\Sigma \\
\hline
-\Sigma F_{(4)}^{(L/R)}X & \Sigma F^{(L/R)}_{(4)}\Sigma
\end{array}\right)
\, ,
\ee
\be
\label{Sup} S^{MN} = (-\mathrm{det}\,\tilde{g})^{-2} \left(
\begin{array}{c|c}
-X^t\mathcal{S} X & -X^t S_{(4)}\mathcal{S}\\
\hline
\mathcal{S} S_{(4)} X & \mathcal{S} S_{(4)}\mathcal{S}
\end{array}\right)
\, ,
\ee
\begin{align}
\label{F2up} &\left(F^{(L/R)}\right)^{MS}\left(F^{(L/R)}\right)_S^{\,\,\,\,\,N} = \\
\nn & \qquad\qquad\qquad = (-\mathrm{det}\,\tilde{g})^{-3}\, \left(
\begin{array}{c|c}
X^tF_{(4)}^{(L/R)}\mathcal{S} F_{(4)}^{(L/R)} X & -X^tF_{(4)}^{(L/R)}\mathcal{S}F_{(4)}^{(L/R)}\mathcal{S}\\
\hline
\Sigma F_{(4)}^{(L/R)} \mathcal{S} F_{(4)}^{(L/R)} X & \mathcal{S} F^{(L/R)}_{(4)}\mathcal{S} F_{(4)}^{(L/R)}\mathcal{S}
\end{array}\right)
\, ,
\end{align}
\be
\label{trF2}\left(F^{(L/R)}\right)^{MN}\left(F^{(L/R)}\right)_{NM} = (-\mathrm{det}\,\tilde{g})^{-2} \text{tr}\left(\mathcal{S} F_{(4)}^{(L/R)}\mathcal{S} F_{(4)}^{(L/R)}\right) \, ,
\ee
where the symmetric and antisymmetric parts are easily identified.
\end{itemize}

\subsection{Probe instanton}

We consider first the case of the probe regime. We recall the expressions of the non-abelian equations of motion in  \eqref{UEoM} and \eqref{nabEoM}
\be
\label{UEoMprobe} \frac{1}{2}\left[D_{(M}\left(V_f(\l,\tau^2)\ka\tau^2\sqrt{-\mathrm{det}\, \tilde{g}}\left(\tilde{g}^{-1}\right)^{(MN)}D_{N)}U^\dagger\right)U - h.c. \right] = J_{CS}^U \, ,
\ee
\begin{align}
\label{nabEoMprobe}  &\hphantom{=} \frac{1}{2}D_N\left[V_f(\l,\tau^2)\sqrt{-\mathrm{det}\, \tilde{g}}\, w^2 F^{(L)NM} \right]  \\
\nn & = \frac{1}{2} V_f(\l,\tau^2) \sqrt{-\mathrm{det}\, \tilde{g}}\left(\tilde{g}^{-1}\right)^{MN} \ka\tau^2 \left(iD_NU^\dagger U -\frac{1}{N_f}\text{Tr}(iD_NU^\dagger U) + h.c.\right) + 2J_{CS}^{(L)} \, .
\end{align}
With the quadratic expansion introduced in Section \ref{Sec:probe}, the abelian gauge field equations \eqref{abEoM} are modified to take the same shape as the non-Abelian part
\begin{align}
\label{abEoMprobe} &\phantom{=} N_f\partial_N\left[V_f(\l,\tau^2)\sqrt{-\mathrm{det}\, \tilde{g} }\, w^2 \hat{F}^{(L)NM}  \right] \\
\nn &=  V_f(\l,\tau^2) \sqrt{-\mathrm{det}\, \tilde{g}}\left(\tilde{g}^{-1}\right)^{MN} \ka\tau^2 \text{Tr}\left(iD_NU^\dagger U + h.c.\right) + 2\hat{J}_{CS}^{(L)} \, ,
\end{align}
where the result \eqref{TrDUUd0} was used to simplify the right-hand side.

\paragraph*{SU(2) ansatz}
Substituting the ansatz  \eqref{ansatzSU2i}-\eqref{ansatzU1}, \eqref{parA}-\eqref{parF} and \eqref{TU2} into the equations of motion   \eqref{UEoMprobe}-\eqref{abEoMprobe} yields the following system of equations for the fields of the ansatz  \eqref{sumans}
\begin{align}
\label{UEoMprobeans} \hphantom{=}& \frac{x^a}{2\xi^3}\s^a\left[\partial_r\left(V_f(\l,\tau^2) \ka\tau^2 \frac{\ex^{3A} }{\sqrt{1+\ex^{-2A}\ka(\partial_r\tau)^2}} \xi^2 \td{A}_r \right)\right. \\
\nonumber &\hphantom{=} \left. + V_f(\l,\tau^2) \ka\tau^2\ex^{3A} \sqrt{1+\ex^{-2A}\ka(\partial_r\tau)^2} \, \left(\partial_{\xi}(\xi^2\td{A}_{\xi}) + 2\td{\phi}_1\td{\phi}_2 \right) \right]  = \frac{i}{2} J_{CS}^U \, ,
\end{align}
\begin{align}
\label{abEoMprobeans} \hphantom{=}& \partial_r\left[V_f(\l,\tau^2) w^2\ex^A \frac{\partial_r\Phi}{\sqrt{1+\ex^{-2A}\ka(\partial_r\tau)^2}}  \right] \\
\nonumber &\hphantom{=} + V_f(\l,\tau^2) w^2\ex^A \sqrt{1+\ex^{-2A}\ka(\partial_r\tau)^2}\, \frac{\partial_{\xi}(\xi^2\partial_{\xi}\Phi)}{\xi^2}   = - \hat{J}_{CS\, 0} \, ,
\end{align}
\begin{align}
\label{nabEoMprobeansr2} \hphantom{=}& \frac{\ex^A}{\sqrt{1+\ex^{-2A}\ka (\partial_r\tau)^2}}V_f(\l,\tau^2) w^2 \frac{x^a}{\xi^3}\left[\partial_{\xi}(\xi^2F_{\xi r})  - (i\phi^*D_r\phi - i\phi D_r\phi^*) \right] \\
\nonumber &\hphantom{=}  = 4 V_f(\l,\tau^2)\frac{x^a}{\xi} \frac{ \ex^{3A} }{\sqrt{1+\ex^{-2A}\ka(\partial_r\tau)^2}}\ka \tau^2 \td{A}_r + 4 J^{(L)\, a}_{CS\, r} \, ,
\end{align}
\begin{align}
\label{nabEoMprobeansi2} & \frac{\e^{iak}x_k}{2\xi^2}\left[V_f\ex^A\sqrt{1+\ex^{-2A}\ka (\partial_r\tau)^2}w^2\left(iD_{\xi}D_{\xi}\td{\phi} - i\td{\phi}\frac{|\phi|^2-1}{\xi^2} + h.c.\right)\right. \\
\nonumber & \left. + \left(iD_r\left(\frac{ V_f\ex^A w^2}{\sqrt{1+\ex^{-2A}\ka (\partial_r\tau)^2}}D_r\td{\phi}\right) + h.c.\right) \right] \\
\nonumber & +\frac{\xi^2\d_{ia}-x_ix_a}{2\xi^3}\left[V_f\ex^A\sqrt{1+\ex^{-2A}\ka (\partial_r\tau)^2}w^2\left(D_{\xi}D_{\xi}\td{\phi}  -\td{\phi}\frac{|\phi|^2-1}{\xi^2} + h.c.\right) \right.\\
\nonumber &\hphantom{=}   \left. + \left(D_r\left(\frac{ V_f\ex^A w^2}{\sqrt{1+\ex^{-2A}\ka (\partial_r\tau)^2}}D_r\td{\phi}\right) + h.c.\right) \right] \\
\nonumber & -\frac{x_ix_a}{2\xi^4}\Bigg[V_f\ex^A\sqrt{1+\ex^{-2A}\ka (\partial_r\tau)^2}w^2\left(2i\phi^*D_{\xi}\phi + h.c. \right)+  \\
\nn &\hphantom{-\frac{x_ix_a}{2\xi^4}\Bigg[V_f\ex^A } \left. + 2\xi^2\partial_r\left(\frac{ V_f\ex^A w^2}{\sqrt{1+\ex^{-2A}\ka (\partial_r\tau)^2}}F_{\xi r}\right) \right] \\
\nonumber &= 4 V_f \ex^{3A} \sqrt{1+\ex^{-2A}\ka(\partial_r\tau)^2}\ka \tau^2 \left(\frac{\xi^2\d_{ia}-x_ix_a}{2\xi^3}(\td{\phi} + \td{\phi}^*) - \frac{x_ix_a}{2\xi^4}(-2\xi^2\td{A}_{\xi})  \right) + 4 J^{(L)\, a}_{CS\, i} \, ,
\end{align}
where we used the covariant quantities  \eqref{covDiv} and \eqref{defFLR}. The right-handed equations are obtained by performing
\be
\label{LtoRans} \phi \to -\phi^* \sp A_{\bar{\mu}} \to - A_{\bar{\mu}} \, .
\ee
It can be checked that  \eqref{abEoMprobeans}, \eqref{nabEoMprobeansr2} and \eqref{nabEoMprobeansi2} yield the same form for the equations of motion as what was found in \cite{Pomarol08} and \cite{Cherman} when the tachyon and the CS contribution are set to 0. We derive the CS contribution in the next subsection.

\subsubsection{TCS contribution}

The general form of the contribution of the TCS term to the equations of motion is presented in Appendix \ref{Sec:CS_EoM}. For the $SU(2)$ ansatz  \eqref{ansatzSU2i}-\eqref{ansatzU1} and \eqref{TU2}, the CS currents $\hat{J}_{CS\, M}^{(L)}$, $J_{CS\, M}^{(L)\, a}$ and $J_{CS}^U$ in the probe approximation are found to be
\begin{align}
\nonumber \hat{J}_{CS\, 0}^{(L)} = \hat{J}_{CS\, 0}^{(R)} &= \frac{- f_1(\tau)}{4\pi^2 M^3\xi^2}\left(\left(1+\frac{f_3(\tau)}{f_1(\tau)}\right)\e^{\bar{\mu}\bar{\nu}}\left(F_{\bar{\mu}\bar{\nu}} + \partial_{\bar{\mu}}(-i\phi^*D_{\bar{\nu}}\phi + h.c.)\right)\right. \\
\nonumber &\hphantom{=}  +\left(3i\frac{f_2(\tau)}{f_1(\tau)}-1-\frac{f_3(\tau)}{f_1(\tau)}\right) \e^{\bar{\mu}\bar{\nu}}\left(\frac{1}{2}F_{\bar{\mu}\bar{\nu}}(\td{\phi}+\td{\phi}^*)^2 - A_{\bar{\mu}}\partial_{\bar{\nu}}(\td{\phi}+\td{\phi}^*)^2\right) \\
\nonumber &\hphantom{=}  + 2\tau'\left( \frac{f_1'(\tau)+f_3'(\tau)}{f_1(\tau)}( \td{A}_{\xi}(|\phi|^2-1) + \frac{i}{2}(\td{\phi}+\td{\phi}^*)(D_{\xi}\td{\phi}-D_{\xi}\td{\phi}^*))\right) \\
\label{Jhans} &\hphantom{= + 2\tau'\bigg(}\left. - 3i\tau'\frac{f_2'(\tau)}{f_1(\tau)}\td{A}_{\xi}(\td{\phi} + \td{\phi}^*)^2   \right)  \, ,
\end{align}
\begin{align}
\nn J_{CS\, r}^{(L)\, a} &= -J_{CS\, r}^{(R)\, a} \\
\label{Jrans} &= \frac{f_1(\tau)}{2\pi^2 M^3 \xi^3}x^a\, \partial_{\xi}\Phi\Bigg(\left(1 + \frac{f_3(\tau)}{f_1(\tau)}\right)(1-|\phi|^2)  + \\
\nn &\hphantom{= \frac{f_1(\tau)}{2\pi^2 M^3 \xi^3}x^a\, \partial_{\xi}\Phi\bigg(} \left. + \left(3i\frac{f_2(\tau)}{f_1(\tau)} - 1 - \frac{f_3(\tau)}{f_1(\tau)}\right)\frac{(\td{\phi} + \td{\phi}^*)^2}{2} \right) \, ,
\end{align}
\begin{align}
\nonumber J_{CS\, i}^{(L)\, a} &= \frac{\xi^2\d_{ia}-x_ix_a}{2\xi^3} \frac{f_1(\tau)\e^{\bar{\mu}\bar{\nu}}}{2\pi^2M^3}\Bigg(-i\left(1+\frac{f_3(\tau)}{f_1(\tau)}\right)\partial_{\bar{\mu}}\Phi D_{\bar{\nu}}\td{\phi} + \\
\nn &\hphantom{= \frac{\xi^2\d_{ia}-x_ix_a}{2\xi^3} \frac{f_1(\tau)\e^{\bar{\mu}\bar{\nu}}}{2\pi^2M^3}\Bigg(}   + 2\left(3i\frac{f_2(\tau)}{f_1(\tau)}-1-\frac{f_3(\tau)}{f_1(\tau)}\right)\td{\phi}\partial_{\bar{\mu}}\Phi \td{A}_{\bar{\nu}} + h.c. \Bigg)+ \\
\nn &\hphantom{=} -\frac{x_ix_a}{2\xi^4}\frac{ f_1(\tau)}{2\pi^2M^3}\Bigg(-2\left(1+\frac{f_3(\tau)}{f_1(\tau)}\right)\partial_r\Phi(|\phi|^2-1) + \\
\nn &\hphantom{= -\frac{x_ix_a}{2\xi^4}\frac{ f_1(\tau)}{2\pi^2M^3}\Bigg(} + \left(3i\frac{f_2(\tau)}{f_1(\tau)}-1-\frac{f_3(\tau)}{f_1(\tau)}\right) \partial_r\Phi(\td{\phi}+\td{\phi}^*)^2  \Bigg) +\\
\label{Jians} &\hphantom{=} +\frac{\e^{iak}x_k}{2\xi^2}\frac{f_1(\tau)}{2\pi^2M^3}\Bigg(\left(1 + \frac{f_3(\tau)}{f_1(\tau)}\right)\e^{\bar{\mu}\bar{\nu}}(\partial_{\bar{\mu}}\Phi D_{\bar{\nu}}\td{\phi} + h.c.) +\\
\nn &\hphantom{= +\frac{\e^{iak}x_k}{2\xi^2}\frac{f_1(\tau)}{2\pi^2M^3}\Bigg( } + \frac{f_1'(\tau)+f_3'(\tau)}{f_1(\tau)}\tau' (\td{\phi}+\td{\phi}^*)\partial_{\xi}\Phi \Bigg) \, .
\end{align}
\begin{align}
\nonumber J_{CS}^{U} &= \s^a\frac{x_a}{2\xi^3}\times\frac{-i}{\pi^2M^3}\left((f_1(\tau)+f_3(\tau)-3if_2(\tau))\e^{\bar{\mu}\bar{\nu}}(\td{\phi} +\td{\phi}^*) \partial_{\bar{\mu}}\Phi(D_{\bar{\nu}}\td{\phi} + D_{\bar{\nu}}\td{\phi}^*) + \right. \\
\label{JUans} &\hphantom{= \s^a\frac{x_a}{2\xi^3}\times\frac{-i}{\pi^2M^3}} +\left. (f_1'(\tau)+f_3'(\tau))\tau'\partial_{\xi}\Phi(|\phi|^2-1)-\frac{3i}{2}f_2'(\tau)\tau'\partial_{\xi}\Phi(\td{\phi}+\td{\phi}^*)^2 \right) \, .
\end{align}
Note that only $f_1(\tau)+f_3(\tau)$ and $f_2(\tau)$ contribute to the equations of motion.

\subsubsection{Full EoMs}

Substituting  \eqref{Jhans}-\eqref{JUans} into  \eqref{UEoMprobeans},\eqref{abEoMprobeans},\eqref{nabEoMprobeansr2} and \eqref{nabEoMprobeansi2} yields the full set of 6 real equations of motion for the 6 fields of the SU(2) ansatz  \eqref{sumans} in the probe instanton limit (where the tachyon modulus $\tau$ is a background field)
\begin{align}
\nonumber \hphantom{=}& \partial_r\left(\ex^{2A}k\,  \ka\tau^2  \xi^2 \td{A}_r \right) + \ex^{2A}h\, \ka\tau^2 \, \left(\partial_{\xi}(\xi^2\td{A}_{\xi})  + \frac{1}{2i}(\td{\phi}^2 -(\td{\phi}^*)^2)\right) \\
\nonumber &= \frac{f_1(\tau)+f_3(\tau)-3if_2(\tau)}{2\pi^2M^3}\e^{\bar{\mu}\bar{\nu}}(\td{\phi} +\td{\phi}^*) \partial_{\bar{\mu}}\Phi(D_{\bar{\nu}}\td{\phi} + D_{\bar{\nu}}\td{\phi}^*) \\
\label{SUCSn} &\hphantom{=} + \frac{1}{2\pi^2M^3}(f_1'(\tau)+f_3'(\tau))\tau'\partial_{\xi}\Phi(|\phi|^2-1)-\frac{3i}{4\pi^2M^3}f_2'(\tau)\tau'\partial_{\xi}\Phi(\td{\phi}+\td {\phi}^*)^2 \, ,
\end{align}
\begin{align}
\nonumber \hphantom{=}& \partial_r\left[k w^2\,  \xi^2 \partial_r\Phi  \right] + h w^2 \, \partial_{\xi}(\xi^2\partial_{\xi}\Phi) \\
\nonumber &= \frac{f_1(\tau)}{4\pi^2 M^3}\left(\left(1+\frac{f_3(\tau)}{f_1(\tau)}\right)\e^{\bar{\mu}\bar{\nu}}\left(F_{\bar{\mu}\bar{\nu}} + \partial_{\bar{\mu}}(-i\phi^*D_{\bar{\nu}}\phi + h.c.)\right) + \right. \\
\nn &\hphantom{= \frac{f_1(\tau)}{2\pi^2 N_f M^3}\bigg(}  +\left(3i\frac{f_2(\tau)}{f_1(\tau)}-1-\frac{f_3(\tau)}{f_1(\tau)}\right) \e^{\bar{\mu}\bar{\nu}}\left(\frac{1}{2}F_{\bar{\mu}\bar{\nu}}(\td{\phi}+\td{\phi}^*)^2 - \td{A}_{\bar{\mu}}\partial_{\bar{\nu}}(\td{\phi}+\td{\phi}^*)^2\right)+ \\
\nn  &\hphantom{= \frac{f_1(\tau)}{2\pi^2 N_f M^3}\bigg(} \left. + 2\tau' \frac{f_1'(\tau)+f_3'(\tau)}{f_1(\tau)}\left( \td{A}_{\xi}(|\phi|^2-1) + \frac{i}{2}(\td{\phi}+\td{\phi}^*)(D_{\xi}\td{\phi}-D_{\xi}\td{\phi}^*)\right)  \right. -
\\
\label{abCSn} &\hphantom{= \frac{f_1(\tau)}{2\pi^2 N_f M^3}\bigg(} \left. - 3i\tau'\frac{f_2'(\tau)}{f_1(\tau)}\td{A}_{\xi}(\td{\phi} + \td{\phi}^*)^2   \right)  \, ,
\end{align}
\begin{align}
\nonumber \hphantom{=}& k w^2 \left[\partial_{\xi}(\xi^2F_{\xi r})  - (i\phi^*D_r\phi - i\phi D_r\phi^*) \right] \\
\nonumber &= 4 \ex^{2A}k\, \ka \tau^2 \xi^2 \td{A}_r \\
\label{nabCSrn} &\hphantom{=}+ \frac{2f_1(\tau)}{\pi^2 M^3 }\, \partial_{\xi}\Phi\left(\left(1 + \frac{f_3(\tau)}{f_1(\tau)}\right)(1-|\phi|^2) + \frac{1}{2}\left(3i\frac{f_2(\tau)}{f_1(\tau)} - 1 - \frac{f_3(\tau)}{f_1(\tau)}\right)(\td{\phi} + \td{\phi}^*)^2 \right) ,
\end{align}
\begin{align}
\nonumber & h w^2\left(D_{\xi}D_{\xi}\td{\phi}  -\td {\phi}\frac{|\phi|^2-1}{\xi^2}\right)  + D_r\left(k w^2D_r\td{\phi}\right) + h.c. \\
\nonumber &= 4 \ex^{2A} h\, \ka \tau^2 \td{\phi} \, + \\
\label{nabCSi1n} &\hphantom{=} + \frac{2f_1(\tau)\e^{\bar{\mu}\bar{\nu}}}{\pi^2M^3}\left(-i\left(1+\frac{f_3(\tau)}{f_1(\tau)}\right)\partial_{\bar{\mu}}\Phi D_{\bar{\nu}}\td{\phi} + \right. \\
\nn &\hphantom{= + \frac{2f_1(\tau)\e^{\bar{\mu}\bar{\nu}}}{\pi^2M^3}\bigg(} \left. + 2\left(3i\frac{f_2(\tau)}{f_1(\tau)}-1-\frac{f_3(\tau)}{f_1(\tau)}\right)\td{\phi}\partial_{\bar{\mu}}\Phi \td{A}_{\bar{\nu}}\right) + h.c. \, ,
\end{align}
\begin{align}
\nonumber & h w^2\left(-i\phi^*D_{\xi}\td{\phi} + h.c. \right)  + \xi^2\partial_r\left(kw^2 F_{r\xi}\right)  \\
\nonumber &= 4 \ex^{2A} h \ka \tau^2 \xi^2\td{A}_{\xi}  \\
\label{nabCSi2n} &\hphantom{=} -  \frac{ 2f_1(\tau)}{\pi^2M^3}\partial_r\Phi\left(\left(1+\frac{f_3(\tau)}{f_1(\tau)}\right)(1-|\phi|^2) + \frac{1}{2} \left(3i\frac{f_2(\tau)}{f_1(\tau)}-1-\frac{f_3(\tau)}{f_1(\tau)}\right)(\td{\phi}+\td{\phi}^*)^2  \right) \, ,
\end{align}
\begin{align}
\nonumber & h w^2\left(iD_{\xi}D_{\xi}\td{\phi} - i\td{\phi}\frac{|\phi|^2-1}{\xi^2}\right)   +iD_r\left(kw^2 D_r\td{\phi}\right) + h.c.  \\
\label{nabCSi3n} &=  \frac{2f_1(\tau)}{\pi^2M^3}\left(\left(1 + \frac{f_3(\tau)}{f_1(\tau)}\right)\e^{\bar{\mu}\bar{\nu}}\partial_{\bar{\mu}}\Phi D_{\bar{\nu}}\td{\phi} + \frac{f_1'(\tau)+f_3'(\tau)}{f_1(\tau)}\tau' \td{\phi}\partial_{\xi}\Phi \right) + h.c. \,\,\,\, .
\end{align}
where we defined
\be
\label{defhk} k(r) \equiv \frac{\ex^A}{\sqrt{1+\ex^{-2A}\ka (\partial_r\tau)^2}}V_f(\la,\tau^2) \sp h(r) \equiv \ex^A \sqrt{1+\ex^{-2A}\ka  (\partial_r\tau)^2} V_f(\la,\tau^2) \,  .
\ee
 \eqref{SUCSn}-\eqref{nabCSi3n} are consistent with the equations of motion derived in \cite{Pomarol08} (equation (19)) and \cite{Cherman} (equations (2.16)-(2.19)) in the limit where the tachyon goes to 0\footnote{Except that $\g$ should be divided by 2 in  (2.19) of \cite{Cherman} and multiplied by -1 in  (2.17) and (2.18). Also, to obtain the same coefficients for $\gamma$ it is necessary to change the convention for the normalization of the Abelian gauge field $\Phi$: $\Phi \to \Phi / \sqrt{2 N_f}$ .}.

\subsubsection{Lorenz gauge}

\label{Sec:EoMLorenz}

We write in this subsection the equations of motion \eqref{SUCSn}-\eqref{nabCSi3n} in the Lorenz gauge, which corresponds to the constraint
\be
\label{Lorenzc} \partial_r A_r + \partial_{\xi} A_{\xi} = 0 \, .
\ee
In this gauge, the equations of motion for the 2-dimensional gauge field $A_{\bar{\mu}}$ can be written in an elliptic form, which is the appropriate form to solve the equations numerically via the heat diffusion method. The full set of equations of motion in Lorenz gauge reads
\begin{align}
\nonumber & \partial_r\left(\ex^{2A}k\,  \ka\tau^2  \xi^2 \partial_r\theta \right)+  \ex^{2A}h\, \ka\tau^2 \partial_{\xi}(\xi^2\partial_{\xi}\theta) \\
\nn &\hphantom{=} + \partial_r\left(\ex^{2A}k\,  \ka\tau^2  \xi^2 A_r \right) + \ex^{2A}h\, \ka\tau^2 \, \left( \partial_{\xi}(\xi^2A_{\xi})  + \frac{1}{2i}(\tilde{\phi}^2 -(\tilde{\phi}^*)^2)\right) \\
\nonumber &= \frac{f_1(\tau)+f_3(\tau)-3if_2(\tau)}{2\pi^2M^3}\e^{\bar{\mu}\bar{\nu}}(\tilde{\phi} +\tilde{\phi}^*) \partial_{\bar{\mu}}\Phi \left(D_{\bar{\nu}}\tilde{\phi} + h.c. \right) \\
\label{SUCSLg} &\hphantom{=} + \frac{1}{2\pi^2M^3}(f_1'(\tau)+f_3'(\tau))\tau'\partial_{\xi}\Phi(|\tilde{\phi}|^2-1)-\frac{3i}{4\pi^2M^3}f_2'(\tau)\tau'\partial_{\xi}\Phi(\tilde{\phi} + \tilde{\phi}^*)^2 \, ,
\end{align}
\begin{align}
\nonumber & \partial_r\left[k w^2\,  \xi^2 \partial_r\Phi  \right] + h w^2 \, \partial_{\xi}(\xi^2\partial_{\xi}\Phi) \\
\nonumber &= \frac{f_1(\tau)}{4\pi^2 M^3}\left(\left(1+\frac{f_3(\tau)}{f_1(\tau)}\right)\e^{\bar{\mu}\bar{\nu}}\left(F_{\bar{\mu}\bar{\nu}} + \partial_{\bar{\mu}}\left(-i\tilde{\phi}^* D_{\bar{\nu}}\tilde{\phi} + h.c. \right)\right) \right. \\
\nonumber &\hphantom{=}  +\left(3i\frac{f_2(\tau)}{f_1(\tau)}-1-\frac{f_3(\tau)}{f_1(\tau)}\right) \e^{\bar{\mu}\bar{\nu}}\left(\frac{1}{2}F_{\bar{\mu}\bar{\nu}}(\tilde{\phi}+\tilde{\phi}^*)^2 - \td{A}_{\bar{\mu}} \partial_{\bar{\nu}}(\tilde{\phi}+\tilde{\phi}^*)^2\right) \\
\nn &\hphantom{=}  + 2\tau' \frac{f_1'(\tau)+f_3'(\tau)}{f_1(\tau)}\left( \td{A}_{\xi} (|\tilde{\phi}|^2-1) + \frac{i}{2}(\tilde{\phi}+\tilde{\phi}^*)\left(D_\xi\tilde{\phi}- h.c. \right) \right)  \\
\label{abCSLg} &\hphantom{=} - \left. 3i\tau'\frac{f_2'(\tau)}{f_1(\tau)}\td{A}_{\xi} (\tilde{\phi} + \tilde{\phi}^*)^2   \right)  \, ,
\end{align}
\begin{align}
\nonumber & k w^2 \left[\xi^2(\partial_{\xi}^2A_r + \partial_r^2A_r) + 2\xi F_{\xi r}  - \left(i\tilde{\phi}^*D_r\tilde{\phi} + h.c. \right) \right] \\
\nonumber &= 4 \ex^{2A}k\, \ka \tau^2 \xi^2 \td{A}_r \\
\label{nabCSrLg} &\hphantom{=}+ \frac{2f_1(\tau)}{\pi^2 M^3 }\, \partial_{\xi}\Phi\left(\left(1 + \frac{f_3(\tau)}{f_1(\tau)}\right)(1-|\tilde{\phi}|^2) + \frac{1}{2}\left(3i\frac{f_2(\tau)}{f_1(\tau)} - 1 - \frac{f_3(\tau)}{f_1(\tau)}\right)(\tilde{\phi} + \tilde{\phi}^*)^2 \right) ,
\end{align}
\begin{align}
\nonumber &  h w^2\left(D_\xi D_\xi\tilde{\phi}  -\tilde{\phi}\frac{|\tilde{\phi}|^2-1}{\xi^2}\right) + D_r\left(k w^2 D_r\tilde{\phi}\right)  + h.c. \\
\nonumber &= 4 \ex^{2A} h\, \ka \tau^2  \tilde{\phi} \\
\nn &\hphantom{=} + \frac{2f_1(\tau)\e^{\bar{\mu}\bar{\nu}}}{\pi^2M^3} \left(-i\left(1+\frac{f_3(\tau)}{f_1(\tau)}\right)\partial_{\bar{\mu}}\Phi D_{\bar{\nu}}\tilde{\phi} + \right. \\
\label{nabCSi1Lg} &\hphantom{+ \frac{=2f_1(\tau)\e^{\bar{\mu}\bar{\nu}}}{\pi^2M^3} \bigg(} \left. + 2\left(3i\frac{f_2(\tau)}{f_1(\tau)}-1-\frac{f_3(\tau)}{f_1(\tau)}\right)\tilde{\phi}\partial_{\bar{\mu}}\Phi \td{A}_{\bar{\nu}} \right) + h.c. \, ,
\end{align}
\begin{align}
\nonumber &\hphantom{=} h w^2\left(-i\tilde{\phi}^*D_\xi\tilde{\phi} + h.c. \right)  + \xi^2 kw^2 (\partial_\xi^2 A_{\xi} + \partial_r^2 A_{\xi})+ \xi^2 (kw^2)' F_{r\xi}  \\
\nonumber &= 4 \ex^{2A} h \ka \tau^2 \xi^2\td{A}_{\xi} \\
\label{nabCSi2Lg} &\hphantom{=} -  \frac{ 2f_1(\tau)}{\pi^2M^3}\partial_r\Phi\left(\left(1+\frac{f_3(\tau)}{f_1(\tau)}\right)(1-|\tilde{\phi}|^2) + \frac{1}{2} \left(3i\frac{f_2(\tau)}{f_1(\tau)}-1-\frac{f_3(\tau)}{f_1(\tau)}\right)(\tilde{\phi}+\tilde{\phi}^*)^2  \right) \, ,
\end{align}
\begin{align}
\nonumber &\hphantom{=} h w^2\left(iD_\xi D_\xi\tilde{\phi} - i\tilde{\phi}\frac{|\tilde{\phi}|^2-1}{\xi^2}\right) +iD_r\left(kw^2 D_r\tilde{\phi}\right)  + h.c.  \\
\label{nabCSi3Lg} &=  \frac{2f_1(\tau)}{\pi^2M^3}  \left(\left(1 + \frac{f_3(\tau)}{f_1(\tau)}\right)\e^{\bar{\mu}\bar{\nu}}\partial_{\bar{\mu}}\Phi D_{\bar{\nu}}\tilde{\phi} + \frac{f_1'(\tau)+f_3'(\tau)}{f_1(\tau)}\tau' \tilde{\phi}\partial_{\xi}\Phi \right) + h.c. \,\,\,\, .
\end{align}

\subsection{Inhomogeneous tachyon}

\label{Sec:EoMx}

This appendix presents the expressions for the equations of motion for the ansatz fields  \eqref{sumans} in the case where the tachyon modulus $\tau$ is allowed to be dynamical.

The equations of motion involve the inverse effective metric which in this case takes the following form
\be
\label{ansgitx} \tilde{g}^{-1} =
\begin{pmatrix}
-\ex^{-2A}  & 0 & 0 & 0 & 0 \\
0 & \D_{rr} & - \frac{x_1}{\xi} \D_{\xi r} & - \frac{x_2}{\xi} \D_{\xi r} & - \frac{x_3}{\xi} \D_{\xi r} \\
0 & - \frac{x_1}{\xi} \D_{\xi r} & \,\, \ex^{-2A} - \left(\frac{x_1}{\xi}\right)^2 \D_{\xi \xi} & - \frac{x_1}{\xi}\frac{x_2}{\xi} \D_{\xi \xi} &  - \frac{x_1}{\xi}\frac{x_3}{\xi} \D_{\xi \xi} \\
0 & - \frac{x_2}{\xi} \D_{\xi r} &  - \frac{x_1}{\xi}\frac{x_2}{\xi} \D_{\xi \xi}& \ex^{-2A}  - \left(\frac{x_2}{\xi}\right)^2 \D_{\xi \xi} & - \frac{x_2}{\xi}\frac{x_3}{\xi} \D_{\xi \xi}\\
0 & - \frac{x_3}{\xi} \D_{\xi r} & - \frac{x_1}{\xi}\frac{x_3}{\xi} \D_{\xi \xi} & - \frac{x_2}{\xi}\frac{x_3}{\xi} \D_{\xi \xi} & \ex^{-2A} - \left(\frac{x_3}{\xi}\right)^2 \D_{\xi \xi}
\end{pmatrix}
 ,
\ee
where we defined the symbol $\Delta$ as
\be
\label{defDxx} \Delta_{\xi \xi} \equiv \frac{\ex^{6A} \ka \, (\partial_{\xi}\tau)^2 }{-\mathrm{det}\, \tilde{g}} = \frac{\ex^{-4A} \ka \, (\partial_{\xi}\tau)^2}{1 + \ex^{-2A}\ka\left((\partial_r\tau)^2+(\partial_{\xi}\tau)^2\right)} \, ,
\ee
\be
\label{defDxr} \Delta_{\xi r} \equiv \frac{\ex^{6A}\ka \, \partial_{\xi}\tau \partial_{r}\tau }{-\mathrm{det}\, \tilde{g}} = \frac{\ex^{-4A}\ka \, \partial_{\xi}\tau \partial_{r}\tau}{1 + \ex^{-2A}\ka\left((\partial_r\tau)^2+(\partial_{\xi}\tau)^2\right)} \, ,
\ee
\be
\label{defDrr} \Delta_{r r} \equiv \frac{\ex^{8A} (1 + \ex^{-2A} \ka \, (\partial_{\xi}\tau)^2) }{-\mathrm{det}\, \tilde{g}} = \frac{\ex^{-2A}  \, (1 + \ex^{-2A} \ka \, (\partial_{\xi}\tau)^2)}{1 + \ex^{-2A}\ka\left((\partial_r\tau)^2+(\partial_{\xi}\tau)^2\right)} \, .
\ee

The equations themselves are then obtained by extremizing the energy  \eqref{Etotx} with respect to small deformations of the ansatz fields  \eqref{sumans}
\begin{align}
\nonumber \hphantom{=}& \partial_r\left( \ka\tau^2 \mathcal{X} \ex^{4A}\D_{rr} \xi^2 \td{A}_r \right) +   \partial_{\xi}(\ka\tau^2 \mathcal{X} \ex^{2A}(1-\ex^{2A}\D_{\xi\xi}) \xi^2 \td{A}_{\xi})  \\
\nonumber &\hphantom{=}  - \,\partial_r(\ka\tau^2 \mathcal{X} \ex^{4A}\D_{\xi r} \xi^2 \td{A}_\xi) - \,\partial_\xi(\ka\tau^2 \mathcal{X} \ex^{4A}\D_{\xi r} \xi^2 \td{A}_r)\\
\nn &\hphantom{=}  + \frac{1}{2i}\ka\tau^2 \mathcal{X} \ex^{2A} (\td{\phi}^2 -(\td{\phi}^*)^2)  \\
\nonumber &= \frac{\e^{\bar{\mu}\bar{\nu}}}{2\pi^2M^3}\left[(f_1(\tau)+f_3(\tau)-3if_2(\tau))(\td{\phi} +\td{\phi}^*) \partial_{\bar{\mu}}\Phi(D_{\bar{\nu}}\td{\phi} + D_{\bar{\nu}}\td{\phi}^*)\right. \\
\label{SUCSx} &\hphantom{= \frac{\e^{\bar{\mu}\bar{\nu}}}{2\pi^2M^3}\bigg[} \left.+ \partial_{\bar{\mu}}\Phi\partial_{\bar{\nu}}\tau\left((f_1'(\tau)+f_3'(\tau))(|\phi|^2-1)-\frac{3i}{2}f_2'(\tau)(\td{\phi}+\td{\phi}^*)^2\right)\right] \, ,
\end{align}
\begin{align}
\nonumber & \partial_r\left[ w^2\mathcal{X}\ex^{2A}\D_{rr} \xi^2 \partial_r\Phi  \right] + \partial_{\xi}\left[w^2\mathcal{X}(1-\ex^{2A}\D_{\xi\xi}) \xi^2\partial_{\xi}\Phi\right]   \\
\nn &\hphantom{=} -\, \partial_r\left[ w^2\mathcal{X}\ex^{2A}\D_{\xi r} \xi^2 \partial_\xi\Phi  \right] -\, \partial_\xi\left[ w^2\mathcal{X}\ex^{2A}\D_{\xi r} \xi^2 \partial_r\Phi  \right] \\
\nonumber &= \frac{f_1(\tau)}{4\pi^2 M^3}\e^{\bar{\mu}\bar{\nu}}\left(\left(1+\frac{f_3(\tau)}{f_1(\tau)}\right)\left(F_{\bar{\mu}\bar{\nu}} + \partial_{\bar{\mu}}(-i\phi^*D_{\bar{\nu}}\phi + h.c.)\right)+ \right. \\
\nonumber &\hphantom{= \frac{f_1(\tau)}{4\pi^2 M^3}\e^{\bar{\mu}\bar{\nu}}}  +\left(3i\frac{f_2(\tau)}{f_1(\tau)}-1-\frac{f_3(\tau)}{f_1(\tau)}\right) \left(\frac{1}{2}F_{\bar{\mu}\bar{\nu}}(\td{\phi}+\td{\phi}^*)^2 - \td{A}_{\bar{\mu}}\partial_{\bar{\nu}}\left((\td{\phi}+\td{\phi}^*)^2\right)\right) + \\
\nn  &\hphantom{= \frac{f_1(\tau)}{2\pi^2 N_f M^3}\e^{\bar{\mu}\bar{\nu}}} \left. + 2\partial_{\bar{\nu}}\tau\left( \frac{f_1'(\tau)+f_3'(\tau)}{f_1(\tau)}( \td{A}_{\bar{\mu}}(|\phi|^2-1) + \frac{i}{2}(\td{\phi}+\td{\phi}^*)(D_{\bar{\mu}}\td{\phi}-D_{\bar{\mu}}\td{\phi}^*)) -\right.\right. \\
\label{abCSx} &\hphantom{= \frac{f_1(\tau)}{2\pi^2 N_f M^3}\e^{\bar{\mu}\bar{\nu}} + 2\partial_{\bar{\nu}}\tau \bigg(}
\left.\left. - 3i\frac{f_2'(\tau)}{2f_1(\tau)}\td{A}_{\bar{\mu}}(\td{\phi} + \td{\phi}^*)^2 \right)  \right)  \, ,
\end{align}
\begin{align}
\nonumber \hphantom{=}&  \partial_{\xi}\Big[w^2\mathcal{X}\ex^{2A}\left[\D_{rr}(1-\ex^{2A}\D_{\xi\xi}) - \ex^{2A}\D_{\xi r}^2\right]\xi^2F_{\xi r}\Big]  - w^2\mathcal{X}\ex^{2A}\D_{rr}(i\phi^*D_r\phi - i\phi D_r\phi^*)+ \\
\nn &\hphantom{=} + w^2\mathcal{X}\ex^{2A}\D_{\xi r}(i\phi^*D_\xi\phi - i\phi D_\xi\phi^*)  \\
\nonumber &= 4 \ka \tau^2 \mathcal{X} \ex^{4A}\D_{rr} \xi^2 \td{A}_r - 4 \ka \tau^2 \mathcal{X} \ex^{4A}\D_{\xi r} \xi^2 \td{A}_{\xi} + \\
\label{nabCSrx} &\hphantom{=}+ \frac{2f_1(\tau)}{\pi^2 M^3 }\, \partial_{\xi}\Phi\left(\left(1 + \frac{f_3(\tau)}{f_1(\tau)}\right)(1-|\phi|^2) + \frac{1}{2}\left(3i\frac{f_2(\tau)}{f_1(\tau)} - 1 - \frac{f_3(\tau)}{f_1(\tau)}\right)(\td{\phi} + \td{\phi}^*)^2 \right) \, ,
\end{align}
\begin{align}
\nonumber &\hphantom{=} D_{\xi}\left[w^2 \mathcal{X}(1-\ex^{2A}\D_{\xi\xi})D_{\xi}\td{\phi}\right] + D_r\left[w^2 \mathcal{X}\ex^{2A}\D_{rr} D_r\td{\phi}\right]- \\
\nn &\hphantom{=} -D_\xi\left[w^2 \mathcal{X}\ex^{2A}\D_{\xi r} D_r\td{\phi}\right] - D_r\left[w^2 \mathcal{X}\ex^{2A}\D_{\xi r} D_\xi\td{\phi}\right]  -w^2 \mathcal{X}\, \td{\phi}\frac{|\phi|^2-1}{\xi^2}   + h.c. \\
\nonumber &= 4 \ka \tau^2 \mathcal{X}\ex^{2A} \td{\phi} \, + \\
\nn &\hphantom{=} + \frac{2f_1(\tau)\e^{\bar{\mu}\bar{\nu}}}{\pi^2M^3}\left(-i\left(1+\frac{f_3(\tau)}{f_1(\tau)}\right)\partial_{\bar{\mu}}\Phi D_{\bar{\nu}}\td{\phi} +\right. \\
\label{nabCSi1x} &\hphantom{= +\frac{2f_1(\tau)\e^{\bar{\mu}\bar{\nu}}}{\pi^2M^3}\bigg(}
\left. + 2\left(3i\frac{f_2(\tau)}{f_1(\tau)}-1-\frac{f_3(\tau)}{f_1(\tau)}\right)\td{\phi}\partial_{\bar{\mu}}\Phi \td{A}_{\bar{\nu}}\right) + h.c. \, ,
\end{align}
\begin{align}
\nonumber &  \partial_{r}\Big[w^2\mathcal{X}\ex^{2A}\left[\D_{rr}(1-\ex^{2A}\D_{\xi\xi}) - \ex^{2A}\D_{\xi r}^2\right]\xi^2F_{r \xi}\Big] -
\\
\nn & - w^2\mathcal{X}(1-\ex^{2A}\D_{\xi\xi})(i\phi^*D_\xi\phi - i\phi D_\xi\phi^*) + w^2\mathcal{X}\ex^{2A}\D_{\xi r}(i\phi^*D_r\phi - i\phi D_r\phi^*)  \\
\nonumber &= 4 \ka \tau^2 \mathcal{X} \ex^{2A}(1-\ex^{2A}\D_{\xi\xi}) \xi^2 \td{A}_\xi - 4 \ka \tau^2 \mathcal{X} \ex^{4A}\D_{\xi r} \xi^2 \td{A}_{r} -  \\
\label{nabCSi2x} &\hphantom{=} -  \frac{ 2f_1(\tau)}{\pi^2M^3}\partial_r\Phi\left(\left(1+\frac{f_3(\tau)}{f_1(\tau)}\right)(1-|\phi|^2) + \frac{1}{2} \left(3i\frac{f_2(\tau)}{f_1(\tau)}-1-\frac{f_3(\tau)}{f_1(\tau)}\right)(\td{\phi}+\td{\phi}^*)^2  \right) \, ,
\end{align}
\begin{align}
\nonumber &\hphantom{=} iD_{\xi}\left[w^2 \mathcal{X}(1-\ex^{2A}\D_{\xi\xi})D_{\xi}\td{\phi}\right] + iD_r\left[w^2 \mathcal{X}\ex^{2A}\D_{rr} D_r\td{\phi}\right] - \\
\nn &\hphantom{=} -iD_\xi\left[w^2 \mathcal{X}\ex^{2A}\D_{\xi r} D_r\td{\phi}\right] - iD_r\left[w^2 \mathcal{X}\ex^{2A}\D_{\xi r} D_\xi\td{\phi}\right]  -iw^2 \mathcal{X}\, \td{\phi}\frac{|\phi|^2-1}{\xi^2}   + h.c. \\
\label{nabCSi3x} &=  \frac{2f_1(\tau)}{\pi^2M^3}\e^{\bar{\mu}\bar{\nu}}\left(\left(1 + \frac{f_3(\tau)}{f_1(\tau)}\right)\partial_{\bar{\mu}}\Phi D_{\bar{\nu}}\td{\phi} + \frac{f_1'(\tau)+f_3'(\tau)}{f_1(\tau)}\partial_{\bar{\nu}}\tau \td{\phi}\partial_{\bar{\mu}}\Phi \right) + h.c. \,\,\,\, ,
\end{align}
\begin{align}
\nn & \partial_r\bigg[ \mathcal{Y} \bigg( \partial_r\tau \mathcal{L}_{\text{DBI}} +\\
\nn &\hphantom{\partial_r\bigg[ \mathcal{Y} } + \ex^{2A}\mathcal{X} \bigg(\ex^{2A}\xi^2\ka(\l)\tau^2\left(\D_{rr,r}\td{A}_r^2 - \D_{\xi\xi,r}  \td{A}_{\xi}^2  - 2\D_{\xi r,r}\td{A}_r\td{A}_{\xi} \right)+  \\
\nonumber &\hphantom{\partial_r\bigg[ \mathcal{Y} f\bigg(}  + w(\l)^2\left(\frac{1}{8}\Big(\D_{rr,r}\left(1 - \ex^{2A}\D_{\xi\xi}  \right) - \ex^{2A}\D_{rr}\D_{\xi\xi,r} - 2\ex^{2A}\D_{\xi r,r}\D_{\xi r}\Big)\xi^2(F_{\bar{\mu} \bar{\nu}})^2+ \right. \\
\nn &\hphantom{\partial_r\bigg[ \mathcal{Y} f\bigg( + w(\l)^2}+ \frac{1}{2}\left(- \D_{\xi\xi,r} \left|\mathrm{D}_{\xi}\phi\right|^2 + \D_{rr,r} \left|\mathrm{D}_{r}\phi\right|^2\right)
 - \frac{1}{2}\D_{\xi r,r} (D_r\phi^*D_{\xi}\phi + h.c.)- \\
\nn &\hphantom{\partial_r\bigg[ \mathcal{Y} f\bigg( + w(\l)^2} \left.\left.\left. \left. - \xi^2 \left(\D_{rr,r}(\partial_r\Phi)^2 - \D_{\xi\xi,r} (\partial_{\xi}\Phi)^2 - 2\D_{\xi r,r}\partial_{\xi}\Phi\partial_r\Phi \right)\right)\,\right)\,\right)\right] + \\
\nn & + \partial_\xi\bigg[ \mathcal{Y} \bigg(\partial_\xi\tau\mathcal{L}_{\text{DBI}}  \\
\nn &\hphantom{+ \partial_\xi\bigg[} + \ex^{2A}\mathcal{X} \bigg(\ex^{2A}\xi^2\ka(\l)\tau^2\left(\D_{rr,\xi}\td{A}_r^2 - \D_{\xi\xi,\xi}  \td{A}_{\xi}^2  - 2\D_{\xi r,\xi}\td{A}_r\td{A}_{\xi} \right)+ \\
\nonumber &\hphantom{+ \partial_\xi\bigg[ + }
\,\,\,\, + w(\l)^2\left(\frac{1}{8}\left(\D_{rr,\xi}\left(1 - \ex^{2A}\D_{\xi\xi}  \right) - \ex^{2A}\D_{rr}\D_{\xi\xi,\xi} - 2\ex^{2A}\D_{\xi r,\xi} \D_{\xi r}\right)\xi^2(F_{\bar{\mu} \bar{\nu}})^2 + \right. \\
\nn &\hphantom{+ \partial_\xi\bigg[ + +w(\l)^2}\,\,\,\, + \frac{1}{2}\left(- \D_{\xi\xi,\xi} \left|\mathrm{D}_{\xi}\phi\right|^2 + \D_{rr,\xi} \left|\mathrm{D}_{r}\phi\right|^2\right)
 - \frac{1}{2}\D_{\xi r,\xi} (D_r\phi^*D_{\xi}\phi + h.c.)- \\
\nn &\hphantom{+ \partial_\xi\bigg[ + +w(\l)^2}
\left.\left.\left. \left. - \xi^2 \left(\D_{rr,\xi}(\partial_r\Phi)^2 - \D_{\xi\xi,\xi} (\partial_{\xi}\Phi)^2 - 2\D_{\xi r,\xi}\partial_{\xi}\Phi\partial_r\Phi \right)\right)\,\right)\,\right)\right] - \\
\nn & - \frac{\d}{\d\tau}\left(\log{V_f}\right)\mathcal{L}_{\text{DBI}} \\
\nn &= 2\mathcal{X}\ex^{2A}\xi^2\ka\tau \left(\ex^{2A}\D_{rr}\td{A}_r^2 + \left(1 - \ex^{2A}\D_{\xi\xi}  \right)\td{A}_{\xi}^2 + \frac{(\phi + \phi^*)^2}{2\xi^2} - 2\ex^{2A}\D_{\xi r}\td{A}_r\td{A}_{\xi} \right) \\
\nn &\hphantom{=} + \frac{1}{\pi^2M^3}  \e^{\bar{\mu}\bar{\nu}}\partial_{\bar{\mu}}\Phi\left[(f_1'(\tau)+f_3'(\tau))\left(\td{A}_{\bar{\nu}} + \frac{1}{2}(-i\phi^*D_{\bar{\nu}}\phi + h.c.)+\frac{1}{4i}\partial_{\bar{\nu}}(\td{\phi}^2-(\td{\phi}^*)^2)\right)\right. \\
\label{EoMt} &\hphantom{= + \frac{1}{\pi^2M^3}  \e^{\bar{\mu}\bar{\nu}}\partial_{\bar{\mu}}\Phi\bigg[} \left. + \frac{1}{2}(3if_2'(\tau)-f_1'(\tau)-f_3'(\tau))(\td{\phi} + \td{\phi}^*)^2 \td{A}_{\bar{\nu}} \right] \, ,
\end{align}
where we defined
\be
\label{defX} \mathcal{X} \equiv \sqrt{1 + \ex^{-2A}\ka\left((\partial_r\tau)^2+(\partial_{\xi}\tau)^2\right)}\, V_f(\l,\tau)\,\ex^A \, ,
\ee
\begin{align}
\nn \mathcal{L}_{\text{DBI}} & \equiv N_f\xi^2\mathcal{X}\ex^{4A}+ \\
\nn &\hphantom{\equiv}
+\mathcal{X} \Bigg(\ex^{2A}\xi^2\ka(\l)\tau^2 \times \\
\nn &\hphantom{\equiv +\mathcal{X} \bigg(}
\times\bigg(\ex^{2A}\D_{rr}\td{A}_r^2 + \left(1 - \ex^{2A}\D_{\xi\xi}  \right)\td{A}_{\xi}^2 + \frac{(\td{\phi} + \td{\phi}^*)^2}{2\xi^2} - 2\ex^{2A}\D_{\xi r}\td{A}_r\td{A}_{\xi} \bigg) +  \\
\nn &\hphantom{\equiv + \mathcal{X}}  + w(\l)^2\bigg(\frac{1}{8}\ex^{2A}\left[\D_{rr}\left(1 - \ex^{2A}\D_{\xi\xi}  \right) -\ex^{2A}\D_{\xi r}^2 \right]\xi^2(F_{\bar{\mu} \bar{\nu}})^2 +  \\
\nn &\hphantom{\equiv + \mathcal{X}  + w(\l)^2}
+ \frac{1}{2}\left(\left(1 - \ex^{2A}\D_{\xi\xi}  \right)\left|\mathrm{D}_{\xi}\phi\right|^2 + \ex^{2A}\D_{rr} \left|\mathrm{D}_{r}\phi\right|^2\right) + \\
\nn & \hphantom{\equiv + \mathcal{X}  + w(\l)^2}
+ \frac{\left(1 - |\phi|^2\right)^2}{4\xi^2} - \frac{1}{2}\ex^{2A}\D_{\xi r} (D_r\phi^*D_{\xi}\phi + h.c.) - \\
\nn & \hphantom{\equiv + \mathcal{X}  + w(\l)^2}
 - \xi^2 \Big(\ex^{2A}\D_{rr}(\partial_r\Phi)^2 + \left(1 - \ex^{2A}\D_{\xi\xi}  \right)(\partial_{\xi}\Phi)^2 - \\
\label{defLx} &\hphantom{\equiv + \mathcal{X}  + w(\l)^2 - \frac{N_f}{2} \xi^2 \Big( \ex^{2A}\D_{rr}(\partial_r\Phi)^2}
 - 2\ex^{2A}\D_{\xi r}\partial_{\xi}\Phi\partial_r\Phi \Big)\bigg)\Bigg) \, ,
\end{align}
\be
\label{defY} \mathcal{Y} \equiv \frac{\ex^{-2A}\ka}{1 + \ex^{-2A}\ka\left((\partial_r\tau)^2+(\partial_{\xi}\tau)^2\right)} \, ,
\ee
and introduced the following condensed notation
\be
\label{notx} \D_{xy,r} \equiv \frac{1}{\mathcal{Y}}\frac{\d \D_{xy}}{\d\, \partial_r\tau}\sp \D_{xy,\xi} \equiv \frac{1}{\mathcal{Y}}\frac{\d \D_{xy}}{\d\, \partial_\xi\tau}\sp x,y\in \{r,\xi\} \, .
\ee
These are equal to
\be
\label{Drrz} \D_{rr,r} = -2\partial_r\tau \D_{rr} \sp \D_{rr,\xi} = 2\partial_{\xi}\tau \left[-\D_{rr} + \ex^{-2A} \right] \, ,
\ee
\be
\label{Dxxz} \D_{\xi\xi,r} = -2\partial_r\tau \D_{\xi\xi} \sp \D_{\xi\xi,\xi} = 2\partial_{\xi}\tau \left[-\D_{\xi\xi} + \ex^{-2A} \right] \, ,
\ee
\be
\label{Dxrz} \D_{\xi r,r} = -2\partial_r\tau \D_{\xi r} + \ex^{-2A}\partial_\xi\tau  \sp \D_{\xi r,\xi} = -2\partial_\xi\tau \D_{\xi r} + \ex^{-2A}\partial_r\tau  \, .
\ee

\section{Asymptotics of the probe instanton solution}

\label{Sec:asymptotics}

We present in this appendix the analytical asymptotics near the boundaries of the 2-dimensional space $(r,\xi)$ of the bulk instanton solution in the probe limit. The leading asymptotics will be necessary to determine the boundary conditions for the numerical method used to compute the instanton solution.

\subsection{UV asymptotics}

\label{Sec:asymUV}

We start by analyzing the asymptotics of the solution near the AdS-like UV boundary, that is in the limit where $r\to 0$, which are obtained by solving the equations of motion  \eqref{SUCSn}-\eqref{nabCSi3n} in this limit. This analysis involves the UV asymptotics of the background  \eqref{v12} and  \eqref{AUV}-\eqref{tUV}. According to the holographic dictionary, the UV asymptotics contain the external sources and vacuum expectation values (vev) for the operators dual to the instanton bulk fields.

At leading order in $r$,  \eqref{abCSn}, \eqref{nabCSi1n}, \eqref{nabCSi2n} and \eqref{nabCSi3n} respectively determine the power-law behavior of the two independent solutions for $\Phi$, $\td{\phi}_1$, $\td{A}_{\xi}$ and $\td{\phi}_2$, which are the components of the gauge fields  \eqref{ansatzSU2i}-\eqref{ansatzU1}, redefined to absorb the tachyon phase as in  \eqref{LtoLtcyl}. These power-laws correspond to the standard result for massless gauge fields on AdS
\be
\label{vphiUV}\varphi = \varphi^{(0)}(\xi)(1 + \cdots ) + r^2 \varphi^{(2)}(\xi)(1 + \cdots) \, ,
\ee
where $\varphi$ represents all the above mentioned fields, the $(0)$ index refers to the source for the dual operator, the $(2)$ index to a term proportional to the vev for that operator and the dots terms that go to 0 at the boundary. Note that this behavior does not apply to $\td{A}_r$ as the equations of motion are first order for $\td{A}_r$ in the $r$ direction.

For the baryon state under study, the chemical potential is set to 0 : $\Phi^{(0)}=0$ and there is no source for the gauge field at the boundary, which for the redefined gauge fields  \eqref{LtoLtcyl} corresponds to
\be
\label{pgUV} \td{A}_{\xi}^{(0)}  = \partial_{\xi} \theta(0,\xi) \sp \td{\phi}_1^{(0)} = \sin \theta(0,\xi) \sp \td{\phi}_2^{(0)} = -\cos \theta(0,\xi) \, ,
\ee
where $\theta$ is the phase in the tachyon ansatz  \eqref{TU2}. Near the boundary, the ansatz fields at leading order therefore behave as
\be
\label{PhiUV}\Phi = \Phi^{(2)}(\xi)\, r^2(1 + \cdots) \, ,
\ee
\be
\label{phi1UV} \td{\phi}_1 = \sin \theta(0,\xi)(1+\cdots) + \td{\phi}_1^{(2)}(\xi)\, r^2(1 + \cdots) \, ,
\ee
\be
\label{phi2UV}\td{\phi}_2 = -\cos \theta(0,\xi)(1+\cdots) + \td{\phi}_2^{(2)}(\xi)\, r^2(1 + \cdots) \, ,
\ee
\be
\label{AxUV}\td{A}_{\xi} = \partial_{\xi} \theta(0,\xi)(1+\cdots) + \td{A}_\xi^{(2)}(\xi)\, r^2(1 + \cdots) \, .
\ee
Then, the constraint  \eqref{SUCSn} imposes that $\td{A}_r$ vanishes at the boundary, and behaves at most linearly in $r$
\be
\label{ArUV1} \td{A}_r = \td{A}^{(1)}_{r}(\xi)\, r \left( 1 + \cdots \right) \, ,
\ee
where $\td{A}_r^{(1)}$ obeys
\be
\label{EOM1l1} \partial_{\xi}(\xi^2\partial_{\xi}\theta(0,\xi)) - \sin(2\theta(0,\xi)) + 4 \xi^2 \td{A}_{r}^{(1)}(\xi) = 0 \, .
\ee
From the behavior of the background  \eqref{v12} and  \eqref{AUV}-\eqref{tUV}, we deduce the order of the next terms in the UV expansion
\be
\label{PhiUV2}\Phi = \Phi^{(2)}(\xi)\, r^2\left(1 + \mathcal{O}\left(\frac{1}{r\log{r\L}}\right)\right) \, ,
\ee
\be
\label{phi1UV2} \td{\phi}_1 = \sin \theta(0,\xi) + \td{\phi}_1^{(2)}(\xi)\, r^2\left(1 + \mathcal{O}\left(\frac{1}{r\log{r\L}}\right)\right) \, ,
\ee
\be
\label{phi2UV2}\td{\phi}_2 = -\cos \theta(0,\xi) + \td{\phi}_2^{(2)}(\xi)\, r^2\left(1 + \mathcal{O}\left(\frac{1}{r\log{r\L}}\right)\right) \, ,
\ee
\be
\label{AxUV2}\td{A}_{\xi} = \partial_{\xi} \theta(0,\xi) + \td{A}_\xi^{(2)}(\xi)\, r^2\left(1 + \mathcal{O}\left(\frac{1}{r\log{r\L}}\right)\right) \, ,
\ee
\be
\label{ArUV2} \td{A}_r = \td{A}^{(1)}_{r}(\xi)\, r \left(1 + \mathcal{O}\left(\frac{1}{r\log{r\L}}\right)\right) \, .
\ee
Note the peculiar feature that the source mode does not receive $r$-dependent corrections when moving away from the boundary. This property is specific to the chiral limit $m_q=0$.

\subsection{Asymptotics at $\xi \to \infty$}

\label{Sec:asymxinf}

We study here the EoMs  \eqref{SUCSn}-\eqref{nabCSi3n} in the limit where $\xi \to \infty$, given the boundary conditions of the second column of Table \ref{tab:bcs}, which impose that the instanton energy is finite. These asymptotics contain information about the tail of the meson cloud far from the baryon, which determines the long range meson-exchange interaction between baryons.

We assume that the fields can be expanded as Taylor series in $1 / \xi$ and analyse the EoMs order by order in $1/\xi$. This yields the following behavior for the instanton fields in the limit $\xi \to \infty$
\be
\label{PhiXinf} \Phi(r,\xi) = \frac{\hat{\Phi}_6(r)}{\xi^6} + \mathcal{O}(\xi^{-7}) \, ,
\ee
\be
\label{phi1Xinf} \td{\phi}_1(r,\xi) = \frac{\hat{\phi}_{1,1}(r)}{\xi} + \frac{\hat{\phi}_{1,2}(r)}{\xi^2}+ \frac{\hat{\phi}_{1,3}(r)}{\xi^3} + \frac{\hat{\phi}_{1,4}(r)}{\xi^4} + \frac{\hat{\phi}_{1,5}(r)}{\xi^5} + \mathcal{O}(\xi^{-6}) \, ,
\ee
\be
\label{phi2Xinf} \td{\phi}_2(r,\xi) = 1 + \frac{\hat{\phi}_{2,1}}{\xi} + \frac{\hat{\phi}_{2,2}}{\xi^2}+ \frac{\hat{\phi}_{2,3}}{\xi^3} + \frac{\hat{\phi}_{2,4}(r)}{\xi^4} + \frac{\hat{\phi}_{2,5}(r)}{\xi^5} + \mathcal{O}(\xi^{-6}) \, ,
\ee
\be
\label{AxXinf} \td{A}_{\xi}(r,\xi) = \frac{\hat{A}_{\xi,2}(r)}{\xi^2} + \frac{\hat{A}_{\xi,3}(r)}{\xi^3}+ \frac{\hat{A}_{\xi,4}(r)}{\xi^4} + \frac{\hat{A}_{\xi,5}(r)}{\xi^5} + \mathcal{O}(\xi^{-6}) \, ,
\ee
\be
\label{ArXinf} \td{A}_{r}(r,\xi) =  \frac{\hat{A}_{r,3}(r)}{\xi^3}+ \frac{\hat{A}_{r,4}(r)}{\xi^4} + \frac{\hat{A}_{r,5}(r)}{\xi^5} + \frac{\hat{A}_{r,6}(r)}{\xi^6} + \mathcal{O}(\xi^{-7}) \, ,
\ee
where $\hat{\phi}_{2,1}, \hat{\phi}_{2,2}$ and $\hat{\phi}_{2,3}$ are constants and $\hat{A}_{r,n}$ is a linear combination (with $r$-dependent coefficients) of $\hat{\phi}_{1,n-2}', \hat{\phi}_{1,n-4}', \cdots ,$ $\hat{A}_{\xi,n-1}', \hat{A}_{\xi,n-3}', \cdots$. The other coefficients obey second order differential equations:
\be
\label{eqp2P} F'' + \frac{(kw^2)'}{kw^2} F' = S_F \sp F = \hat{\phi}_{2,n}, \hat{\Phi}_n \, ,
\ee
\be
\label{eqp1Ax} G'' + \frac{(kw^2)'}{kw^2} G' - 4\frac{\ex^{2A}h\kappa \tau^2}{kw^2}G = S_G \sp G = \hat{\phi}_{1,n}, \hat{A}_{\xi,n} \, ,
\ee
where $k$ and $h$ are defined in  \eqref{defhk} and $S_G$ and $S_F$ are source terms that depend on lower order coefficients. In particular
\be
\label{SFG0} S_{\hat{\phi}_{1,1}} = S_{\hat{\phi}_{1,2}} = S_{\hat{A}_{\xi,2}} = S_{\hat{A}_{\xi,3}} = 0 \, .
\ee
Note that, because of the UV boundary conditions of Table \ref{tab:bcsLg}, the boundary values of the coefficients $\hat{\phi}_{1,n}, \hat{\phi}_{2,n}$ and $\hat{A}_{\xi,n}$ can all be written explicitly in terms of the asymptotics of $\theta(0,\xi)$ at $\xi\to\infty$. In particular, this implies that $\hat{\phi}_{2,1} = 0$.

Another interesting observation is that the source terms are such that the solution at $\xi \to \infty$ can be divided into two independent parity sets
\be
\label{evenXinf}\Phi \, \text{even} \sp \phi_1 \, \text{odd} \sp \phi_2 \, \text{even up to r-independent terms} \sp A_{\xi} \, \text{even} \sp A_r \, \text{odd} \, ,
\ee
or
\be
\label{oddXinf}\Phi \, \text{odd} \sp \phi_1 \, \text{even} \sp \phi_2 \, \text{odd up to r-independent terms} \sp A_{\xi} \, \text{odd} \sp A_r \, \text{even} \, .
\ee

\subsubsection{IR behavior of the large $\xi$ coefficients}

\label{Sec:xinfIR}

The homogeneous solutions of \eqref{eqp2P} are of the form
\be
\label{soleqp2P} F(r) = c_1 + c_2 \int \frac{\mathrm{d}r}{k(r)w(r)^2} \, ,
\ee
where $c_{1,2}$ are two independent constants. From  \eqref{v15} and \eqref{kIR}, it is clear that the second term in  \eqref{soleqp2P} is singular in the IR. For regularity in the IR $c_2$ should therefore be set to 0 for every coefficients obeying  \eqref{eqp2P}. Note that in the UV, $c_2$ parametrizes the freedom for the vev term that goes like $\sim r^2$ as $r\to 0$. Comparing with the UV expansions \eqref{PhiUV}-\eqref{AxUV}, we see that this constraint imposes that $\Phi^{(2)}$ and $\td{\phi}_2^{(2)}$ are fixed in terms of $\theta(0,\xi), \td{\phi}_1^{(2)}$ and $\td{A}_{\xi}^{(2)}$.

Near the IR boundary, the homogeneous solutions of  \eqref{eqp1Ax} are found to behave as
\begin{align}
\nn G(r) &= c'_1 \exp{\left(a_{\text{IR}} \tau_0^2\left(\frac{r}{R}\right)^{2C_\tau}\left(1+\mathcal{O}(r^{-2})\right)\right)} \\
\label{soleqp1Ax} &\hphantom{=} + c'_2 \exp{\left(-\frac{2 C_\tau}{3a_{\text{IR}}(C_\tau-1)}\left(\frac{\ka_{IR}}{w_{IR}}\right)^2\tau_0^2 \left(\frac{r}{R}\right)^{2C_\tau-2}\left(1+\mathcal{O}(r^{-2})\right)\right)} \, .
\end{align}
The first solution diverges in the IR so for all coefficients obeying  \eqref{eqp1Ax}, $c'_1$ should be set to 0. In terms of the UV expansion of  \eqref{PhiUV}-\eqref{AxUV}, this should impose $\td{\phi}_1^{(2)}$ and $\td{A}_{\xi}^{(2)}$ in terms of $\theta(0,\xi)$.

\subsection{IR asymptotics}

\label{bcIR}

We consider in this subsection the asymptotic behavior of the fields near the IR boundary, that is in the limit where $r\to \infty$. For each field, there exist generically two independent solutions in the IR as in  \eqref{soleqp2P} and \eqref{soleqp1Ax}, only one of which is regular. The asymptotics we present in the following are the unique regular asymptotics.

From the asymptotics of the background  \eqref{v15} and \eqref{laIR}-\eqref{tauIR} we find the functions $k(r)$ and $h(r)$ defined in  \eqref{defhk} to behave in the IR as
\begin{align}
\label{kIR} k(r) &\sim \frac{W_{\text{IR}}R}{\tau_0 C_\tau}\left(\sqrt{\frac{3}{2}}\ka_{IR}\right)^{-1/2}\ex^{2A_c+(5/3)\l_c}\left(\frac{r}{R}\right)^{3/2 - C_\tau}\times\\
\nn &\hphantom{\sim \frac{W_{\text{IR}}R}{\l_0^2\tau_0 C_\tau}\left(\sqrt{\frac{3}{2}}\frac{\l_0^{4/3}}{\kappa_0\bar{\kappa}_0}\right)^{-1/2}}
\times \exp{\left(-\tau_0^2\left(\frac{r}{R}\right)^{2C_\tau}\left(1+\mathcal{O}(r^{-2})\right) + 2 \frac{r^2}{R^2}\right)} \, ,
\end{align}
\begin{align}
\label{hIR} h(r) &\sim \frac{W_{\text{IR}}\tau_0 C_\tau}{ R }\left(\sqrt{\frac{3}{2}}\ka_{IR}\right)^{1/2} \ex^{(1/3)\l_c}\left(\frac{r}{R}\right)^{C_\tau - 1/2} \times \\
\nn &\hphantom{sim \frac{W_{\text{IR}}\tau_0 C_\tau}{\l_0^2 R }\left(\sqrt{\frac{3}{2}}\frac{\l_0^{4/3}}{\kappa_0\bar{\kappa}_0}\right)^{1/2}}
\times \exp{\left(-\tau_0^2\left(\frac{r}{R}\right)^{2C_\tau}\left(1+\mathcal{O}(r^{-2})\right) + 2 \frac{r^2}{R^2}\right)} \, .
\end{align}

The IR behavior of the fields is obtained by solving the equations of motion  \eqref{SUCSn}-\eqref{nabCSi3n} in the IR limit. In this limit the CS terms become negligible, which simplifies a lot the EoMs.

The IR behavior of $\Phi$, $\td{\phi}_1$ and $\td{A}_{\xi}$ is found to be separable in $\xi$ and $r$
\be
\label{PhIR} \Phi(r) = \frac{1}{r} \a_\Phi(\xi) \left(1 + \mathcal{O}\left( r^{-2C_\tau} \right) \right) \, ,
\ee
\be
\label{p1IR} \td{\phi}_1(r) = \a_1(\xi) \exp{\left(-\zeta \left(\frac{r}{R}\right)^{2C_\tau-2}\left(1+\mathcal{O}(r^{-2})\right)\right)} \, ,
\ee
\be
\label{AxIR} \td{A}_{\xi}(r) = \a_{\xi}(\xi) \exp{\left(-\zeta \left(\frac{r}{R}\right)^{2C_\tau-2}\left(1+\mathcal{O}(r^{-2})\right)\right)} \, ,
\ee
where we defined the constant
\be
\zeta \equiv \frac{2}{3}\frac{C_\tau}{a_{\text{IR}}(C_\tau-1)}\left(\frac{\ka_{IR}}{w_{IR}}\right)^2 \tau_0^2 \, .
\ee
and $\a_{\Phi}(\xi)$ obeys an independent linear ODE
\be
\label{eqaPh} \a''(\xi) + \frac{2}{\xi}\a'(\xi) - \frac{1}{\b}\a(\xi) = 0 \, ,
\ee
with
\be
\label{defbet} \b \equiv \sqrt{\frac{3}{8}} \frac{\ex^{-2A_c}\ka_{IR}C_\tau R}{ a_{\text{IR}}} \, .
\ee
 \eqref{eqaPh} is compatible with the condition that $\Phi$ goes to 0 at $\xi \to \infty$ and admits a unique solution up to overall normalization once the boundary condition that $\partial_\xi\Phi$ goes to 0 at $\xi\to 0$ is imposed.

From  \eqref{p1IR} and \eqref{AxIR}, the constraint  \eqref{SUCSn} in the IR limit then fixes the leading behavior for $\td{A}_r$ to be
\be
\label{ArIR} \td{A}_r(r) = \frac{\a_r(\xi)}{r} \exp{\left(-\zeta \left(\frac{r}{R}\right)^{2C_\tau-2}\left(1+\mathcal{O}(r^{-2})\right)\right)} \, ,
\ee

\begin{figure}[h]
\begin{center}
\begin{overpic}
[scale=0.9]{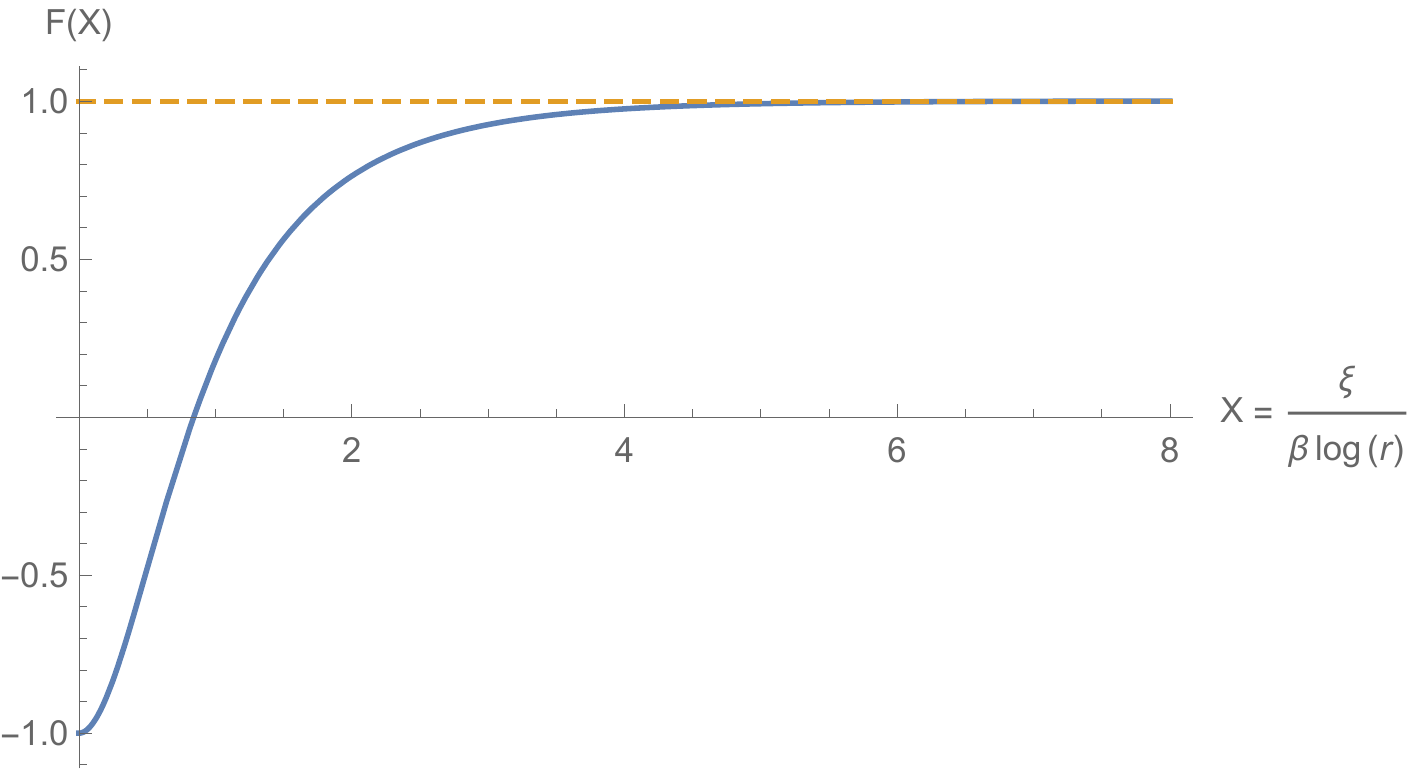}
\end{overpic}
\caption{Numerical solution to  \protect\eqref{ODE_F} with the boundary conditions of  \protect\eqref{bcxF}. }
\label{fig:plot_F}
\end{center}
\end{figure}

We finally discuss the $\td{\phi}_2$ field, whose IR behavior is more subtle as the $r$ and $\xi$ dependence cannot be factorized. Instead, the relevant ansatz is found to be of the form
\be
\label{IRansph2t} \td{\phi}_2(r,\xi) = F\left(\frac{\xi}{\b \log{r}}\right) \, ,
\ee
where $\b$ is defined in  \eqref{defbet} and the function $F$ obeys a second order non-linear ODE
\be
\label{ODE_F} F''(X) - F(X)\frac{F(X)^2-1}{X^2} + F'(X) = 0 \sp X\equiv \frac{\xi}{\b \log{r}} \, .
\ee
It can be checked numerically that  \eqref{ODE_F} admits a unique solution compatible with the boundary conditions
\be
\label{bcxF} F(0) = -1 \sp F(\infty) = 1 \, .
\ee
The plot of the corresponding solution is shown in figure \ref{fig:plot_F}

\subsection{Asymptotics at $\xi = 0$}

\label{Sec:asymx0}

We present here the asymptotics of the instanton solution near the center of the baryon at $\xi = 0$. We assume that the fields are regular at $\xi = 0$ and can be expanded as Taylor series. Then, solving the EoMs  \eqref{SUCSn}-\eqref{nabCSi3n} order by order in $\xi$ with the condition that the gauge fields  \eqref{ansatzSU2i}-\eqref{ansatzU1} are well defined at $\xi = 0$, yields the following asymptotics up to order $\mathcal{O}(\xi^4)$
\be
\label{PhiX0} \Phi(r,\xi) = \bar{\Phi}_0(r) + \bar{\Phi}_2(r)\xi^2 + \bar{\Phi}_4(r) \xi^4 + \mathcal{O}(\xi^6) \, ,
\ee
\be
\label{phi1X0} \td{\phi}_1(r,\xi) = \bar{\phi}_{1,0}(r)\xi + \bar{\phi}_{1,3}(r)\xi^3 + \mathcal{O}(\xi^5) \, ,
\ee
\be
\label{phi2X0} \td{\phi}_2(r,\xi) = -1 + \bar{\phi}_{2,2}(r)\xi^2 + \bar{\phi}_{2,4}(r)\xi^4 + \mathcal{O}(\xi^6) \, ,
\ee
\be
\label{AxX0} \td{A}_{\xi}(r,\xi) = \bar{A}_{\xi,0}(r) + \bar{A}_{\xi,2}(r)\xi^2 + \bar{A}_{\xi,4}(r) \xi^4 + \mathcal{O}(\xi^6) \, ,
\ee
\be
\label{ArX0} \td{A}_{r}(r,\xi) = \bar{A}_{r,1}(r)\xi + \bar{A}_{r,3}(r)\xi^3 + \mathcal{O}(\xi^5) \, ,
\ee
where $\Phi, \td{\phi}_2$ and $\td{A}_{\xi}$ are even and $\td{\phi}_1$ and $\td{A}_r$ are odd. All the coefficients of the expansions are expressed in terms of the 4 functions $\bar{\Phi}_0, \bar{\phi}_{1,0}, \bar{\phi}_{2,0}$ and $\bar{A}_{r,0}$. Note that, as in the $\xi\to\infty$ limit, the UV boundary conditions of Table \ref{tab:bcsLg} imply that the boundary values of the coefficients $\bar{\phi}_{1,n}, \bar{\phi}_{2,n}$ and $\bar{A}_{\xi,n}$ can all be written explicitly in terms of the asymptotics of $\theta(0,\xi)$ at $\xi\to 0$.

\newpage

\end{document}